\documentclass[10pt,aps,prd,twocolumn,floats,floatfix,showpacs,superscriptaddress,nofootinbib]{revtex4-2}
\usepackage[utf8]{inputenc}               		

\usepackage{graphicx,mathtools,amssymb,amsmath,amsthm,amsfonts,epsfig,epsf}
\usepackage[outdir=./]{epstopdf}
\usepackage[caption=false]{subfig}
\usepackage[dvipsnames]{xcolor}
\usepackage[linktocpage,colorlinks=true,allcolors=blue]{hyperref}  %
\hypersetup{
	pdfstartview=FitH,       
	linkcolor=blue,         
	citecolor=BrickRed,          
	urlcolor=blue,           
}

\hypersetup{pdfstartview=}
\usepackage{tensor}	
\usepackage{csquotes}
\usepackage{mathrsfs}
\usepackage[caption=false]{subfig}

\begin{document}

\title{\boldmath Self-resonance preheating in deformed attractor models: oscillon formation and evolution}

\author{Bao-Min Gu}
\affiliation{Department of physics, Nanchang University,
	Nanchang 330031, China}
\affiliation{Center for Relativistic Astrophysics and High Energy Physics, Nanchang University, Nanchang, 330031, China}

\author{Yu-Peng Zhang}
\email{zhangyupeng@lzu.edu.cn}
\thanks{Contact author.}
\affiliation{Key Laboratory of Quantum Theory and Applications of MoE, Lanzhou Center for Theoretical Physics, \\Key Laboratory of Theoretical Physics of Gansu Province, \\Gansu Provincial Research Center for Basic Disciplines of Quantum Physics, Lanzhou University, Lanzhou 730000, China}
\affiliation{Institute of Theoretical Physics, \&  Research Center of Gravitation, Lanzhou University, Lanzhou 730000, China}

\author{Fu-Wen Shu}
\email{shufuwen@ncu.edu.cn}
\thanks{Contact author.}
\affiliation{Department of physics, Nanchang University, Nanchang 330031, China}
\affiliation{Center for Relativistic Astrophysics and High Energy Physics, Nanchang University, Nanchang, 330031, China}

\author{Yu-Xiao Liu}
\email{liuyx@lzu.edu.cn}
\thanks{Contact author.}
\affiliation{Key Laboratory of Quantum Theory and Applications of MoE, Lanzhou Center for Theoretical Physics, \\Key Laboratory of Theoretical Physics of Gansu Province, \\Gansu Provincial Research Center for Basic Disciplines of Quantum Physics, Lanzhou University, Lanzhou 730000, China}
\affiliation{Institute of Theoretical Physics, \&  Research Center of Gravitation, Lanzhou University, Lanzhou 730000, China}

\begin{abstract}
It is well known that, in potentials that are quadratic near the minimum but shallower away—such as small-$\alpha$ ($\alpha \ll M_P^2$) attractors, the inflaton condensate fragments into localized compact objects known as oscillons during self-resonance preheating. In this work, we investigate the self-resonance in deformed $\alpha$-attractor T-model with a Gaussian feature near the minimum, distant from inflation's end. Linear analysis reveals altered resonance bands and deformed Floquet charts dependent on feature parameters, primarily through the localized modification of $V''(\phi)$. In fully nonlinear lattice simulations, we find that the potential feature has only a modest influence on gradient-energy transfer during resonance, whereas the subsequent evolution becomes strongly dependent on the Gaussian feature’s dimensionless amplitude $h$. Increasing $|h|$ suppresses the formation and persistence of compact oscillons, resulting in fewer objects and a smaller fraction of energy stored in them. The total-energy and internal-gradient-energy diagnostics give consistent lifetime trends. Within the directly resolved range, the oscillon lifetime decreases nonlinearly with increasing $|h|$, with negative $h$ generally producing shorter lifetimes than positive $h$ of the same magnitude. The gravitational wave emission is dominated by the resonant stage and becomes strongly suppressed after oscillon formation. We further find an enhancement in the high-frequency tail of the spectrum, sourced by the potential feature. Although a complete reheating description requires external couplings and higher-resolution simulations, clear qualitative differences in the post-inflationary expansion history already emerge within our simulated time window. These results highlight the important role of potential features in shaping reheating dynamics and their cosmological implications, and provide a deeper understanding of preheating dynamics and the properties of oscillons.
\end{abstract}

\maketitle

\section{Introduction}
Inflationary cosmology~\cite{Starobinsky:1980te,Guth:1980zm,Linde:1981mu,Albrecht:1982wi,Linde:1983gd} has emerged as a compelling framework for the origin of the universe and has achieved remarkable success to date. Although a direct detection of primordial gravitational waves—the definitive smoking-gun signature of inflation—remains elusive, the paradigm is strongly supported by a broad range of observations and is widely regarded as the standard cosmological scenario. In addition to resolving several conceptual shortcomings of the conventional hot Big Bang model, inflation predicts quantum vacuum fluctuations that naturally account for the small-scale anisotropies observed in the cosmic microwave background and provide the primordial seeds for large-scale structure formation in the late universe. Recent observations from the Planck satellite~\cite{Planck:2018vyg,Planck:2018jri} and the Atacama Cosmology Telescope (ACT)~\cite{ACT:2025fju,ACT:2025tim} offer strong and independent support for the inflationary paradigm. 

Inflation posits that the early universe underwent a phase of quasi–exponential expansion, which in its simplest realization can be described by a canonical single scalar field minimally coupled to gravity within general relativity. During inflation, the inflaton field slowly rolls from a relatively flat region of its potential toward the potential minimum. Throughout this phase, the energy density of the universe is dominated by the inflaton potential, yielding an equation-of-state (EoS) parameter close to $-1$, analogous to the accelerated expansion driven by dark energy of the present universe. To successfully transition to the subsequent radiation dominated era and enable Big-Bang nucleosynthesis (BBN), the energy stored in the inflaton potential must be transferred to relativistic particles through an intermediate stage known as reheating. 

Early studies proposed perturbative decay as the mechanism for reheating~\cite{Abbott:1982hn,Dolgov:1989us,Traschen:1990sw}. However, this description is valid only in the regime of weak couplings and, moreover, violates the fluctuation dissipation theorem~\cite{Boyanovsky:1993xf,Boyanovsky:1994me}. Consequently, perturbative decay alone is now understood to be insufficient for a complete description of the reheating process, which is highly non-perturbative. In contrast, the subsequently proposed mechanism of parametric resonance provides a more natural and successful description of particle production~\cite{Kofman:1994rk,Shtanov1994ce,Kofman:1997yn}. In the presence of external coupling(s), the matter field(s) remains negligible in the initial stage of parametric resonance and the inflaton undergoes coherent oscillations with decaying amplitude around the minimum of its potential. During the inflaton oscillations, the Fourier modes of the matter fields experience a time-dependent effective frequency induced by their coupling to the inflaton. For suitable inflaton dynamics and model parameters, the evolution of this effective frequency violates the adiabaticity condition, leading to exponential amplification of the matter field modes and consequently, explosive particle production. Once the occupation number of the matter field grows sufficiently large, its backreaction on the background becomes important and eventually shuts off the resonance. Interestingly, preheating can occur even in the absence of external couplings. In this case however, the inflaton does not decay into new particle species; instead, the inflaton transfers energy into its own fluctuations through nonlinear self-interactions, resulting in the formation of spatially localized, dense, and long-lived oscillons~\cite{Gleiser:1993pt,Amin:2010xe,Amin:2010dc,Amin:2011hj}. This mechanism is known as self-resonance. For reviews of reheating, see Refs.~\cite{Bassett:2005xm,Allahverdi:2010xz,Lozanov:2019jxc,Amin:2014eta}.

In recent years, it has been established that oscillon formation requires the inflationary potential to satisfy certain conditions: it should be approximately quadratic around the minimum, while becoming shallower than quadratic away from the minimum; including gravitational effects would further enhance the formation efficiency~\cite{Kou:2019bbc,Aurrekoetxea:2023jwd}. It has been demonstrated that a broad class of inflationary models, such as monodromy inflation~\cite{Amin:2011hj,Zhou:2013tsa,Levkov:2023ncb,Jia:2024fmo}, string-inspired models~\cite{Liu:2017hua,Antusch:2017flz,Kasuya:2020szy}, $\alpha$-attractors~\cite{Lozanov:2017hjm,Antusch:2017flz,Zhang:2020ntm,Mahbub:2023faw,Shafi:2024jig}, hilltop~\cite{Antusch:2015ziz,Antusch:2019qrr}, and non-canonical scenarios~\cite{Amin:2013ika,Sang:2020kpd}, can give rise to oscillon formation. Since oscillon formation and their characteristic properties vary among different models, studying oscillon would provide valuable insight into the dynamics of the reheating process of these models. Oscillons are soliton-like, non-topological defect configurations that are not protected by any symmetry and are therefore metastable. Nevertheless, their lifetimes can be extremely long. Oscillons have attracted broad interest for several reasons. In cosmology, oscillons behave effectively as non-relativistic matter with EoS close to zero, hence they decay slower than radiation. During preheating, the copious production of oscillons can drive the universe into a transient early matter-dominated (eMD) phase, provided that their lifetimes are sufficiently long. Such an eMD phase would have significant impact on the expansion history of the universe and thus the inflationary observables~\cite{Lozanov:2016hid,Antusch:2020iyq,Soman:2024zor,Antusch:2025ewc}. Moreover, because oscillons are spatially localized configurations, they carry considerable gradient energy with spatial anisotropies. The formation and decay of oscillons may act as an efficient source of a stochastic gravitational-wave background~\cite{Zhou:2013tsa,Antusch:2017vga,Figueroa:2017vfa,Liu:2018rrt,Amin:2018xfe,Lozanov:2019ylm,Sang:2019ndv,Li:2020qnk,Hiramatsu:2020obh,Kou:2021bij,Krajewski:2022ezo,Lozanov:2022yoy,Sui:2024grm}. 
Recent studies have further suggested that oscillons may collapse into primordial black holes (PBHs)~\cite{Cotner:2018vug,Cotner:2019ykd,Muia:2019coe,Kou:2019bbc,Nazari:2020fmk,Padilla:2024iyr,Kasai:2025coe}. If sufficiently long-lived, oscillons may also constitute viable dark matter candidates~\cite{Garcia:2022vwm,vanDissel:2023zva}. In field theory and particle physics, oscillons provide a paradigmatic example of long-lived, localized excitation in strongly nonlinear, non-equilibrium field theories, offering insights into energy localization, slow thermalization, and the emergence of effective particle-like degrees of freedom across high-energy, nonlinear, and condensed-matter physics \cite{Gleiser:2006te,Fodor:2019ftc,OlleAguilera:2022ewm,Zhou:2024mea}. These considerations motivated widespread interest in studying the formation mechanism and dynamical properties of oscillons, such as shape, decay, and lifetime
~\cite{Gleiser:2008ty,Fodor:2008du,Gleiser:2009ys,Hertzberg:2010yz,Amin:2010dc,Mukaida:2016hwd,Ibe:2019vyo,Amin:2019ums,Olle:2019kbo,Antusch:2019qrr,Zhang:2020bec,Zhang:2020ntm,Olle:2020qqy,Cyncynates:2021rtf,Piani:2023aof,Piani:2025dpy,Drees:2025iue,Li:2025ioq,vanDissel:2025xqn,Li:2025xtf,Evslin:2025hjt}.

In this paper, we study the preheating process realized by self-resonance of deformed T-model $\alpha$-attractors. The model consists of a standard T-model potential supplemented by a Gaussian-type feature. As an exploratory study, this work is motivated by two main aspects. First, inflationary models with features in the potential have attracted growing interest, as they naturally lead to rich and nontrivial cosmological dynamics. In particular, suitably chosen features during inflation can substantially enhance curvature perturbations, opening a viable window for PBH production~\cite{Garcia-Bellido:2017mdw,Ezquiaga:2017fvi,Motohashi:2017kbs,Kannike:2017bxn,Ballesteros:2017fsr,Ozsoy:2018flq,Cicoli:2018asa,Mishra:2019pzq,Inomata:2021uqj,Inomata:2021tpx,Karam:2022nym,Gu:2022pbo,Gu:2023mmd}. Different classes of features can generate PBHs across a broad mass range, from asteroid masses to stellar scales, with correspondingly distinct mass functions and abundances. While current observational bounds severely constrain scenarios in which PBHs make up the entirety of dark matter~\cite{Khlopov:2008qy,Carr:2009jm,Carr:2017jsz,Sasaki:2018dmp,Green:2020jor,Carr:2020xqk,Carr:2020gox,Villanueva:2021spv,Auffinger:2022khh,Escriva:2022duf}, PBHs remain viable as a subdominant dark matter component and as possible progenitors of black holes involved in certain compact binary mergers~\cite{Sasaki:2016jop,Vaskonen:2019jpv,Clesse:2020ghq,DeLuca:2020qqa,DeLuca:2020sae,Andres:2024wqk,
Yuan:2025avq,DeLuca:2025fln}. 
Although the Gaussian feature is introduced phenomenologically, it should be regarded as a minimal smooth template for a localized deformation of the potential, rather than as a unique microscopic prediction. Such a deformation can arise when heavy degrees of freedom coupled to the inflaton are integrated out. For instance, a heavy field with an inflaton-dependent mass generates a Coleman--Weinberg correction~\cite{Coleman:1973jx}, which becomes localized if the heavy-particle spectrum changes appreciably only over a finite interval in field space. A concrete realization of this mechanism was recently constructed in an $\alpha$-attractor model, where a simple polynomial dependence of a heavy-scalar mass on the noncanonical field is mapped, through the kinetic-pole field redefinition, into a quantum correction localized near the minimum of the canonical inflaton potential~\cite{Inada:2026feq}. Although the resulting profile is not Gaussian, it explicitly demonstrates that localized deformations capable of modifying inflaton self-resonance can originate from heavy ultraviolet degrees of freedom. Non-perturbative effects in axion and axion-monodromy models provide another well-motivated source of feature structure in the inflaton potential, typically in the form of periodic modulations. Although such periodic wiggles differ from the single localized Gaussian profile considered here, they illustrate that smooth inflationary potentials can naturally acquire subleading field-dependent structures from ultraviolet physics~\cite{Flauger:2009ab}. Motivated by these considerations, we undertake an exploratory study of the consequences of placing the potential feature in the post-inflationary reheating phase rather than during inflation, and investigate how it affects the details of the reheating process, including its dynamics and energy transfer. Reheating dynamics in the presence of features in the potential were previously studied in reference~\cite{Saha:2024lil}, where a quadratic base potential supplemented by a quadratic-quadratic type of inflaton-daughter coupling was considered. It was shown that the reheating process can be substantially modified, leading to a number of nontrivial and interesting results; for details, we refer the reader to Ref.~\cite{Saha:2024lil}. The second motivation of this work is related to oscillon physics. Despite the new constraints from the latest ACT observations~\cite{ACT:2025fju,ACT:2025tim}, $\alpha$-attractor models remain an attractive and well-motivated class of inflationary theories. Previous studies have shown that oscillon formation during reheating occurs for small $\alpha$, but is absent for large $\alpha$~\cite{Hasegawa:2017iay,Kim:2021ipz}. 
In this work, we concentrate on the small-$\alpha$ ($\alpha\ll M_P^2$) regime and explore how a localized feature in the potential affects oscillon formation and subsequent dynamics. Taking $h$ as the principal deformation parameter, we systematically investigate how the sign and strength of the feature modify inflaton self-resonance, fragmentation, and the properties and evolution of the resulting oscillons. The location and width parameters are kept fixed in the main analysis, with their effects briefly discussed in the appendix~\ref{sigma_phis}.

\begin{figure*}[t]
	\centering
	\includegraphics[width=7cm]{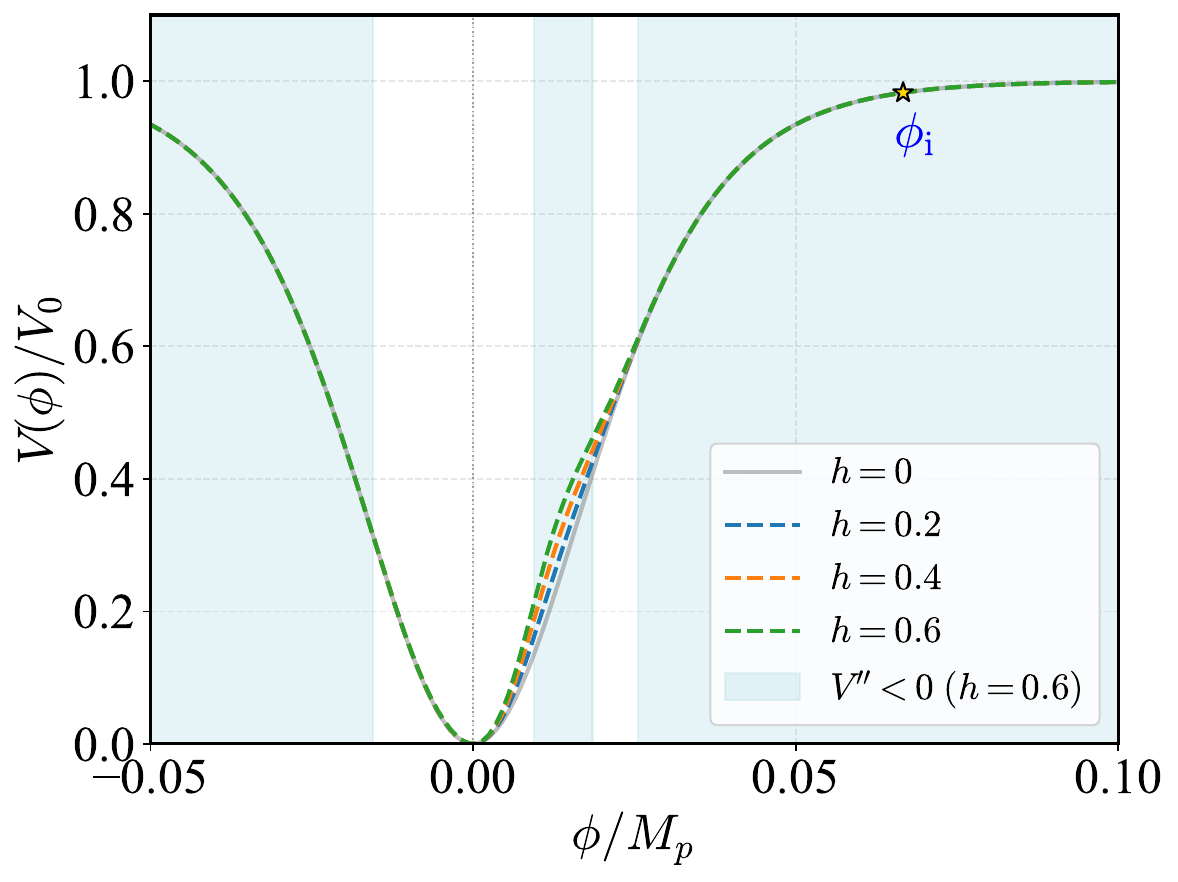}
	\includegraphics[width=7cm]{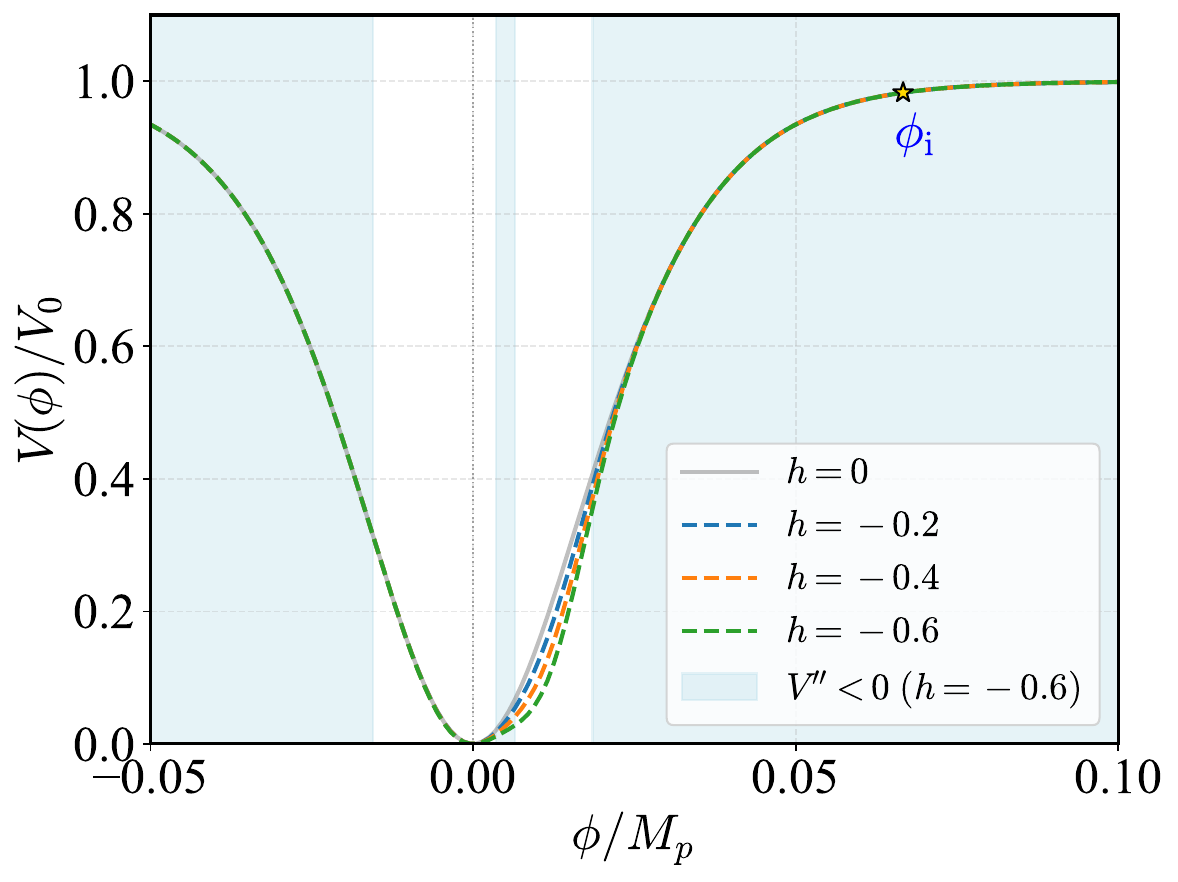}
	\caption{Plots of the deformed T-model potential for the parameters in table~\ref{tab1}. The end of inflation $\phi_{\mathrm{i}}$ is defined by $\epsilon_V(\phi_{\mathrm{i}})=1$. The blue shaded regions are those with $V''<0$ indicating the field space where tachyonic instability may occur.}
	\label{plot_potential}
\end{figure*}

This paper is organized as follows. In section~\ref{sec1}, we present the inflationary dynamics and introduce the deformed T-model together with the model setup. In section~\ref{sec2}, we investigate parametric resonance within the framework of linear perturbation theory. In particular we show the results of Floquet analysis. In section~\ref{sec3}, we show the main results of this paper, including the lattice simulations, the oscillon dynamics and properties, and a brief discussion on the implication for the evolution of the universe. The conclusions and outlook are presented in section~\ref{sec4}. Supplementary analyses are provided in appendices~\ref{appsec}--\ref{app_threshold}.

\section{The Inflationary Dynamics}\label{sec1}

To study the post-inflationary dynamics, we consider general relativity minimally coupled to a canonical scalar field, the inflaton, described by the action
\begin{align}
	S=\int \mathrm{d}^4 x \sqrt{-g} \left(\frac{M_P^2}{2}R-\frac{1}{2}(\partial\phi)^2-V(\phi)\right).
\end{align}
We work in a spatially flat Friedmann–Robertson–Walker background,
\begin{equation}
	\mathrm{d}s^2 = -\mathrm{d}t^2 +a^2(t) \delta_{ij}\mathrm{d}x^i \mathrm{d}x^j,
\end{equation}
with $a(t)$ the scale factor. Under this ansatz, the background evolution then obeys
\begin{eqnarray}
	\ddot{\phi}+3H\dot{\phi}+V_\phi &=& 0, \label{phi_EoM}\\
\frac{1}{2}\dot{\phi}^2 +V(\phi) &=& 3M_P^2 H^2, \label{Fried_eq} \\
-\frac{1}{2}\dot{\phi}^2 &=& M_P^2\dot{H}, \label{acc_eq}
\end{eqnarray}
where dot denotes the derivative with respect to $t$, $H=\dot{a}/a$ is the Hubble parameter. During slow-roll inflation, the scale factor grows exponentially and the Hubble parameter $H$ is nearly a constant, leading to a quasi-de Sitter period of expansion. This can be realized if the slow-roll parameters are small, i.e., 
\begin{eqnarray}
	\epsilon &\equiv& -\frac{\dot{H}}{H^2}\simeq \epsilon_V \equiv \frac{1}{2}M_P^2\left(\frac{V_\phi}{V}\right)^2 \ll 1, \\
	\eta &\equiv& \frac{\dot{\epsilon}}{H\epsilon}\simeq \eta_V \equiv \frac{V_{\phi\phi}}{V} \ll 1.
\end{eqnarray} 
Using the slow-roll parameter $\epsilon_V$, the duration of inflation can be calculated by
\begin{align}
	N_{\text{tot}} = \int_{t_{\text{e}}}^{t_*} H \mathrm{d}t \simeq 
	\int_{\phi_{\mathrm{e}}}^{\phi_*}  \frac{\mathrm{d}\phi}{\sqrt{\epsilon_V}},   
\end{align} 
where the subscripts ``*'' and ``e'' mark the horizon exit of the CMB scales ($k_*=0.05~\mathrm{Mpc}^{-1}$) and the end of inflation, respectively.
Typically, the total duration of inflation is required to satisfy $N_{\text{tot}}\gtrsim 50$. 
This lower bound arises from two considerations: (i) resolving the horizon, flatness, and other puzzles of the hot big bang scenario, and (ii) satisfying the constraints on the scalar spectral index and the tensor-to-scalar ratio from CMB observations, which generally favor $N_{\text{tot}}\gtrsim 50$. In addition, the amplitude of the scalar power spectrum $A_\mathrm{s}$ and the spectral index $n_{\mathrm{s}}$ are severely constrained by the CMB observations \cite{Planck:2018vyg, Planck:2018jri, ACT:2025fju, ACT:2025tim}. They are computed by
\begin{align}
	A_\mathrm{s} &= \frac{H_*^2}{8\pi^2 \epsilon_*} \simeq \frac{1}{12\pi^2}\frac{V^3(\phi_*)}{V_{\phi}^2 (\phi_*)}, 
	\\
	n_{\mathrm{s}} &= 1-6\epsilon_V(\phi_*)+2\eta_V(\phi_*).
\end{align}
These constraints restrict the model parameters as well as the field values at horizon exit and at the end of inflation.

The model considered in this work has the form 
\begin{equation}
V(\phi)=V_b(\phi) 
\left[1+h \exp\left(-\frac{(\phi-\phi_\text{s})^2}{2\sigma^2}\right)\right], \label{deformed_model}
\end{equation}
where the parameters $h$, $\phi_\text{s}$, and $\sigma$ characterize the height, the position, and the width of the feature, respectively. The potential $V_b(\phi)$ is chosen to be consistent with observations. We restrict the height $h$ and the width $\sigma$ to be small enough and the position $\phi_{\mathrm{s}}$ to be after the end of inflation, such that the CMB observations are well described by the base potential $V_{\mathrm{b}}(\phi)$. In this work we consider the T-model of $\alpha$-attractors as the base potential,
\begin{align}
	V_{\mathrm{b}}(\phi)=V_0 \tanh^{2}\left(\frac{\phi}{\sqrt{6\alpha}}\right).
	\label{T_model_pot}
\end{align}
We show the plot of this potential in Fig.~\ref{plot_potential}. 
The parameter sets used in this paper are listed in table~\ref{tab1}. They are in good agreement with latest observational constraints, as can be seen in appendix~\ref{appsec}.

\begin{table}[t]
	\centering
	\setlength{\tabcolsep}{3pt}
	\begin{tabular}{c | c c   c  c  c}
		\hline
		\hline
		Sets & $V_0~[M_p^4]$ & $\alpha~[M_p^2]$ & $h$ & $\sigma~[M_p]$  & $\phi_s~[M_p]$ \\
		\hline
		$P_0$ &$1.14\times 10^{-14}$ & $10^{-4}$ & $ 0 $ & $0.006$ & $0.008$ \\
		$P_1$ &$1.14\times 10^{-14}$ & $10^{-4}$ & $0<h\leqslant0.6$ & $0.006$ & $0.008$ \\
		$P_2$ &$1.14\times 10^{-14}$ & $10^{-4}$ & $-0.6 \leqslant h<0$ & $0.006$ & $0.008$ \\
		\hline
		\hline
	\end{tabular}
	\caption{The parameter sets used in this paper. In the main text, we focus on the feature amplitude $h$, while the effects of the width and position parameters are briefly discussed in the appendix~\ref{sigma_phis}. The feature is centered at $\phi_{\mathrm{s}}=0.008\,M_P$, between the end of inflation value $\phi_{\mathrm{end}}=0.06665\,M_P$ and the potential minimum at $\phi=0$. The inflaton first crosses the feature while rolling toward the minimum and traverses it repeatedly during the first few oscillations.}\label{tab1}
\end{table}

\section{Perturbative approach of preheating}\label{sec2}
\subsection{Perturbative equations}
We first study the self-resonance perturbatively. 
The inflaton field can be decomposed into a combination of homogeneous background field and a fluctuation part, $\phi(t,\vec{x}) = \phi(t) + \delta\phi(t,\vec{x})$. The homogeneous field $\phi(t)$ obeys equation~(\ref{phi_EoM}). The fluctuation part can be analyzed more conveniently in Fourier space, by expanding it as 
\begin{equation}
	\delta\phi(t,\vec{x}) = \int \frac{d^3k}{(2\pi)^3} \delta\phi_k(t) e^{-i\vec{k}\cdot\vec{x}}.
\end{equation}
The Fourier mode $\delta\phi_k(t)$ then satisfies 
\begin{equation}
	\delta\ddot{\phi}_k + 3H \delta\dot{\phi}_k + \left(\frac{k^2}{a^2} + V''(\phi(t))\right) \delta\phi_k =0,
	\label{pert_EoM}
\end{equation}
where $V''(\phi(t)) = \partial^2V(\phi(t))/\partial\phi(t)^2$. Note that the metric perturbations are neglected. During the early stages of preheating, the fluctuation $\delta\phi(t,\vec{x})$ remains small, i.e. $\delta\phi(t,\vec{x})\ll\phi(t)$. The dynamics of preheating can be well described by equation~(\ref{pert_EoM}). As preheating proceeds, the energy of the inflaton field is continuously transferred to the fluctuation modes, which—under suitable conditions—can be resonantly amplified. Once the fluctuations are amplified to amplitudes comparable to the background, the perturbative expansion would breakdown and the backreaction becomes important. At this stage, equation~(\ref{pert_EoM}) no longer provides an accurate description of preheating, which must instead be studied by fully nonlinear methods. Nevertheless, in this section we employ a perturbative approach to gain qualitative insight into the dynamics of preheating, while a more precise and detailed nonlinear analysis will be presented in section~\ref{sec3}.

\begin{figure}[t]
	\centering
	\includegraphics[width=7cm]{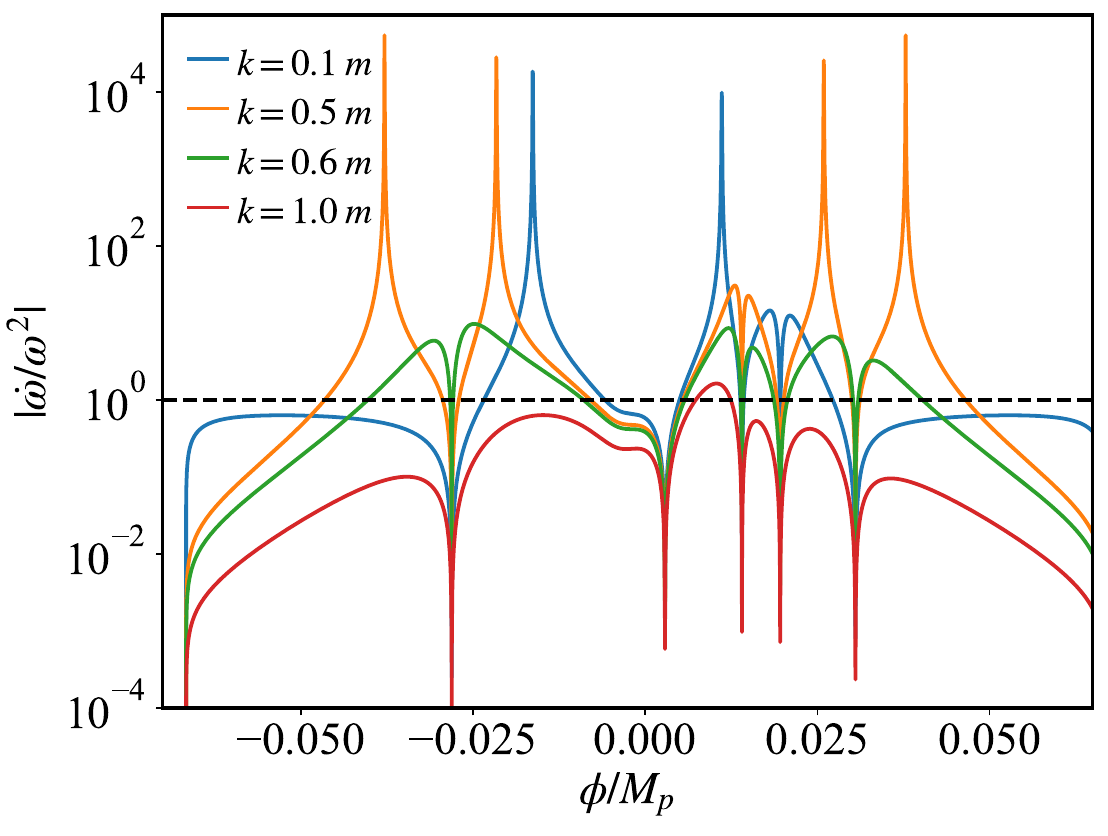}
	\includegraphics[width=7cm]{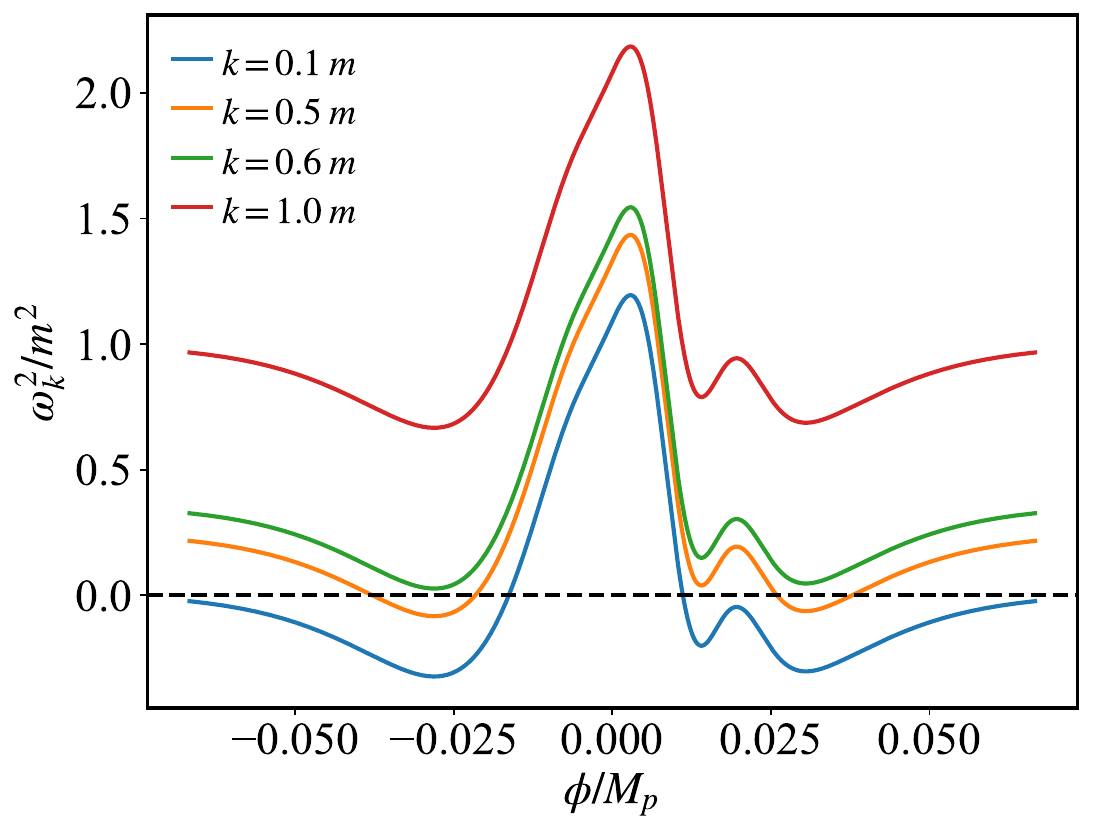}
	\caption{The violation of adiabatic condition for parametric resonance and the tachyonic instability condition for deformed T-model ($h=0.2$).}
	\label{T_Floquet_condi}
\end{figure}

\begin{figure*}[t]
	\centering
	\includegraphics[width=14cm, trim={0 0 0 0}, clip]{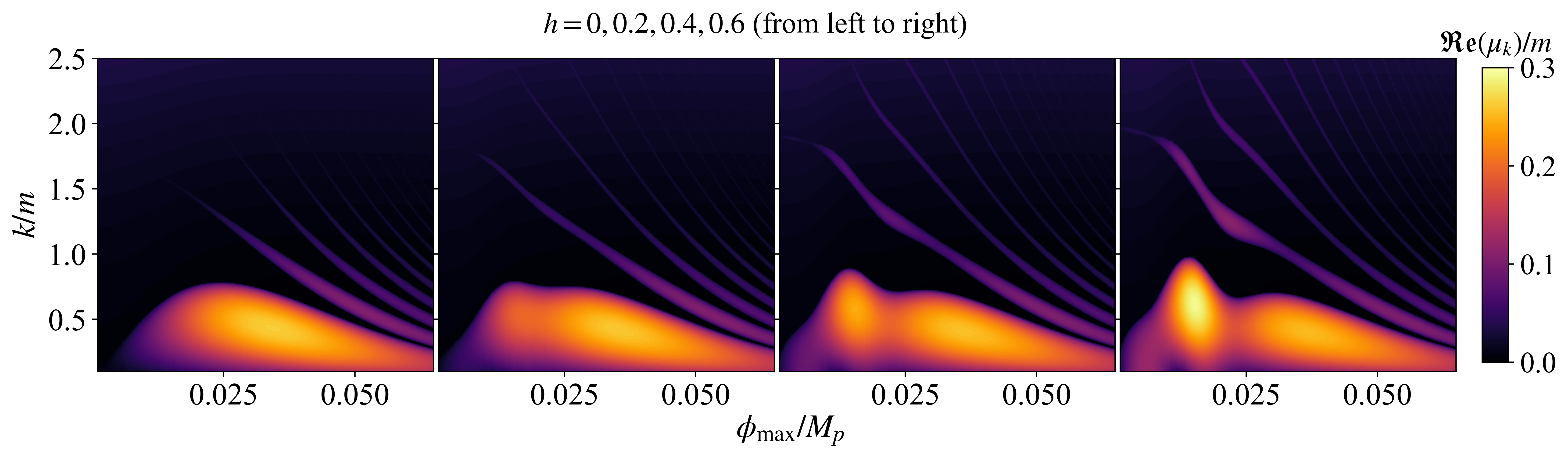}
	\includegraphics[width=14cm, trim={0 0 0 0}, clip]{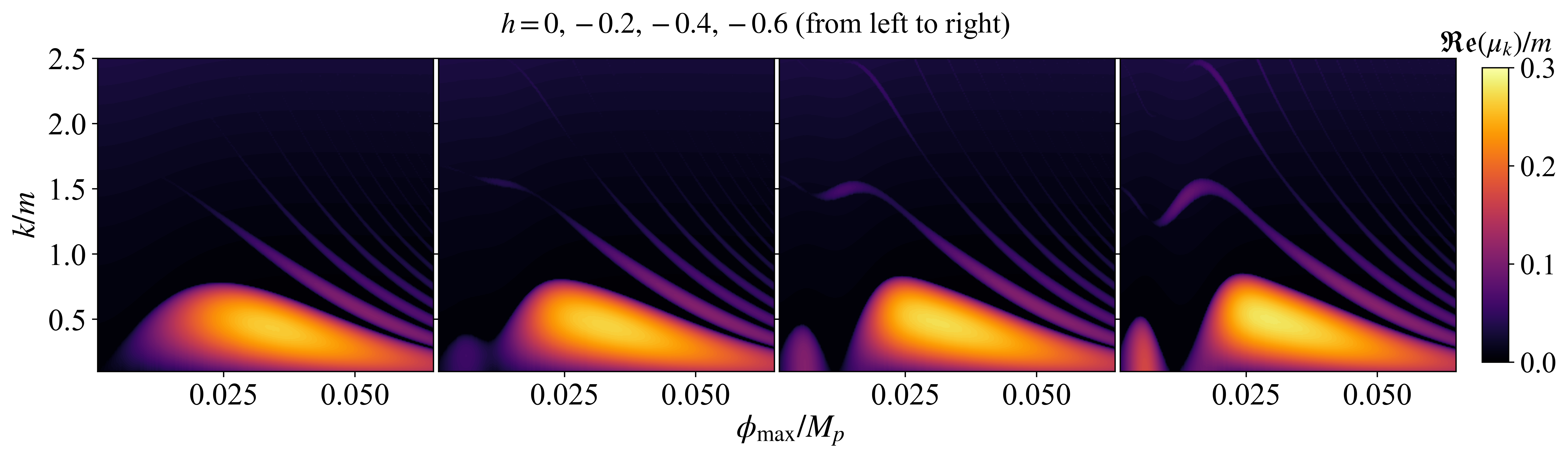}
	\includegraphics[width=6.3cm]{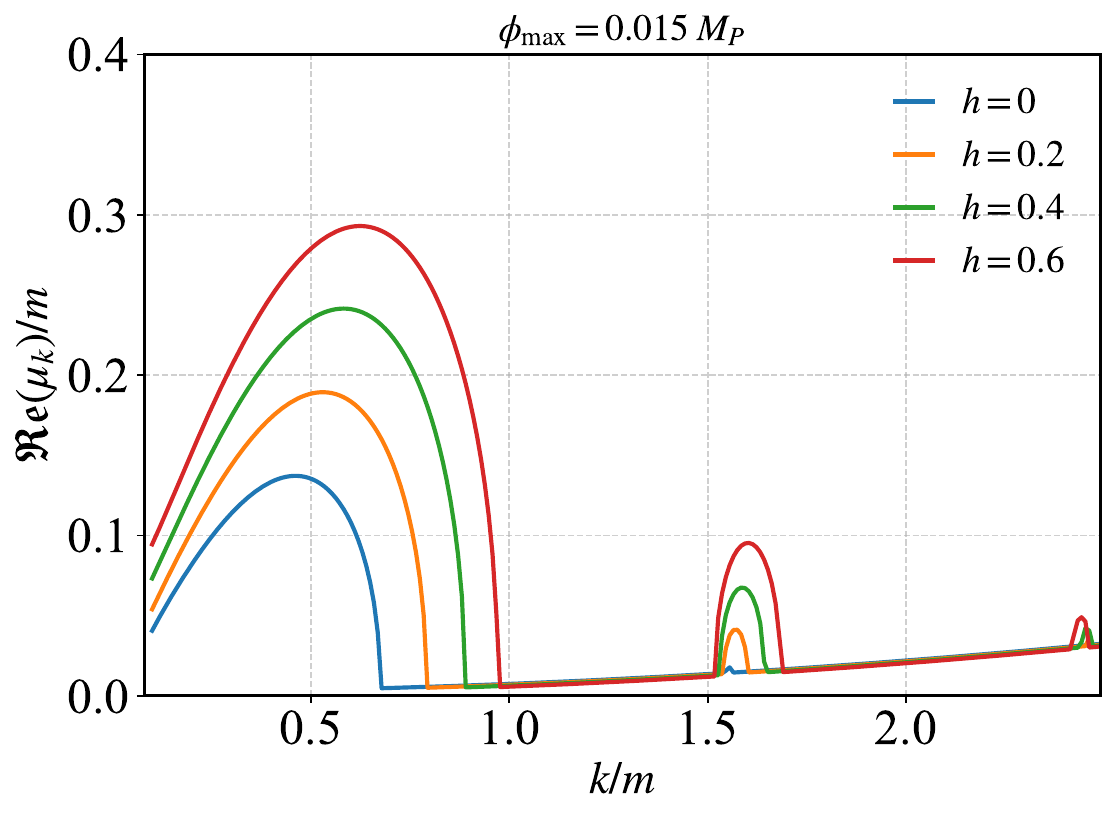}
	\includegraphics[width=6.3cm]{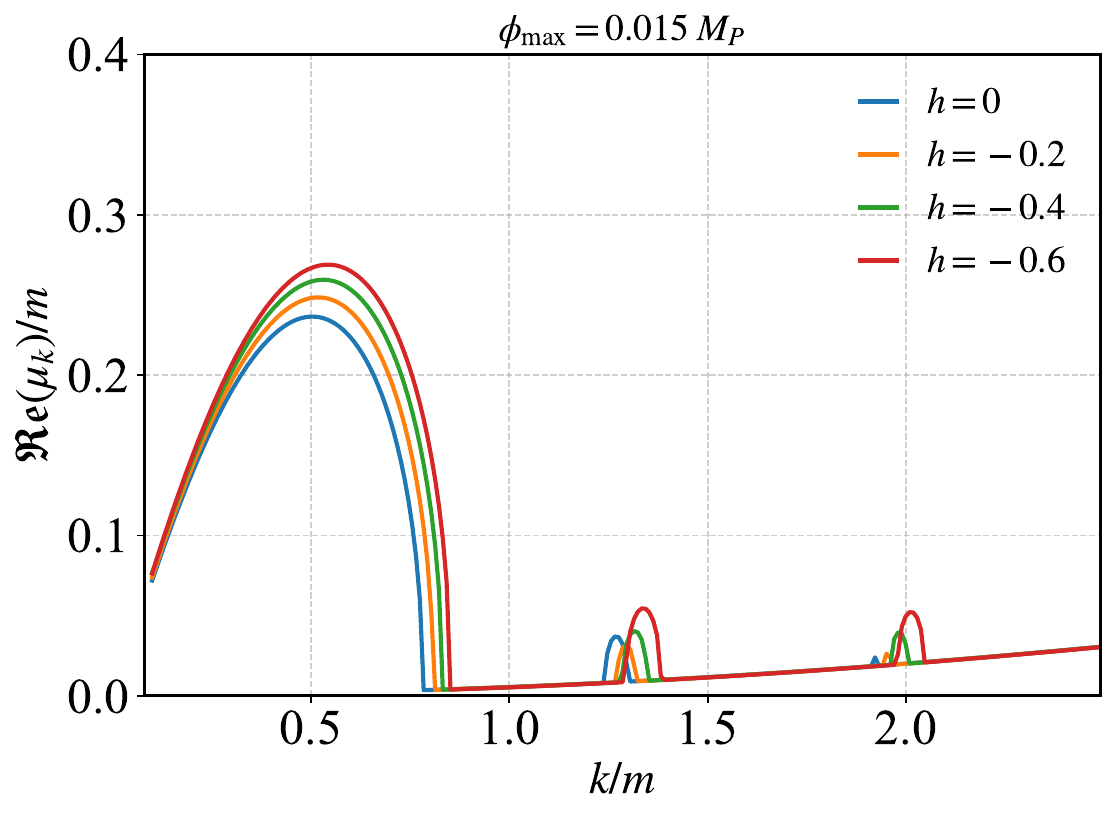}
	\includegraphics[width=6.3cm,trim={0 0 0.2cm 0}, clip]{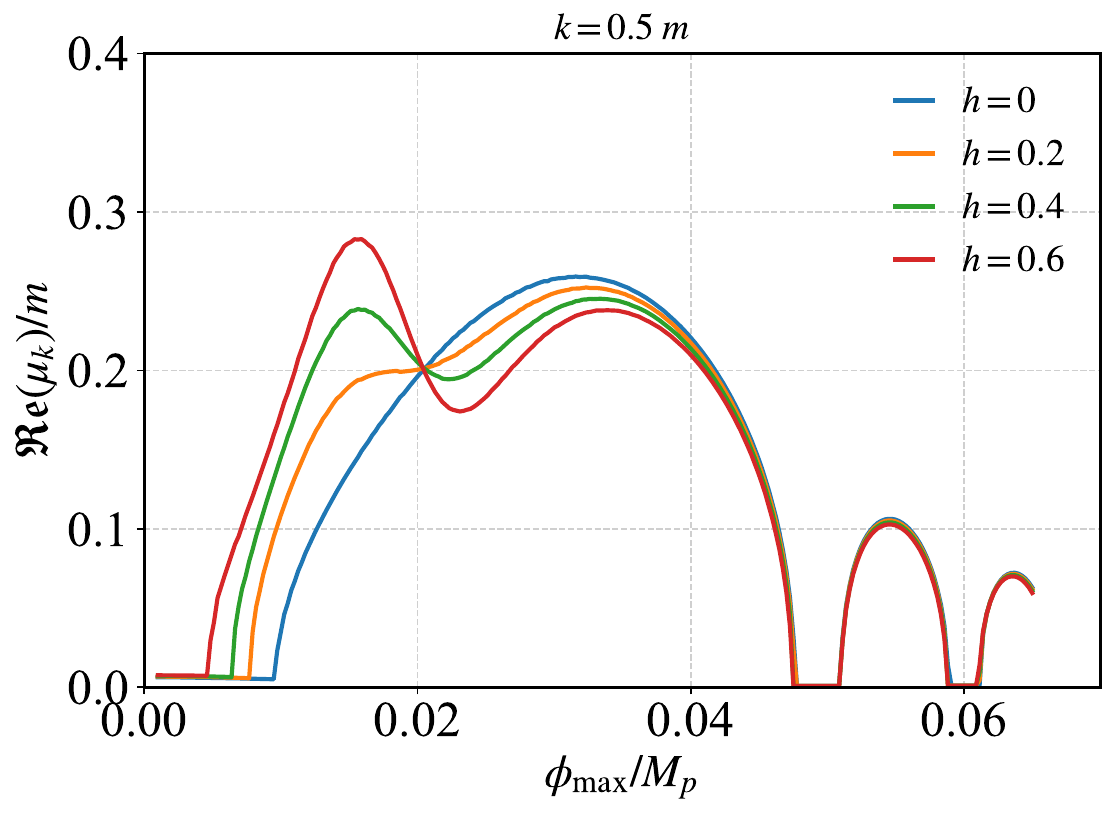}
	\includegraphics[width=6.3cm,trim={0 0 0.2cm 0}, clip]{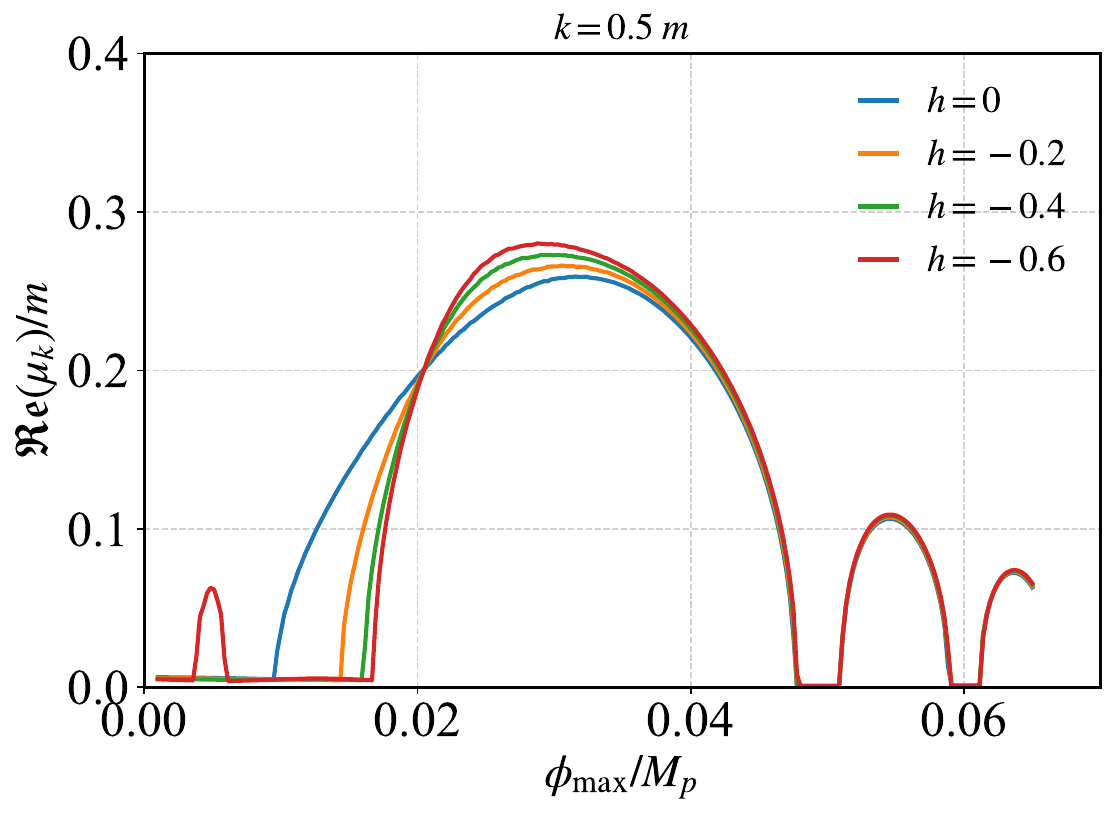}
	\caption{\textit{Upper two rows}: Floquet charts for the models with the $P_1$ and $P_2$ parameter sets. \textit{Lower two rows}: Floquet exponents of different modes at a fixed background oscillation amplitude $\phi_{\text{max}}=0.015\,M_P$, together with that of the mode $k=0.5\,m$ as a function of $\phi_{\text{max}}$. Here, $\phi_{\text{max}}$ denotes the positive turning point of the homogeneous background oscillation, and each Floquet exponent is calculated over one complete oscillation period. Cosmic expansion is neglected.
	}
	\label{Floquet_h}
\end{figure*}

\subsection{Floquet analysis}
Equation~(\ref{pert_EoM}) describes the evolution of $\delta\phi_k(t)$ with a 
time-dependent frequency defined by
\begin{equation}
	\omega_k^2(t) = \frac{k^2}{a^2} + V''(\phi(t)).
\end{equation}
Self-resonance arises when the fluctuation modes $\delta\phi_k(t)$ exhibit exponential growth over a broad range of wavenumbers. This phenomenon can be triggered by distinct mechanisms, most notably parametric resonance and tachyonic preheating. Parametric resonance occurs when the adiabaticity condition is violated, $|\dot{\omega}(t)/\omega^2(t)|>1$, signaling the breakdown of the WKB approximation~\cite{Kofman:1994rk,Kofman:1997yn}. Tachyonic preheating, on the other hand, takes place when the inflaton traverses a region of the potential with negative effective mass squared, i.e., $\omega^2(t)<0$~\cite{Kofman:1997yn,Felder:2000hj}. This behavior occurs in certain inflationary models, such as the hybrid inflation~\cite{Copeland:2002ku} and $\alpha$-attractors~\cite{Kallosh:2013hoa,Kallosh:2013yoa}. In Fig.~\ref{T_Floquet_condi}, it is evident that both parametric resonance and tachyonic instability coexist in the model under consideration.

Now let us study the solutions of equation~(\ref{pert_EoM}) in detail. According to the Floquet theorem~\cite{ASENS_1883_2_12__47_0}, the general solution of $\delta\phi_k(t)$ can be given by
\begin{equation}
	\delta\phi_k(t) = P_+(t) e^{\mu_k t} + P_-(t) e^{-\mu_k t},
	\label{Floquet_solution}
\end{equation}
where $P_{\pm}(t)$ are periodic functions with the frequency determined by $\partial^2 V(\phi(t))/\partial\phi^2$.
This can be understood by Taylor expanding the potential around the minimum and neglecting the terms higher than quadratic order of $\delta\phi$. The remaining mass term explains the oscillation nature of $P_{\pm}(t)$. The parameter $\mu_k$ is the Floquet exponent, which characterizes the growth rate of the fluctuations. For imaginary $\mu_k$, the fluctuation $\delta\phi_k$ simply oscillates periodically and there is no instability. For $\mathfrak{Re}(\mu_k)>0$, however, the corresponding fluctuation mode grows exponentially. The continuously distributed modes form instability bands in parameter space. The resonance is efficient if $\mathfrak{Re}(\mu_k)/H \gg 1$, for which a significant fraction of the inflaton energy is transferred into fluctuations. In general, the Floquet exponent $\mu_k$ can be evaluated numerically. The details are given in appendix~\ref{Floquet_computation}.

\begin{figure}[t]
	\centering
	\includegraphics[width=8.8cm, trim={0 0 0 0}, clip]{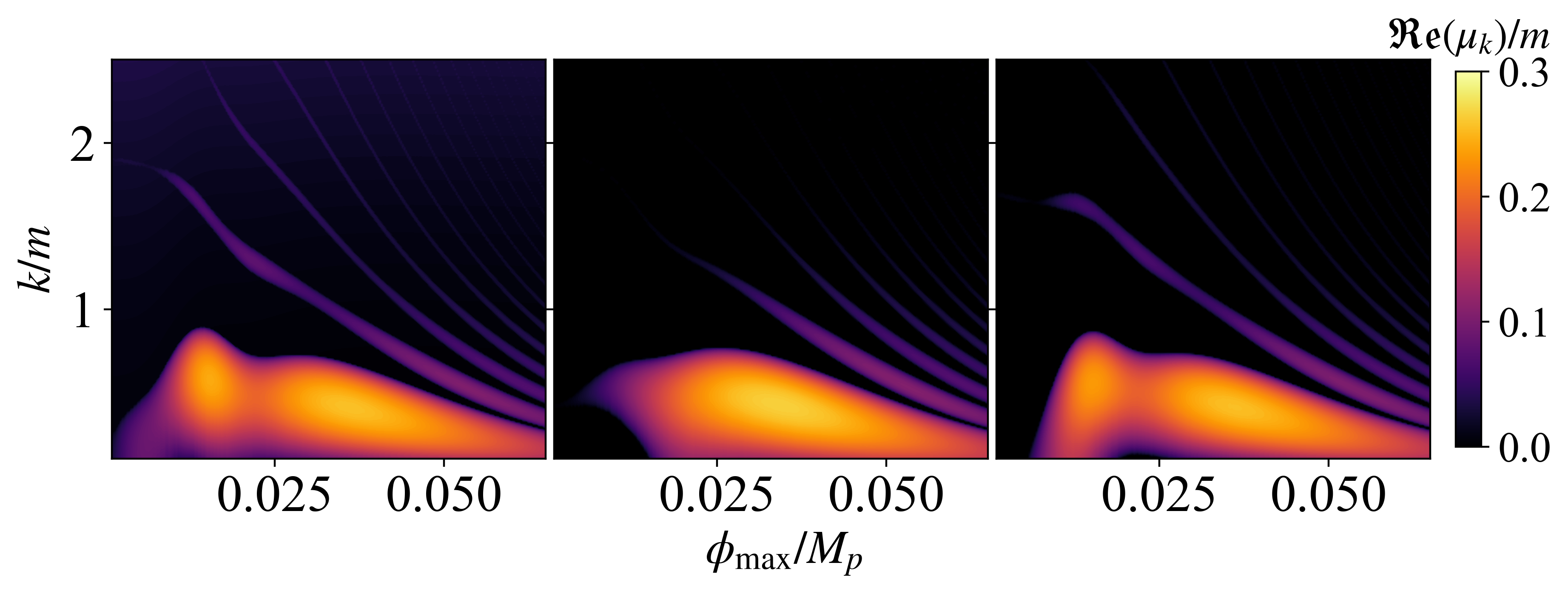}
	\includegraphics[width=8.8cm, trim={0 0 0 0}, clip]{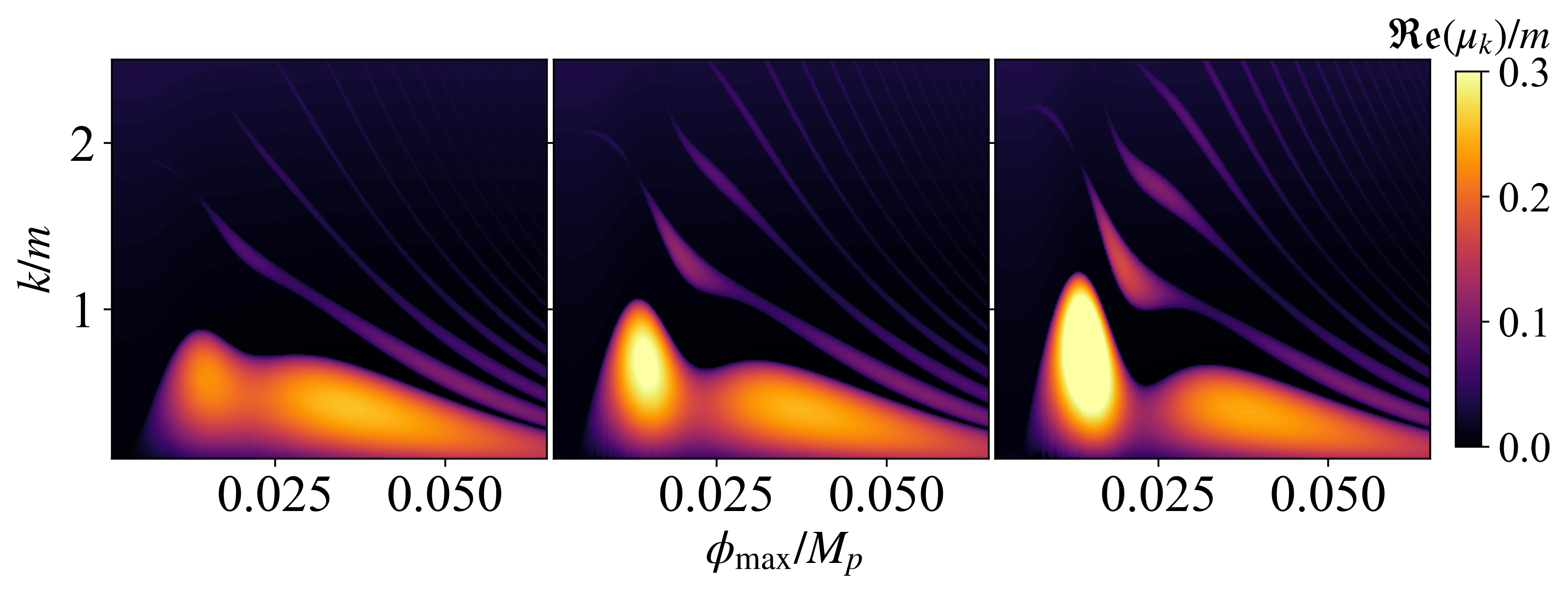}
		\caption{
				\textit{Upper row}: Diagnostic Floquet charts illustrating the origin of the modified resonance structure induced by the Gaussian feature. The left panel shows the full result for the $P_1$ parameter set with $h=0.4$. The middle panel is obtained using the exact background trajectory $\phi(t)$ of the $h=0.4$ model while replacing the effective mass term $V''(\phi)$ with that of the featureless T-model ($h=0$). The right panel is computed using the background trajectory $\phi(t)$ of the standard T-model ($h=0$) while retaining the feature-modified effective mass term $V''(\phi)$ with $h=0.4$.
				\textit{Lower row}: The Floquet charts for the models with symmetric Gaussian features placed at $\phi=\pm\phi_s$, preserving the original $Z_2$ symmetry of the T-model potential. From left to right, the panels correspond to $h=0.2$, $0.4$, and $0.6$, respectively. The parameters $\sigma$ and $\phi_s$ are the same as those adopted in Table I.
			}
	\label{anatomy_symmetric}
\end{figure}

Figure~\ref{Floquet_h} shows the Floquet exponents of the deformed T-model. Here, $\phi_{\mathrm{max}}$ denotes the positive turning point of the homogeneous background oscillation, whereas $\phi(t)$ denotes the instantaneous field value. For each $\phi_{\mathrm{max}}$, the Floquet exponent is calculated over one complete oscillation period. The upper two rows show that the instability bands are deformed relative to those of the original T-model. Positive $h$ (the $P_1$ set) modifies the strongest broad resonance band and creates a new resonance region over a range of background oscillation amplitudes for which the inflaton trajectory probes the potential feature. Negative $h$ slightly narrows the amplitude range supporting broad resonance, while extending the band toward higher momenta and thereby allowing higher-momentum modes to be excited. A narrow additional instability band also appears for trajectories probing the potential feature. This band is confined to $k\lesssim0.5,m$ and has smaller Floquet exponents than the primary band. These effects become more pronounced as $\lvert h\rvert$ increases. The lower two rows show the Floquet exponents of different modes at a fixed amplitude $\phi_{\max}=0.015,M_P$, together with that of the mode $k=0.5,m$ as a function of $\phi_{\mathrm{max}}$. The enhancement of the low-momentum resonance is more pronounced for $h>0$ than for $h<0$.

The Gaussian feature affects parametric resonance through two distinct channels: it modifies the background trajectory $\phi(t)$ and alters the effective mass term $m_{\rm eff}^2=V''(\phi)$ governing the fluctuation dynamics. To determine which effect is responsible for the deformation of the instability bands, we perform a decomposition analysis. We construct two auxiliary Floquet calculations. In the first, we solve equation~(\ref{pert_EoM}) with the T-model ($h=0$) trajectory $\phi_0(t)$ combined with the feature-modified effective mass $V_h''(\phi_0)$ ($h\neq 0$). In the second, we use the exact background trajectory $\phi_h(t)$ ($h\neq 0$) and the standard T-model effective mass $V_T''(\phi_h)$ ($h=0$). The corresponding Floquet charts are shown in the upper row of Fig.~\ref{anatomy_symmetric}. We find that the construction employing the feature-modified effective mass term reproduces the instability structure of the full model with remarkable accuracy, including the emergence of the additional high-$k$ resonance bands. In contrast, the construction employing the exact feature-modified background trajectory but the featureless effective mass term exhibits substantial deviations from the full result. This comparison demonstrates that the deformation of the instability bands is driven predominantly by the localized modification of $V''(\phi)$, while changes in the background-field trajectory have only a minor impact, affecting primarily the resonance of low-$k$ modes when the background oscillation amplitude is small. The dominance of the effective-mass contribution can be understood from the fluctuation equation, in which the resonance structure is controlled directly by the time-dependent frequency $\omega_k^2=k^2/a^2+V''(\phi)$. The Gaussian feature introduces a localized and rapid variation in $V''(\phi)$ as the inflaton traverses the feature region, leading to a strongly nonadiabatic modulation of $\omega_k$. This effect is responsible for the deformation of the instability bands.

A related issue concerns the explicit breaking of the $Z_2$ symmetry of the underlying T-model potential. Since the Gaussian feature is introduced only on the positive side of the potential, the inflaton encounters the feature once per oscillation cycle. It is therefore natural to ask whether the modified resonance structure is associated with this asymmetry or simply with the presence of the feature itself. To address this question, we additionally consider a symmetric configuration in which identical Gaussian features are placed at both $\pm\phi_s$. The corresponding Floquet charts are shown in the lower row of Fig.~\ref{anatomy_symmetric}.
Compared with the one-sided feature configuration shown in Fig.~\ref{Floquet_h}, the symmetric feature produces a stronger modification of the Floquet instability structure, with the additional high-$k$ resonance bands becoming significantly more prominent. This indicates that the emergence of the new resonance bands is not tied to the explicit breaking of the $Z_2$ symmetry or to the fact that the inflaton encounters the feature only once per oscillation cycle. Instead, it originates from the localized modification of the effective mass term $V''(\phi)$ induced by the feature itself. The stronger deformation observed in the symmetric case can be understood from the fact that the inflaton traverses two feature regions during each oscillation period, leading to a more frequent modulation of $V''(\phi)$ and consequently a stronger resonant response. Thus, explicit $Z_2$ symmetry breaking quantitatively affects the efficiency of the resonance, but is not essential to the physical
mechanism responsible for the deformation of the instability bands.

\section{Non-perturbative dynamics of preheating: Lattice simulation}\label{sec3}
\subsection{Setup and initialization}\label{subsec1}

As noted above, perturbative analyses based on equation~(\ref{pert_EoM}) are inherently limited: once fluctuations grow large enough, the expansion ceases to be valid and nonlinear effects dominate, rendering perturbative predictions unreliable. In this regime, equations~(\ref{phi_EoM})–(\ref{acc_eq}) together with equation~(\ref{pert_EoM}) are no longer applicable, and the scalar field dynamics must instead be governed by the following nonlinear equations:
\begin{eqnarray}
	\ddot{\phi} + 3H\dot{\phi} - \frac{1}{a^2} \nabla^2 \phi  +V_{\phi}&=&0,  \label{nonlinear_eq}\\
	\frac{1}{2}\dot{\phi}^2 + \frac{1}{2a^2}\left(\nabla\phi\right)^2 +V &=& 3M_P^2 H^2, \label{nonlinear_fried}
\end{eqnarray}
where $\phi=\phi(t,\vec{x})$ and $\nabla$ is the spatial gradient operator. 
Throughout the lattice simulations, the scalar field is evolved on a spatially flat FLRW background determined by the volume-averaged energy density. Scalar and vector metric perturbations are neglected, while tensor perturbations are evolved linearly without backreaction. The oscillons considered here are supported primarily by scalar self-interactions rather than gravity and are therefore not expected to enter the strong-gravity regime; neglecting metric perturbations is the standard approximation in lattice studies of this type. Previous simulations found the backreaction of metric perturbations to be subdominant in the preheating models considered~\cite{Huang:2011gf}, whereas fully relativistic simulations are required when gravitational effects become important, particularly for gravitational binding or collapse~\cite{Kou:2019bbc,Aurrekoetxea:2023jwd}.

In this section, we use the publicly available code  ${\cal C}$\texttt{osmo}${\cal L}$\texttt{attice}~\cite{Figueroa:2020rrl,Figueroa:2021yhd,Baeza2025tme} to study preheating from a fully nonlinear, lattice-based perspective.  ${\cal C}$\texttt{osmo}${\cal L}$\texttt{attice} evolves a system within a finite discretized box using dimensionless quantities, such as model parameters, fields, and the potentials. For this purpose, all quantities must first be rescaled to dimensionless form, which we achieve by introducing a set of scaling parameters ($m$, $\phi_\mathrm{i}$), with $m$ being the inflaton mass defined by $m^2=\left.\frac{\partial^2 V}{\partial\phi^2}\right|_{\phi=0}$ and $\phi_\mathrm{i}$ the field value at the end of inflation marking the onset of preheating. The coordinates and fields are rescaled as
\begin{equation}
\tilde{t}=mt, \quad \tilde{x}=m x,\quad \tilde{\phi}=\frac{\phi}{\phi_\text{i}}.
\end{equation}
For convenience, we also use a new dimensionless parameter $\lambda = M_P/\sqrt{6\alpha}$ instead of $\alpha$. With these new variables and parameters, the kinetic, gradient, and potential energies are rescaled to be
\begin{align}
	\tilde{E}_{\mathrm{K}} = &~ \frac{1}{m^2\phi^2_{\mathrm{i}}} \frac{1}{2}\dot{\phi}^2
	= \frac{1}{2}\left(\partial_{\tilde{t}}\tilde{\phi}\right)^2,
	\\
	\tilde{E}_{\mathrm{G}} = &~ \frac{1}{m^2\phi^2_{\mathrm{i}}} 
	\frac{1}{2}\left(\frac{\nabla{\phi}}{a}\right)^2
	= \frac{1}{2}\left(\frac{\tilde{\nabla}\tilde{\phi}}{a}\right)^2,
	\\
	\tilde{V} = &~ \frac{1}{{m^2\phi_\text{i}^2}} V(\phi)
	=\frac{M_p^2}{2\lambda^2\phi_\text{i}^2}\tanh^2\left(\lambda\frac{\phi_\text{i}}{M_p}\tilde{\phi}\right)
     \\
    &~\times \left\{ 1+h \exp\left[-\frac{{\phi_{\text{i}}}^2}{2\sigma^2}
	\left(\tilde{\phi}-\frac{\phi_\text{s}}{\phi_{\text{i}}}\right)^2\right]\right\},
\end{align}
where $\tilde{\nabla}=\nabla/m$. Accordingly, the rescaled energy density and pressure are
\begin{eqnarray}
	&&\tilde{\rho}=\frac{1}{m^2\phi_{\mathrm{i}}^2}\rho
	=\tilde{E}_{\mathrm{K}}+\tilde{E}_{\mathrm{G}}+\tilde{V}, \label{energy_density}
	\\
	&&\tilde{p}=\frac{1}{m^2\phi_{\mathrm{i}}^2}p=
	\tilde{E}_{\mathrm{K}}-\frac{1}{3}\tilde{E}_{\mathrm{G}}-\tilde{V}. \label{pressure}
\end{eqnarray}
In ${\cal C}$\texttt{osmo}${\cal L}$\texttt{attice}, we evolve the dimensionless version of equations~(\ref{nonlinear_eq}) and (\ref{nonlinear_fried}),
\begin{eqnarray}
	\partial^2_{\tilde{t}}\tilde{\phi} + 3\tilde{H}\partial_{\tilde{t}}\tilde{\phi} - \frac{1}{a^2} \tilde{\nabla}^2 \tilde{\phi}  + \tilde{V}_{\tilde{\phi}}&=&0,  \label{scalar_dimsionless}\\
	\frac{1}{2}\left(\partial_{\tilde{t}}\tilde{\phi}\right)^2 
	+ \frac{1}{2a^2}\left(\tilde{\nabla}\tilde{\phi}\right)^2 +\tilde{V} &=& 3\frac{M_P^2}{\phi^2_{\mathrm{i}}} \tilde{H}^2, \label{fried_dimsionless}
\end{eqnarray}
where $\tilde{H}= H/m$. 

To initialize the simulations, we need to impose initial conditions for equations~(\ref{scalar_dimsionless}) and (\ref{fried_dimsionless}). In this work, we start at $\epsilon_V(\phi_{\mathrm{end}})=1$ and identify the initial lattice value with the field value at the end of inflation, $\phi_{\mathrm{i}}\equiv\phi_{\mathrm{end}}$, while the initial momentum $\dot{\phi}_{\mathrm{i}}$ is determined numerically from the homogeneous mode equation~(\ref{phi_EoM}). In our setup, the characteristic region of the potential lies far from $\phi_{\mathrm{i}}$, such that the correction term in equation (\ref{deformed_model}) is exponentially suppressed. This allows us to adopt the original T-model ($h=0$) values of $(\phi_{\mathrm{i}},\dot{\phi}_{\mathrm{i}})$ directly for the parameter sets listed in table~\ref{tab1}. The values of the initial conditions employed in our simulations are fixed at $\phi_{\mathrm{i}}=0.06665\,M_P$ with velocity $\dot{\phi}_{\mathrm{i}}=-1.656\times 10^{-8}\,M_P^2$. The main results presented in this work are based on simulations with $N^3=384^3$ grids and infrared cutoff $k_{\rm IR}=0.1\,m$, which yields a box size $L=20\pi \,m^{-1}$ and a lattice spacing $\mathrm{d}x \approx 0.1635\,m^{-1}$. The ultra-violet cutoff is $k_{\rm UV}=\frac{\sqrt{3}}{2}N k_{\rm IR} \approx 33\,m$. The choice of this setup is explained in the convergence test in appendix~\ref{CT}. All lattice simulations presented are performed using a second-order velocity-verlet algorithm.

\subsection{Fields and fluctuations}\label{subsec2}
In this subsection we analyze the data of lattice simulations, in particular those related to oscillon formation. The output data contains averages, spectra, and snapshots. Here, the average denotes the volume average of the physical quantity over the lattice. For example, the scalar field value is given by the volume average of the scalar field at all points of the lattice,
\begin{equation}
	\langle \tilde{\phi}(\tilde{t},\vec{\tilde{x}}) \rangle = 
	\frac{1}{N^3}\int_{\text{volume}} \mathrm{d}^3\tilde{x} \, 
	\tilde{\phi}(\tilde{t},\vec{\tilde{x}})
	=  \frac{1}{N^3} \sum_{\vec{\tilde{x}}} \tilde{\phi}(\tilde{t},\vec{\tilde{x}}).
\end{equation}
The symbol $\langle...\rangle$ denotes volume average. For notational simplicity, we omit this symbol in what follows, with quantities such as field values, fluctuations, and energy densities understood as the volume-averaged one.

\begin{figure}[t]
	\centering
	\includegraphics[width=8cm]{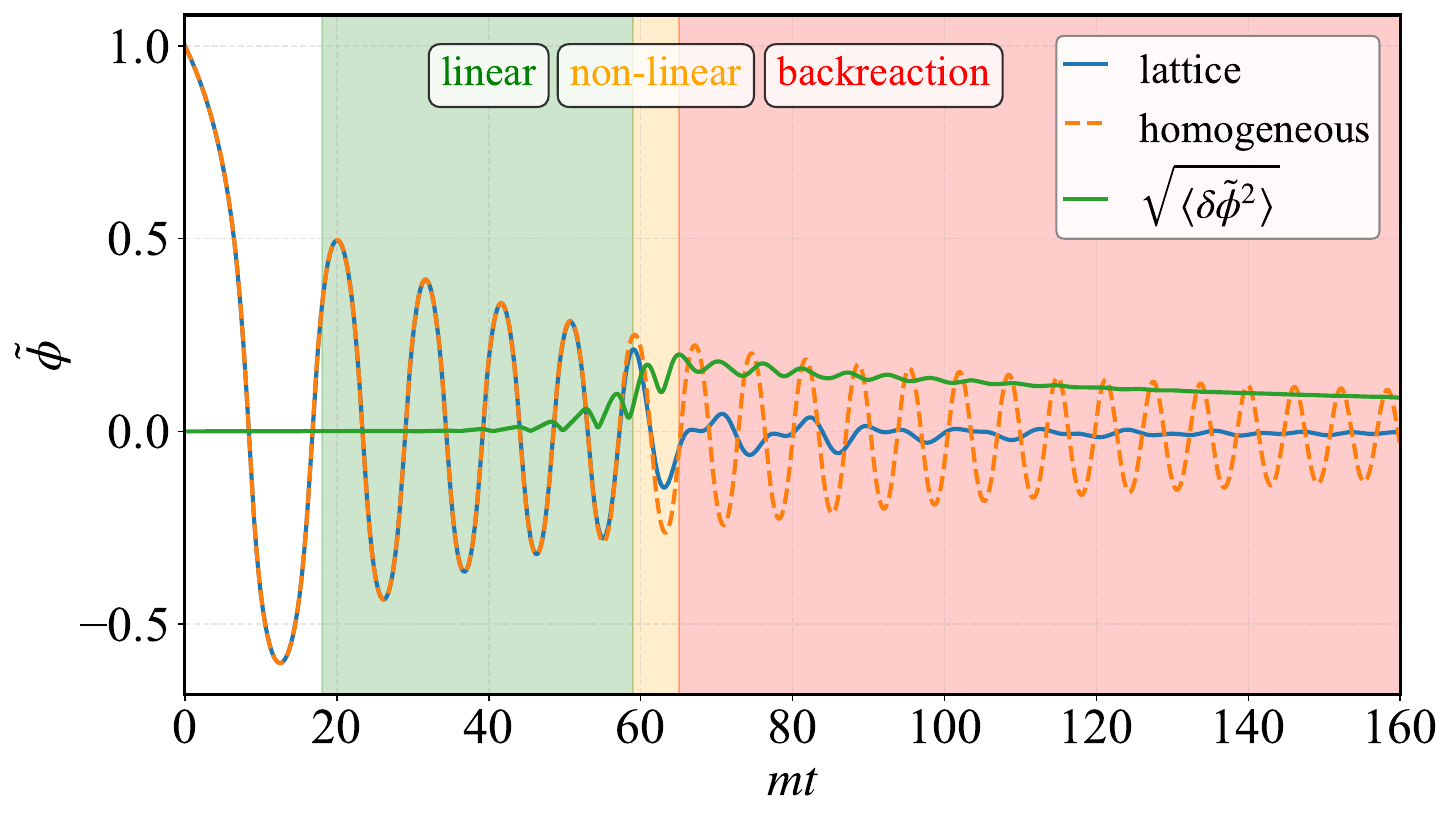}
	\caption{Comparison of the field evolution result obtained from the nonlinear equation~(\ref{scalar_dimsionless}) and the homogeneous equation~(\ref{phi_EoM}), and the r.m.s $\sqrt{\langle \delta\tilde{\phi}^2 \rangle}$ (green), for the parameter set $P_1$ with $h=0.4$.}
	\label{fields_variance}
\end{figure}
\begin{figure}
	\centering
	\includegraphics[width=6.5cm]{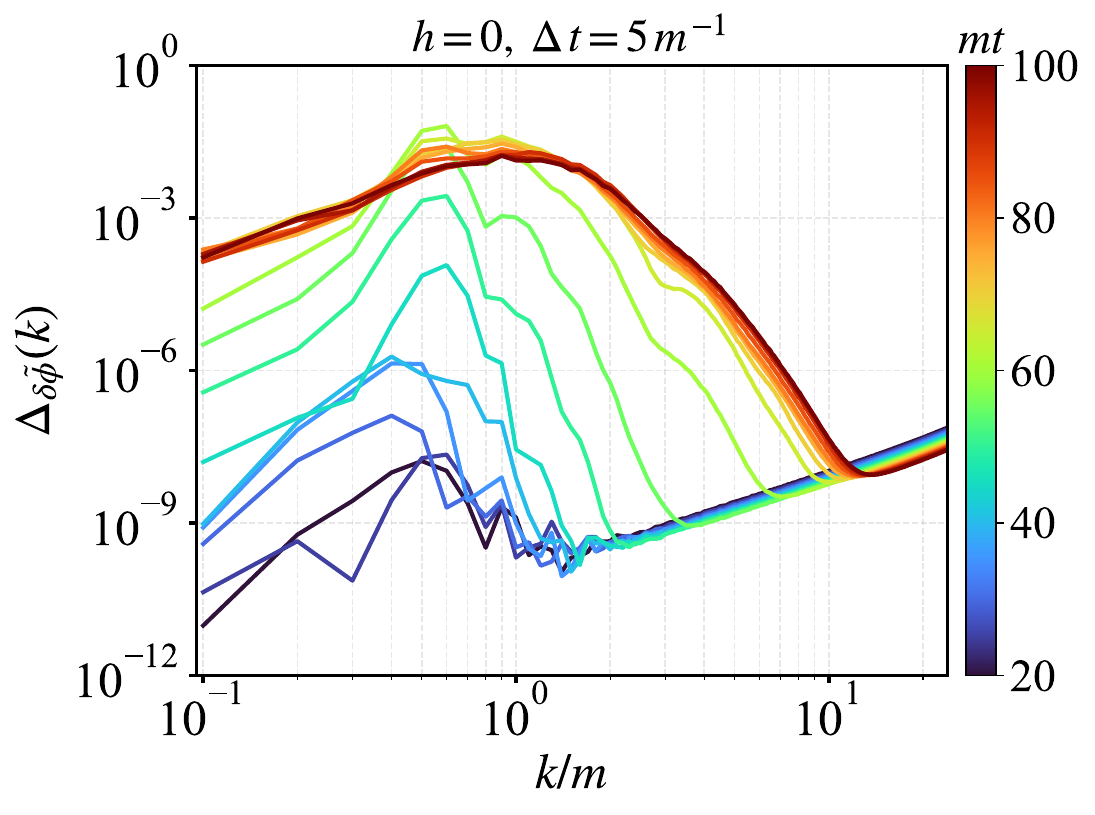}
	\includegraphics[width=6.5cm]{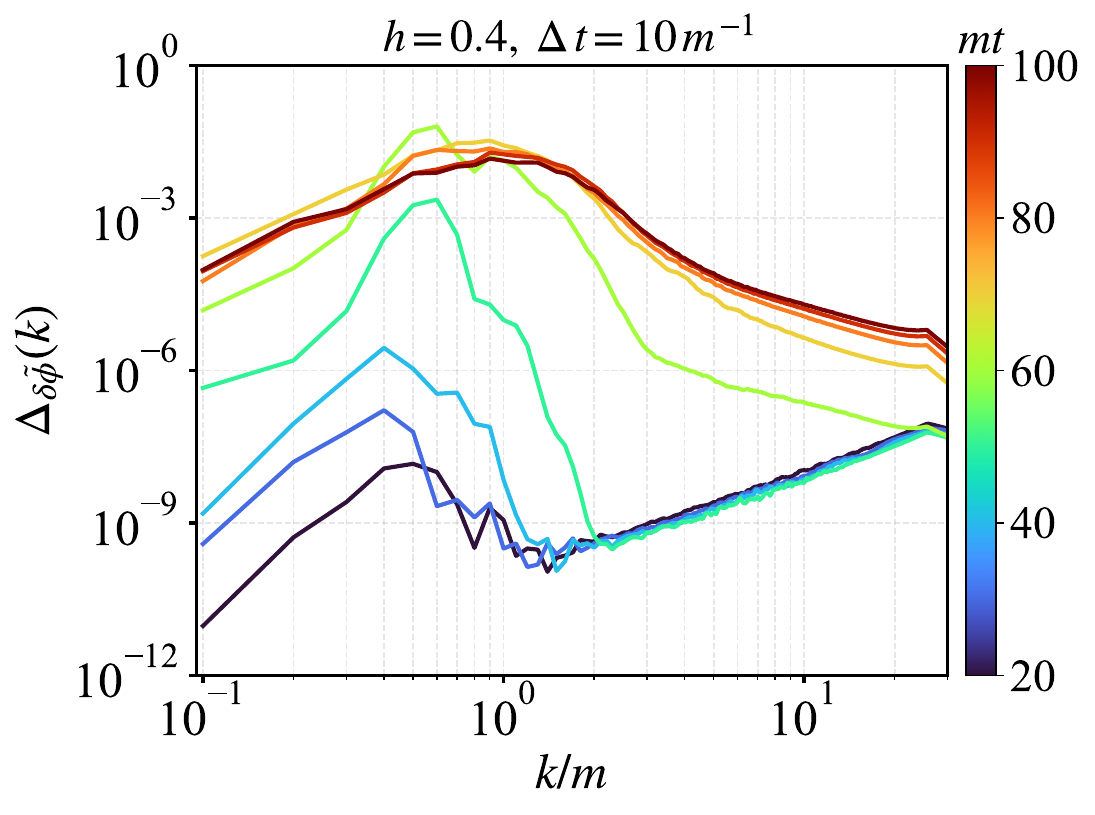}
	\includegraphics[width=6.5cm]{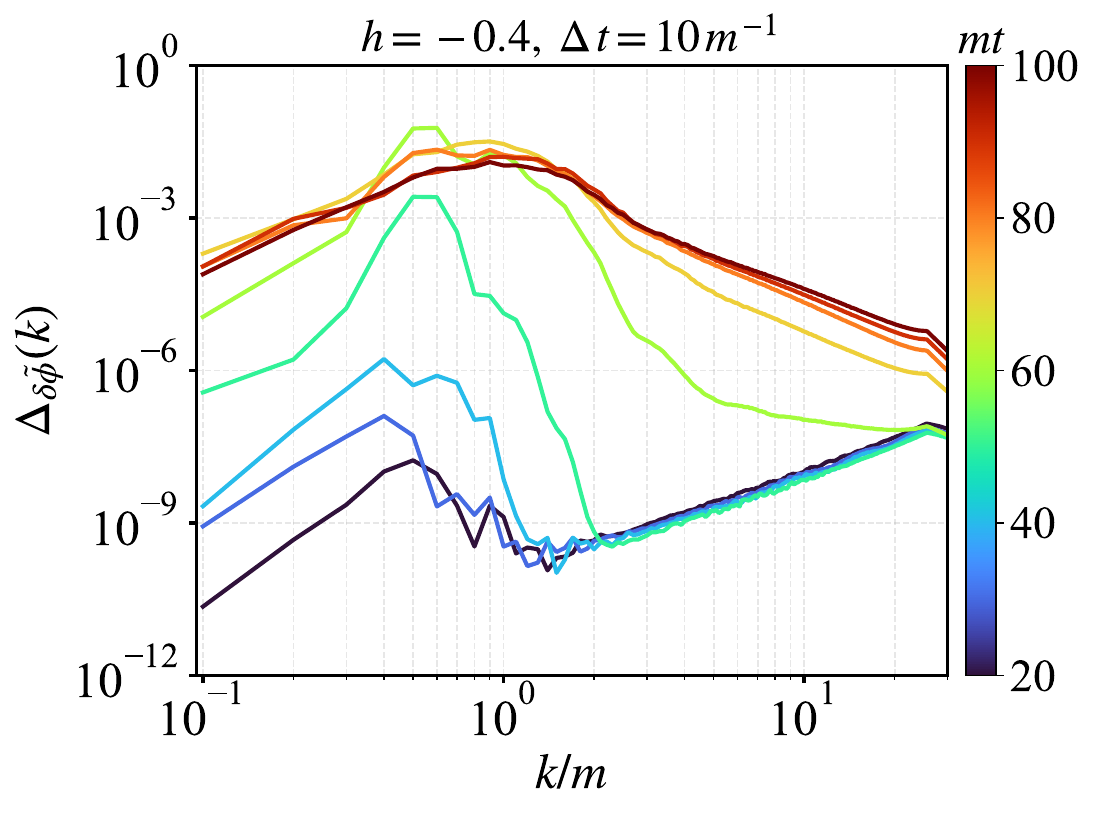}
	\includegraphics[width=6.5cm]{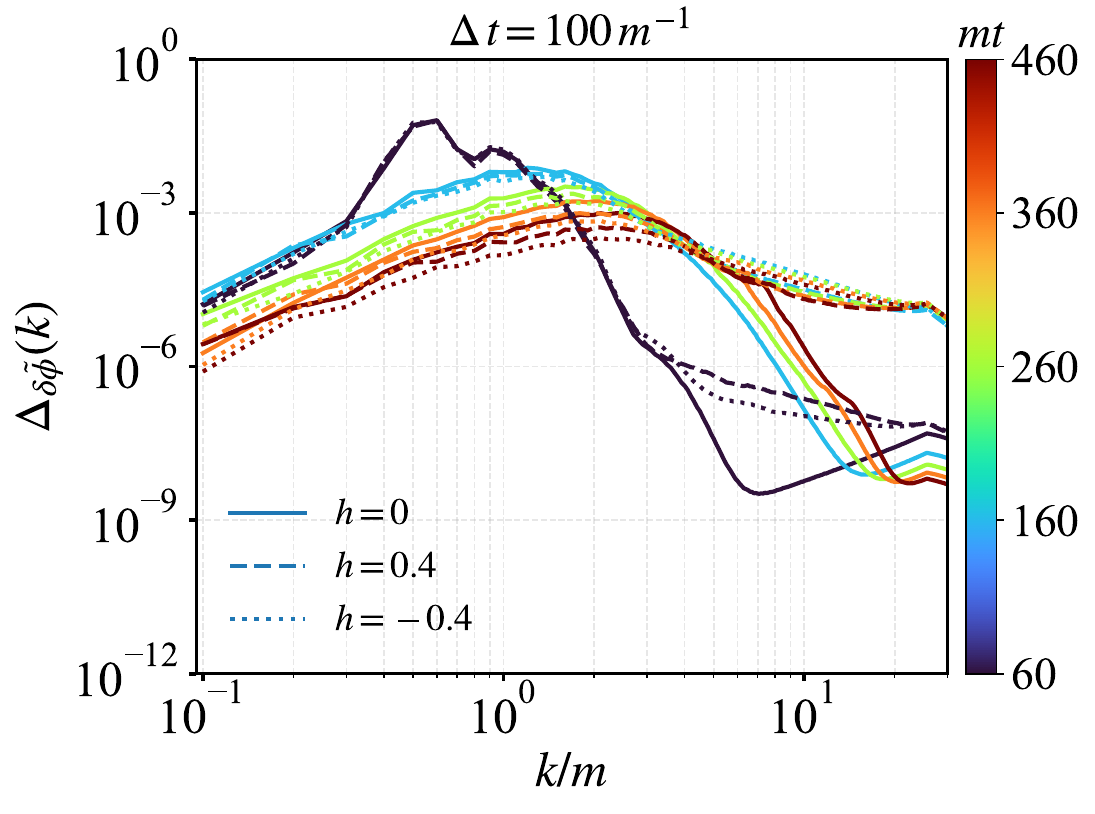}
	\caption{Dimensionless power spectra of the scalar-field fluctuations. From top to bottom, the panels correspond to $h=0$, $h=0.4$, $h=-0.4$, and a comparison of the three cases.}
	\label{power_spectra}
\end{figure}

Now let us analyze the evolution of the scalar field and its fluctuations in lattice. As an example, in Fig.~\ref{fields_variance}  we show the comparison of field evolution obtained from the nonlinear equation~(\ref{scalar_dimsionless}) (the evolution of the field fluctuations for other parameter sets are shown in Fig.~\ref{6variance} in appendix) and that from the homogeneous equation~(\ref{phi_EoM}), as well as the fluctuation, which is defined by the root mean square (r.m.s) of the scalar field,
\begin{equation}
	\sqrt{\langle \delta\tilde\phi^2 \rangle} = \sqrt{\langle \tilde\phi(t,\vec{x})^2 \rangle - \langle \tilde{\phi}(t,\vec{x})\rangle^2}.
\end{equation}
We observe that the two approaches exhibit a clear discrepancy. Although they coincide in the early stage (linear regime) of preheating, once fluctuations grow to amplitudes comparable to the background field, the perturbative expansion breaks down and nonlinear effects become significant. In the nonlinear regime, fluctuations continue to grow until they significantly affect the background evolution, at which backreaction shuts off the resonance.

Combining the evolution of the fluctuations in Fig.~\ref{6variance}, we find that the growth of fluctuations in resonance stage is similar for all parameters. For all parameter sets, the fluctuations grow and peak within the time interval $60\,m^{-1} \lesssim t \lesssim 70\,m^{-1}$. 
Once the fluctuations peak, the subsequent backreaction stage exhibits a faster decay for $|h|\neq 0$, and the suppression becomes stronger as $|h|$ increases, hence the amplitude of fluctuations for $h\neq 0$ is slightly smaller than that of the original $T$-model ($h=0$). 
Part of these features are likewise reflected in the power spectra~\ref{power_spectra}.  We see that the peak amplitudes of the power spectra are nearly identical and occur in the time interval $60\,m^{-1} \lesssim t \lesssim 70\,m^{-1}$. Resonant amplification first occurs for low-momentum modes and subsequently extends to higher momenta. Notably, in the $h=0$ case the resonance is largely confined to modes with $k\lesssim 10\,m$, whereas for $h\neq0$ higher-momentum modes are also excited after the resonance stage. We thus conclude that the characteristic structures of the potential primarily influence high-momentum (small-scale) modes, and that this effect is largely confined in the backreaction stage.

\begin{figure}[t]
	\centering
	\includegraphics[width=6cm]{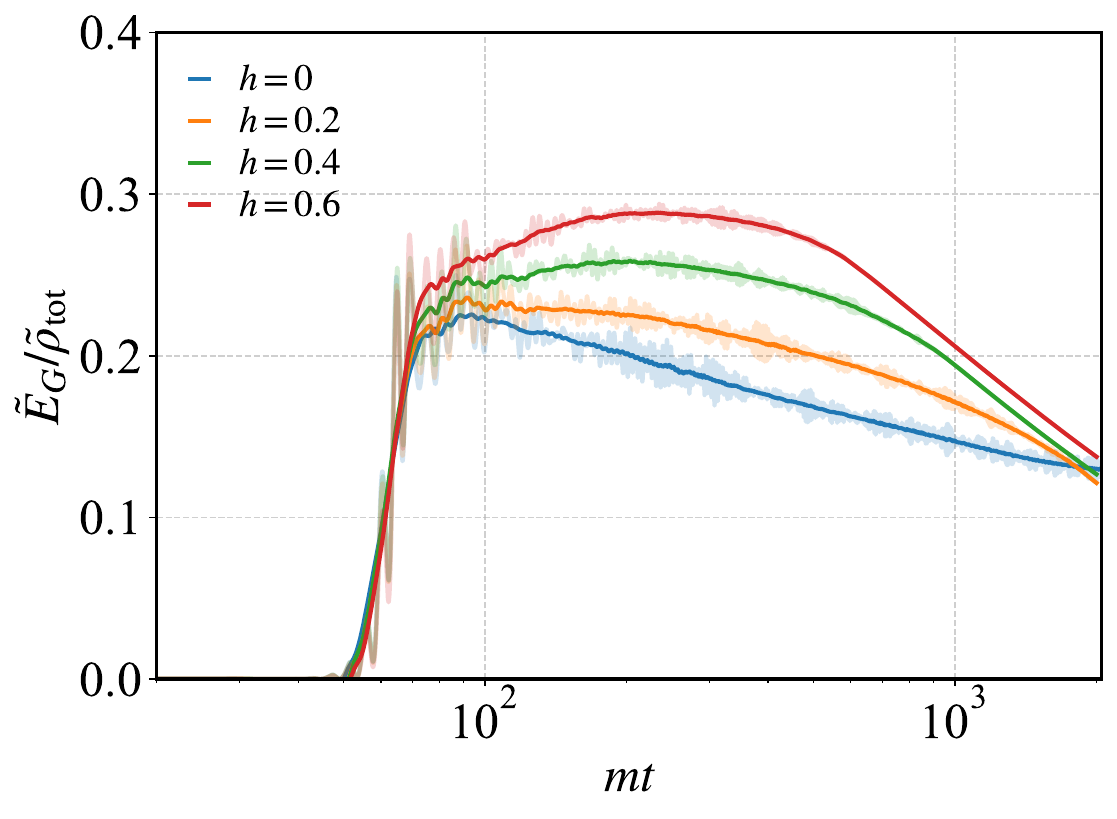}
	\includegraphics[width=6cm]{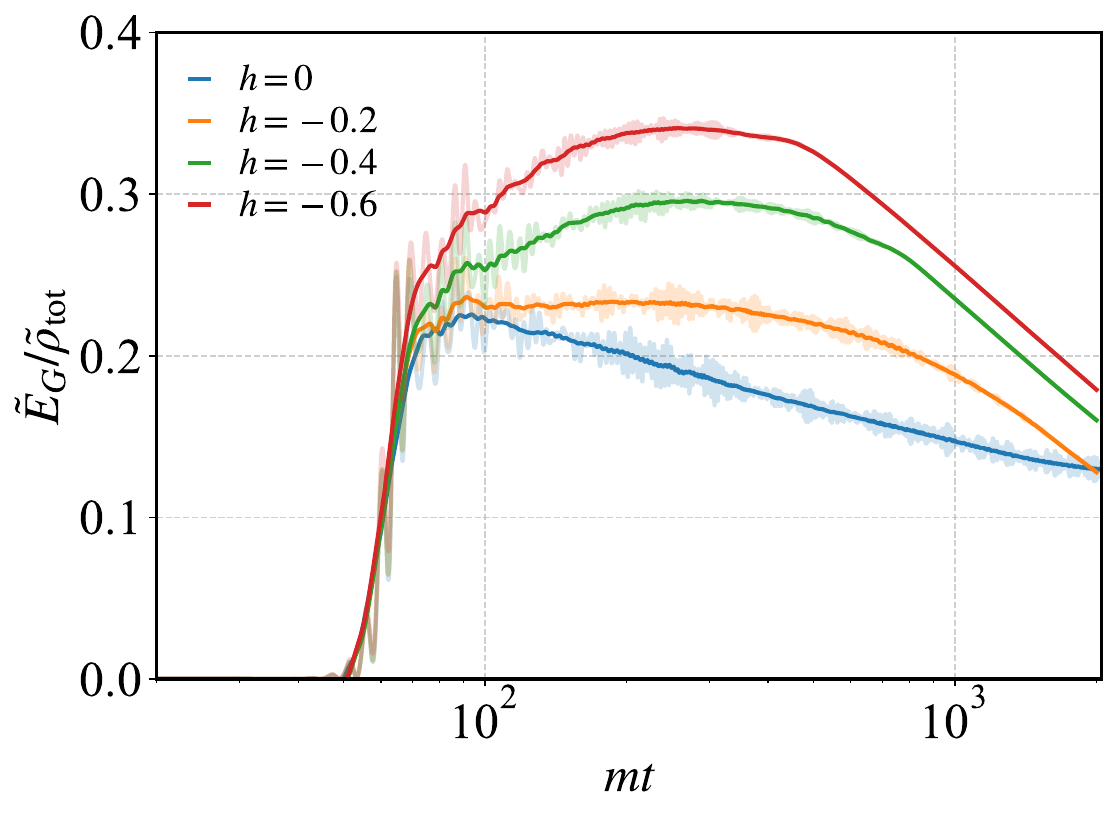}
	\caption{The evolution of gradient energy for $P_1$ (upper) and $P_2$ (lower) set of parameters. The lighter curves show the raw volume-averaged fraction of gradient energy, while the darker curves represent the result after applying an oscillation-averaged smoothing procedure.}
	\label{gradient_fraction}
\end{figure}

\subsection{Formation and decay of oscillons}
\subsubsection{gradient energy}
Now let us analyze the energy transfer in preheating. During the period of self-resonance, the inflaton transfers energy to its own fluctuations, leading to their amplification and the formation of a spatially anisotropic energy distribution.  As an example, we show the fraction of the volume averaged gradient energy in Fig.~\ref{gradient_fraction}. We see that the kinetic and potential energies of the inflaton field are efficiently transferred into gradient energy, leading to its exponential growth before the resonance terminates. These results reveal several characteristic features. 
\begin{itemize}
	\item The fraction of gradient energy reaches a maximum of approximately $20\%$-$35\%$ (for the parameters considered), the same magnitude with contributions from the kinetic and potential sectors. The peak of gradient energy fraction increases with $|h|$, and for the same $|h|$ it is larger for negative $h$ than for positive $h$. hence, the presence of potential feature accelerates the fragmentation of the homogeneous inflaton mode and enhances its energy transfer.
	\item Across different models, the gradient energy fraction exhibits distinct behaviors at different stages. In the resonance stage ($t\lesssim 70\,m^{-1}$), the growth of the gradient energy fraction is nearly identical for all models, showing no significant differences. After the resonance stage, however, clear model-dependent features emerge, with larger values of $h$ leading to a more pronounced growth of the gradient energy fraction. In the backreaction stage, the gradient energy fraction in the $h=0$ case decays as a power law in time, whereas for $h\neq 0$ it displays a qualitatively different evolution. For larger values of $|h|$, the gradient energy fraction undergoes another period of growth, allowing it to reach a higher peak, but also decay faster. It was suggested that the decay of the total gradient energy can be employed to define the lifetime of oscillons. However, in this work, we will introduce more reliable new methods to determine the lifetime of oscillons later.
	\item For sufficiently large $|h|\gtrsim 0.3$, the gradient energy fraction exhibits a power-law decay in time beyond a characteristic epoch, appearing as straight lines with identical slopes, see Fig.~\ref{gradient_fraction}. The onset of this regime depends on $|h|$, occurring at earlier times for larger values of $|h|$. As will be shown below, this behavior is closely related to the decay of gradient energy distributed inside and outside oscillons.
\end{itemize}

\subsubsection{Properties of oscillons}
\begin{figure*}[t]
	\centering
\begin{minipage}{0.23\textwidth}
	\centering
	\includegraphics[width=\linewidth]{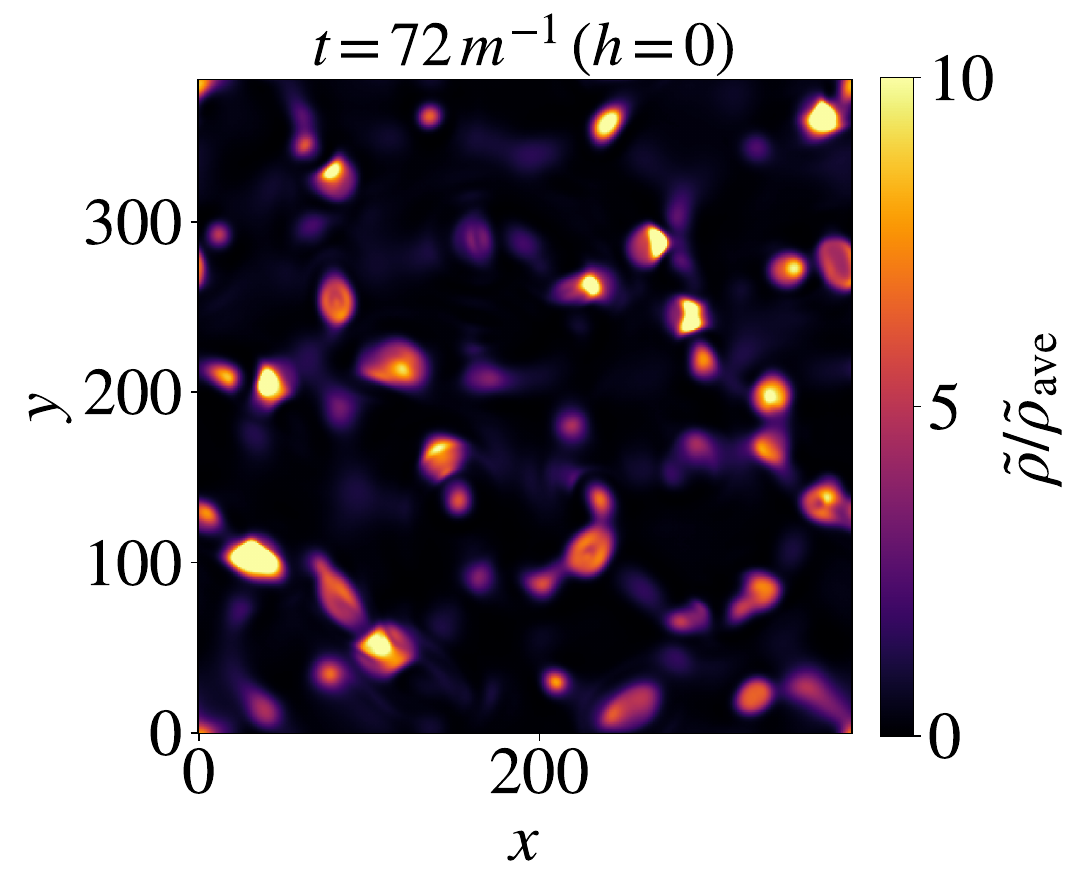}
\end{minipage}
\begin{minipage}{0.23\textwidth}
	\centering
	\includegraphics[width=\linewidth]{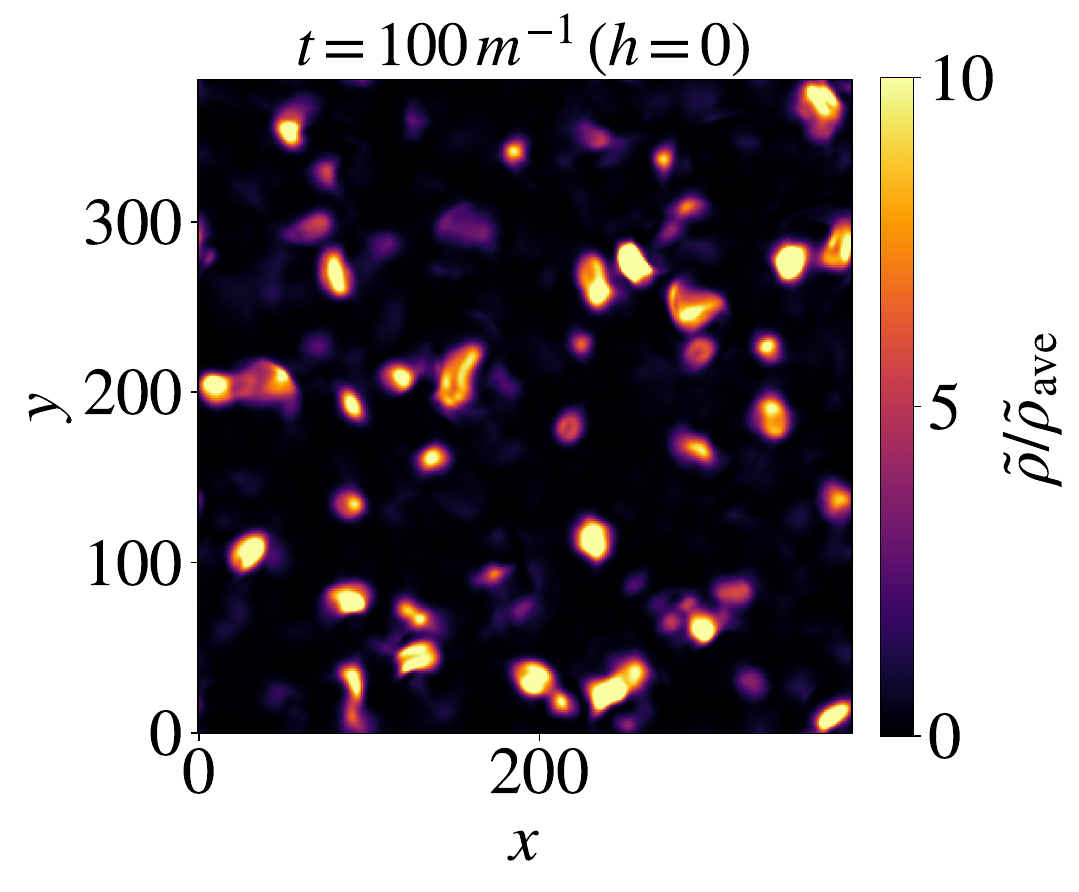}
\end{minipage}
\begin{minipage}{0.23\textwidth}
	\centering
	\includegraphics[width=\linewidth]{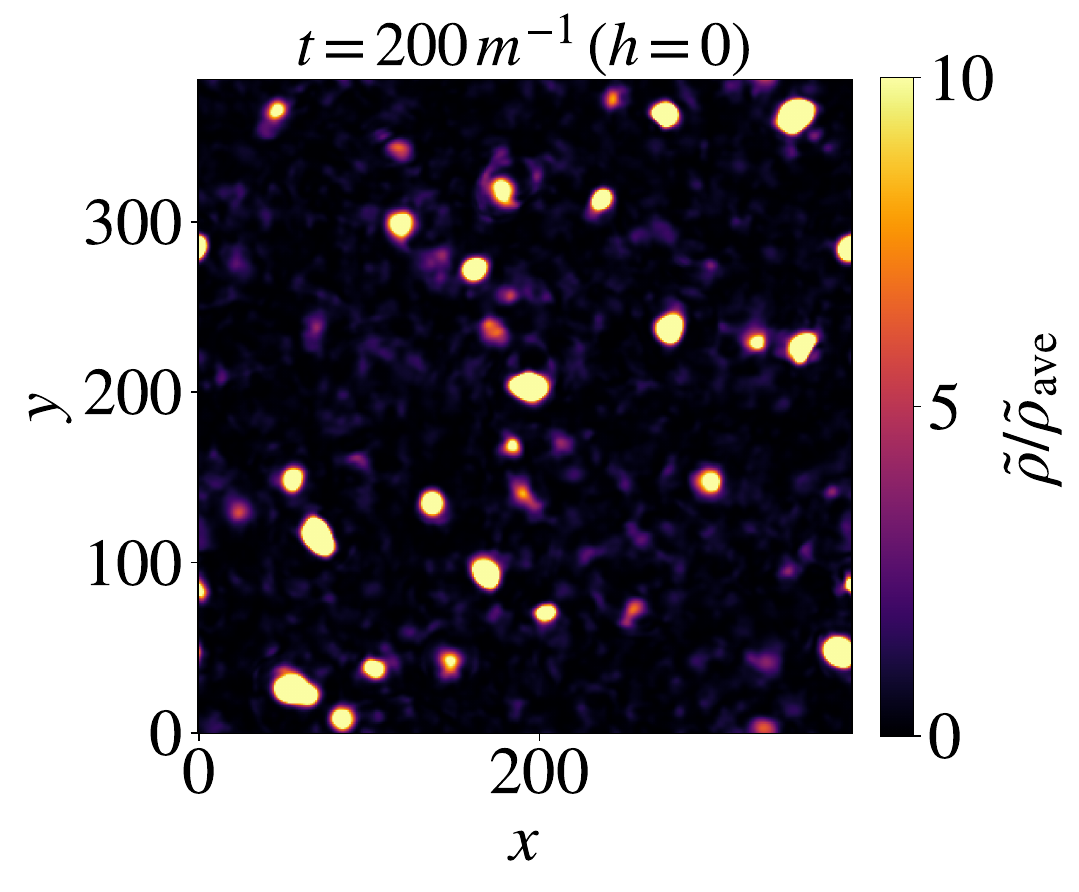}
\end{minipage}
\begin{minipage}{0.23\textwidth}
	\centering
	\includegraphics[width=\linewidth]{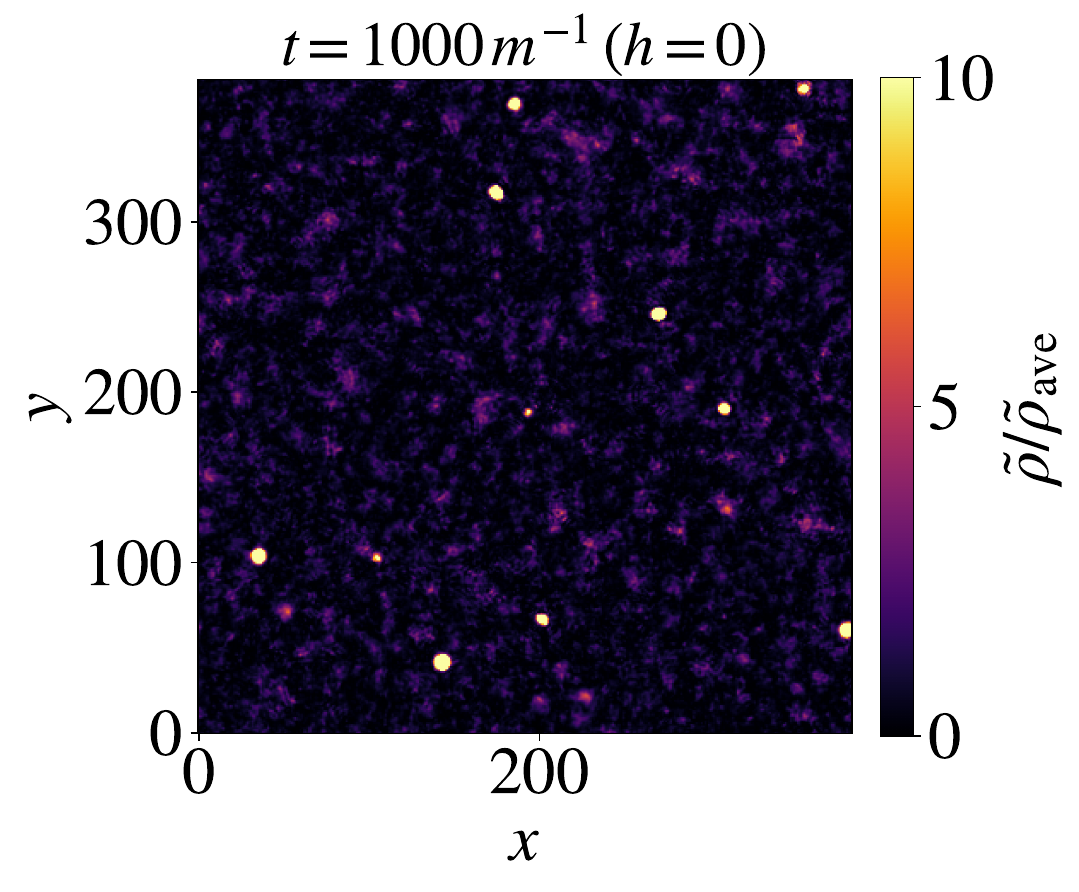}
\end{minipage}
\vspace{2mm}

\begin{minipage}{0.23\textwidth}
	\centering
	\includegraphics[width=\linewidth]{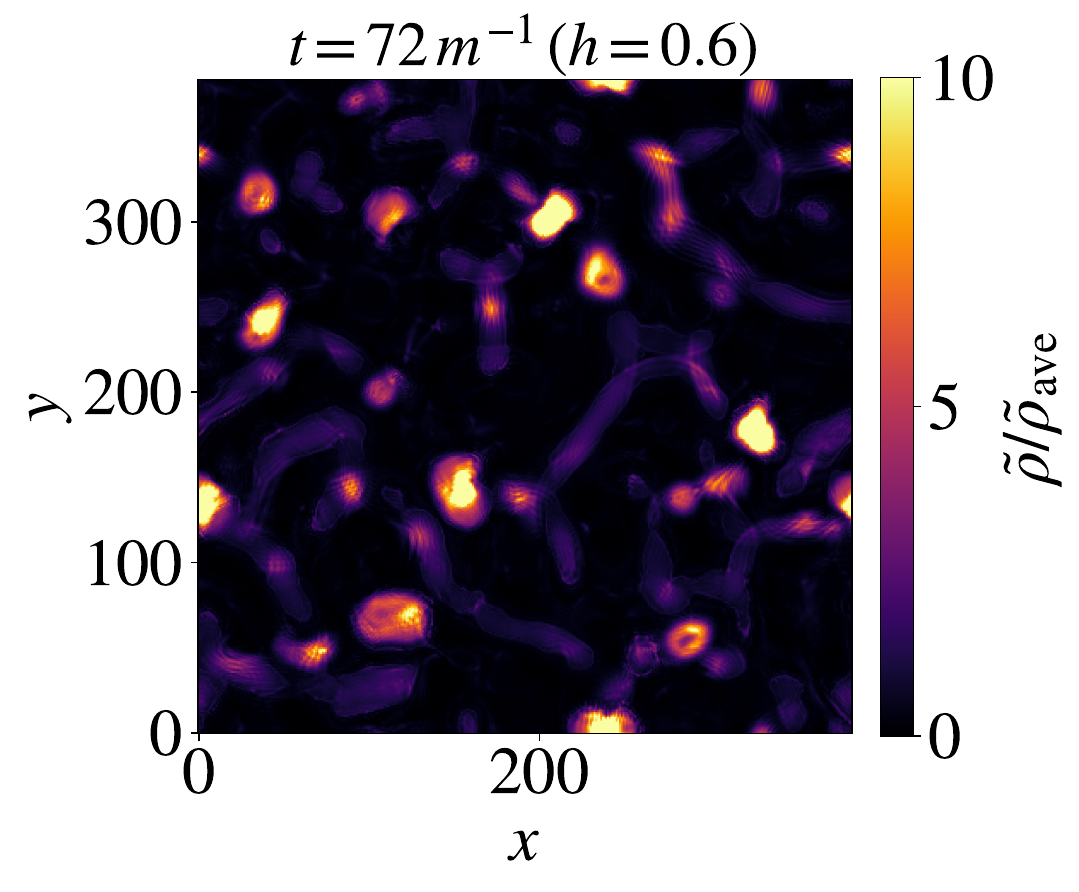}
\end{minipage}
\begin{minipage}{0.23\textwidth}
	\centering
	\includegraphics[width=\linewidth]{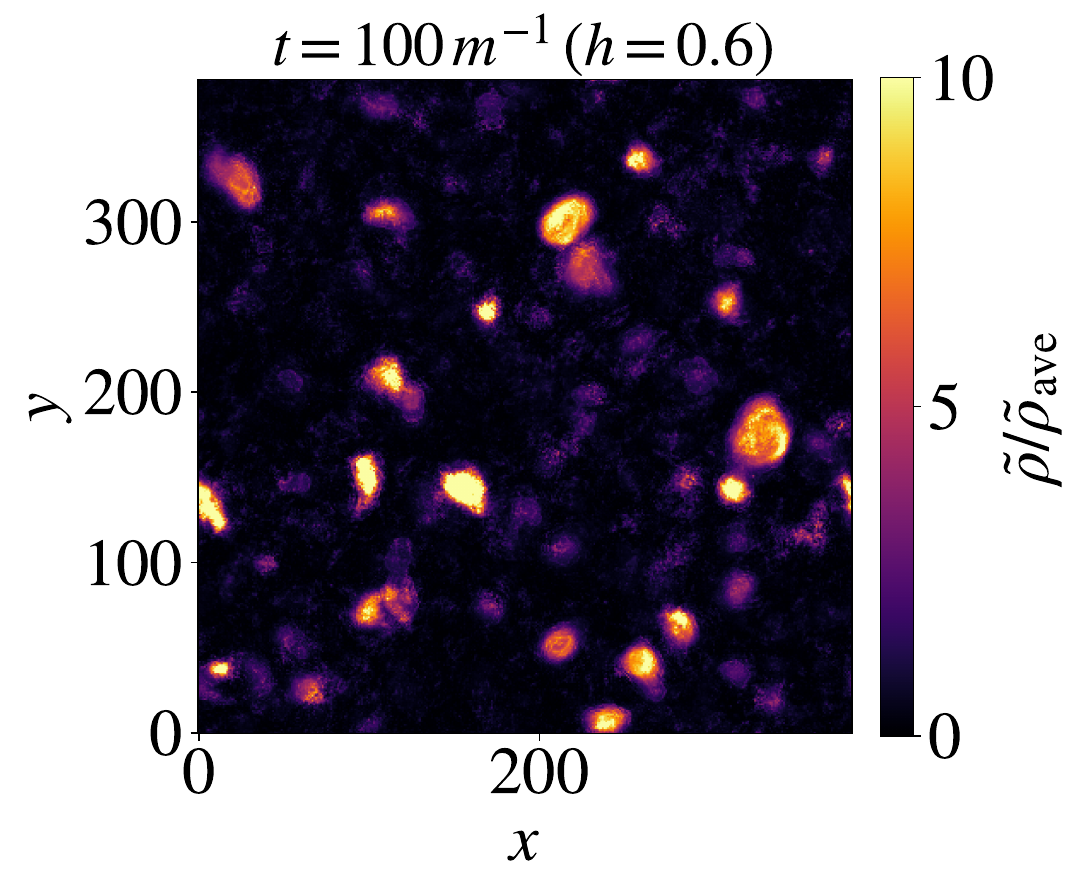}
\end{minipage}
\begin{minipage}{0.23\textwidth}
	\centering
	\includegraphics[width=\linewidth]{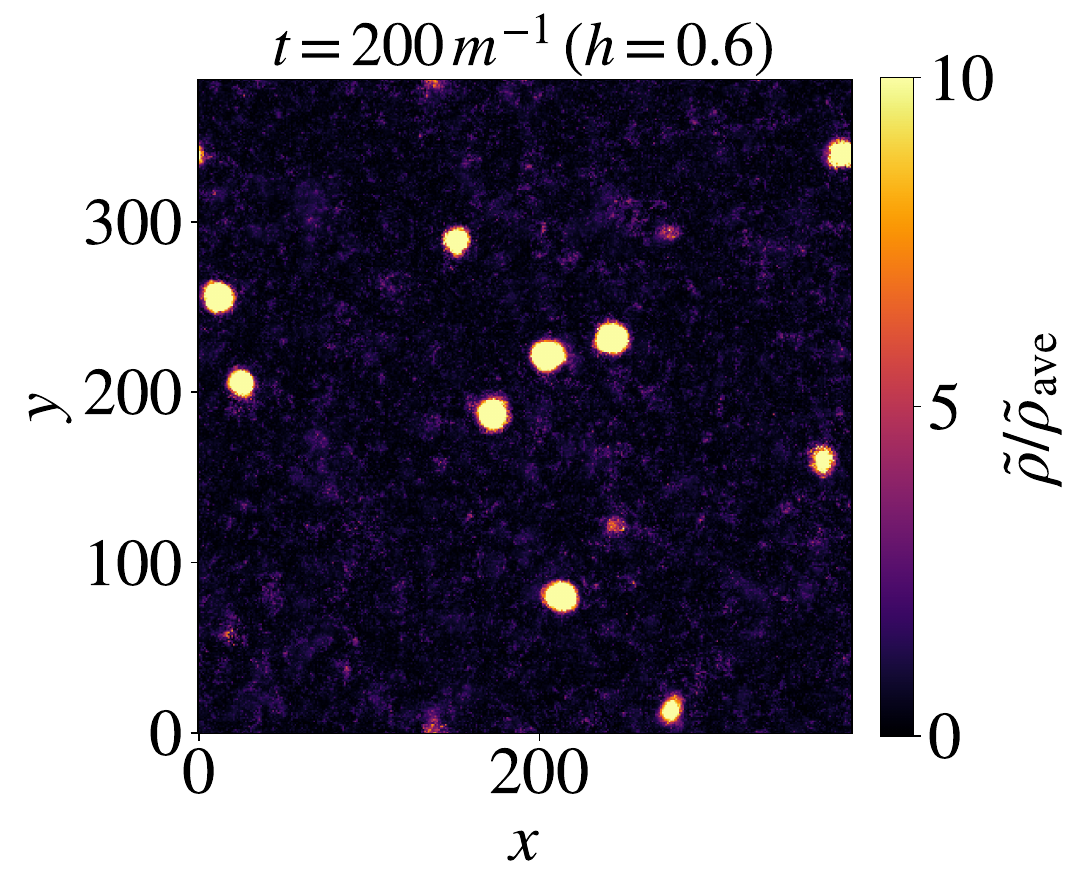}
\end{minipage}
\begin{minipage}{0.23\textwidth}
	\centering
	\includegraphics[width=\linewidth]{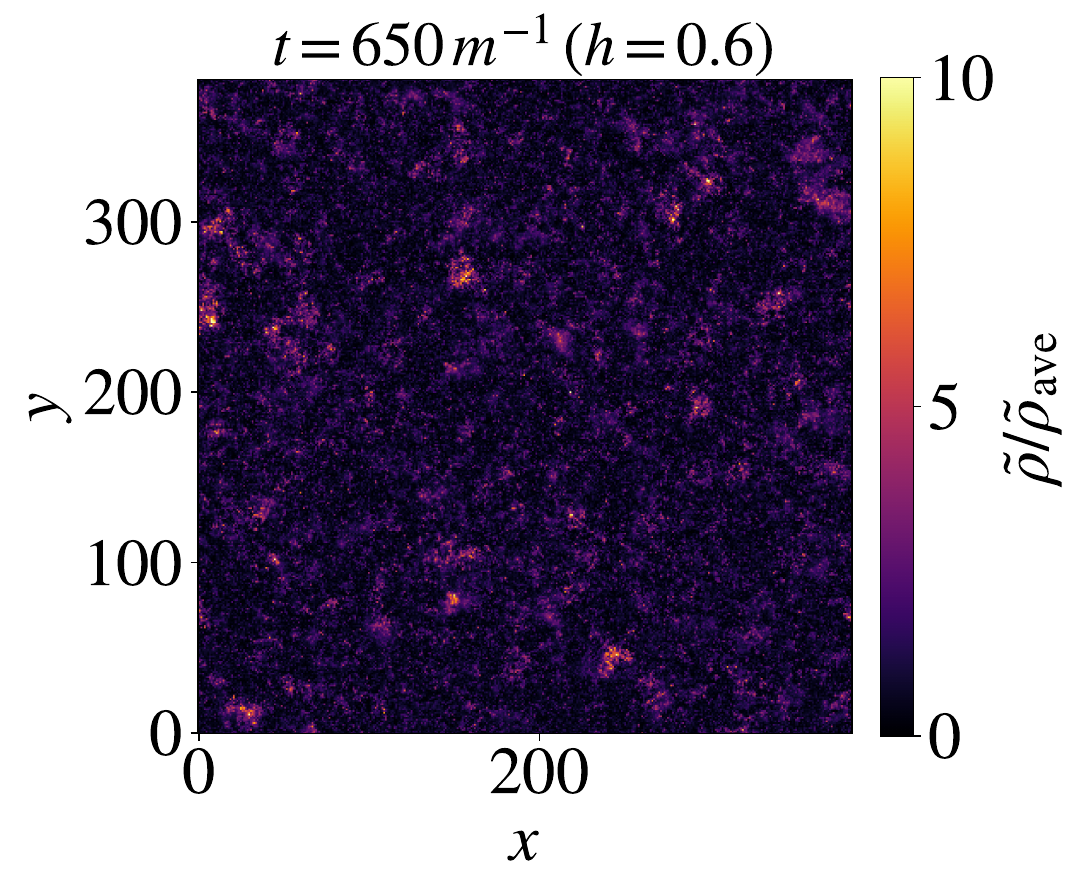}
\end{minipage}
\vspace{2mm}

\begin{minipage}{0.23\textwidth}
	\centering
	\includegraphics[width=\linewidth]{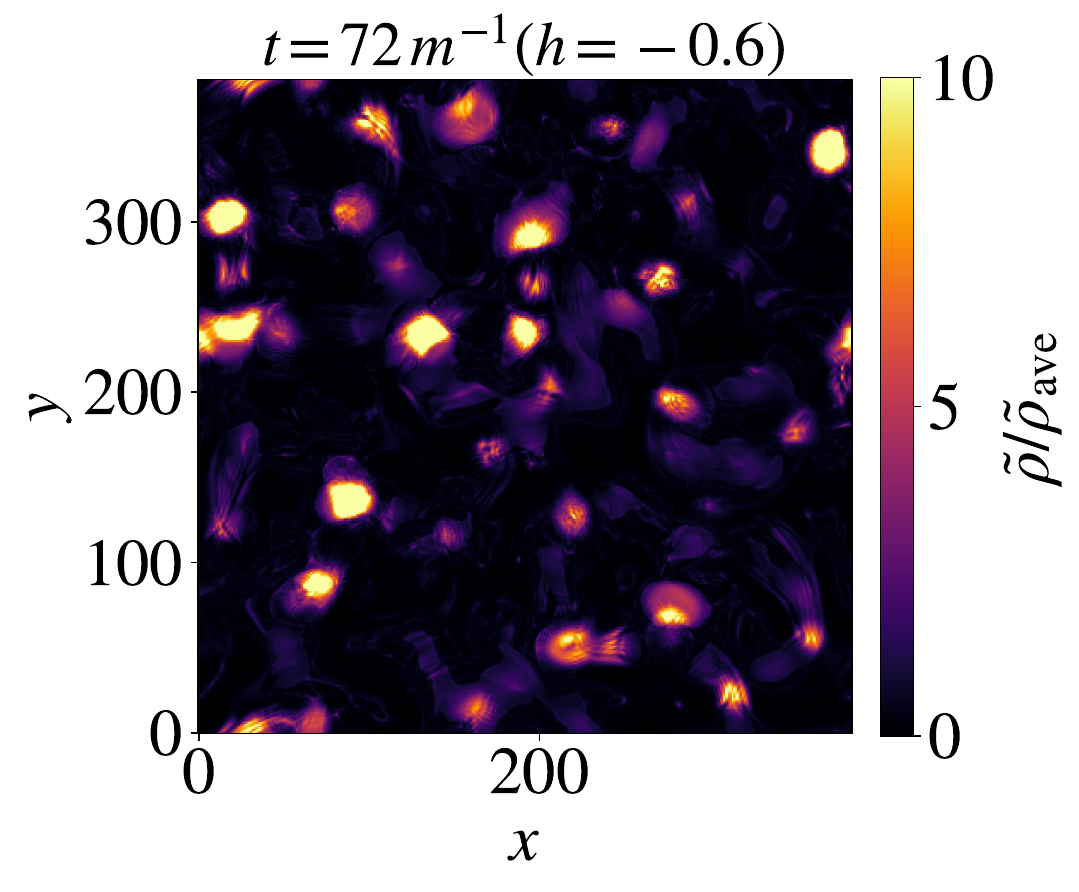}
\end{minipage}
\begin{minipage}{0.23\textwidth}
	\centering
	\includegraphics[width=\linewidth]{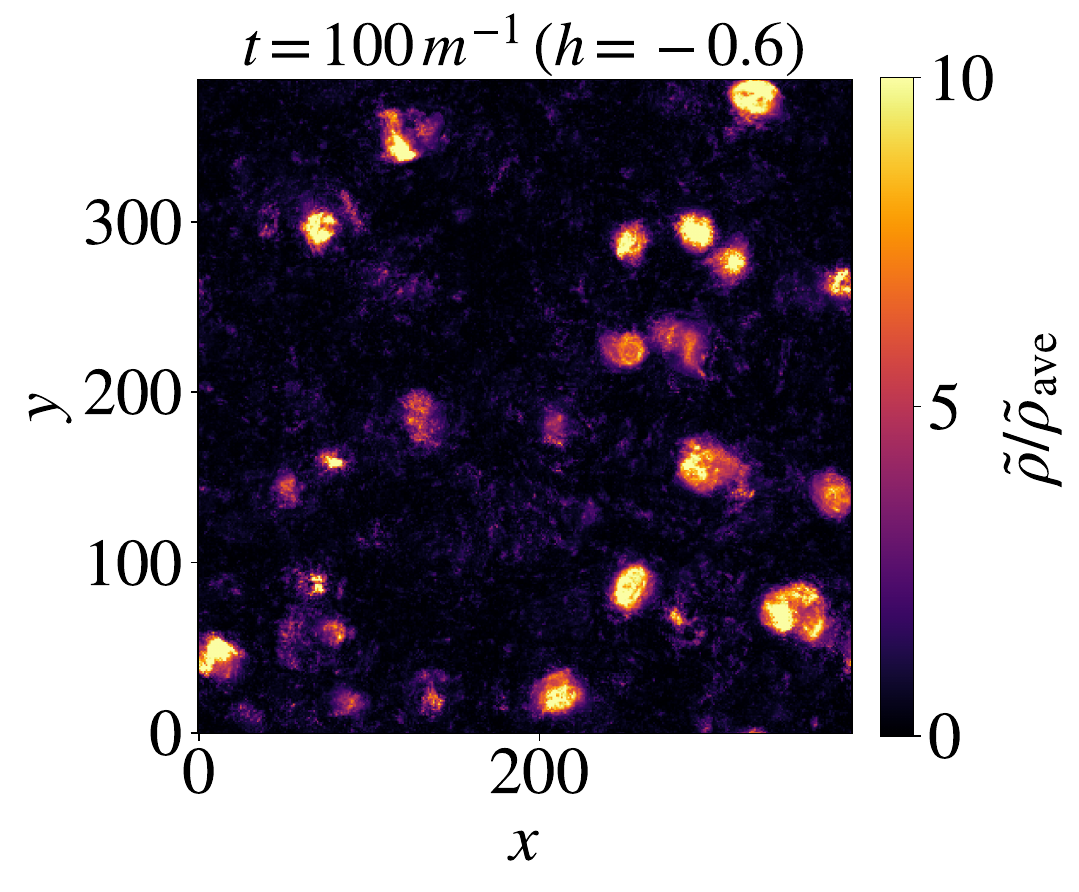}
\end{minipage}
\begin{minipage}{0.23\textwidth}
	\centering
	\includegraphics[width=\linewidth]{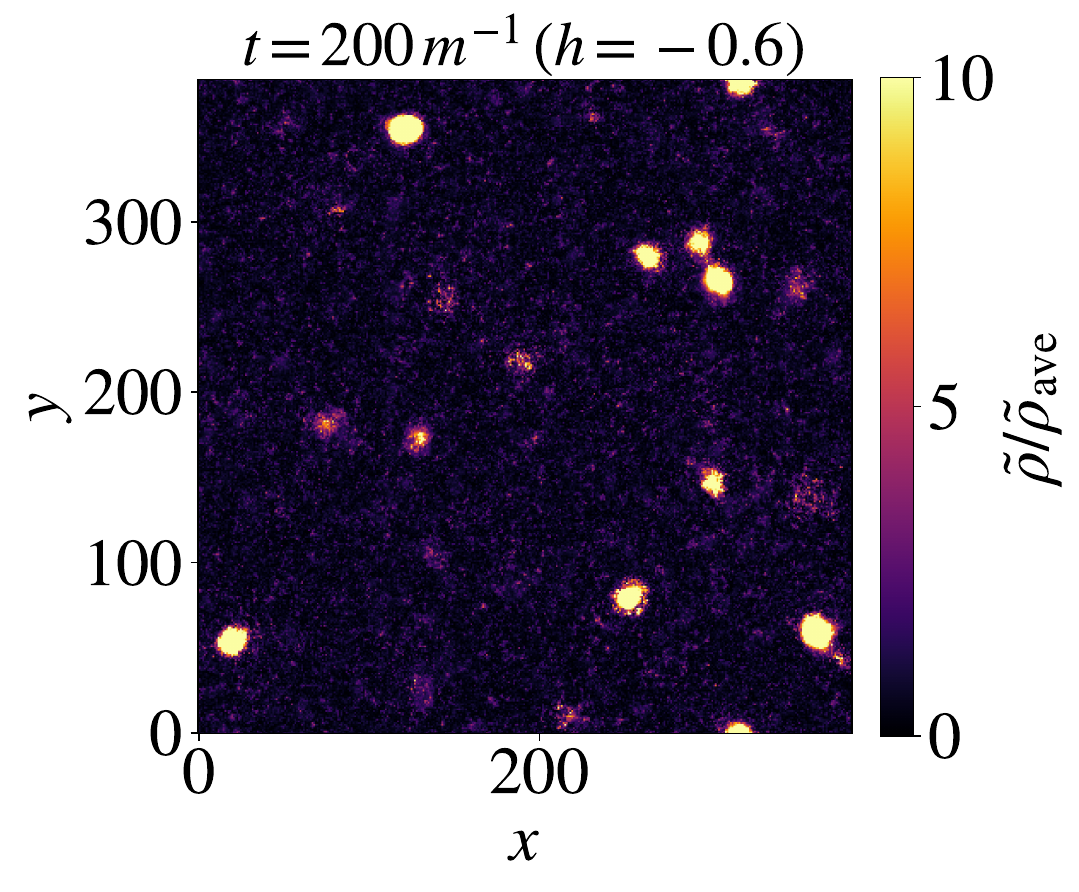}
\end{minipage}
\begin{minipage}{0.23\textwidth}
	\centering
	\includegraphics[width=\linewidth]{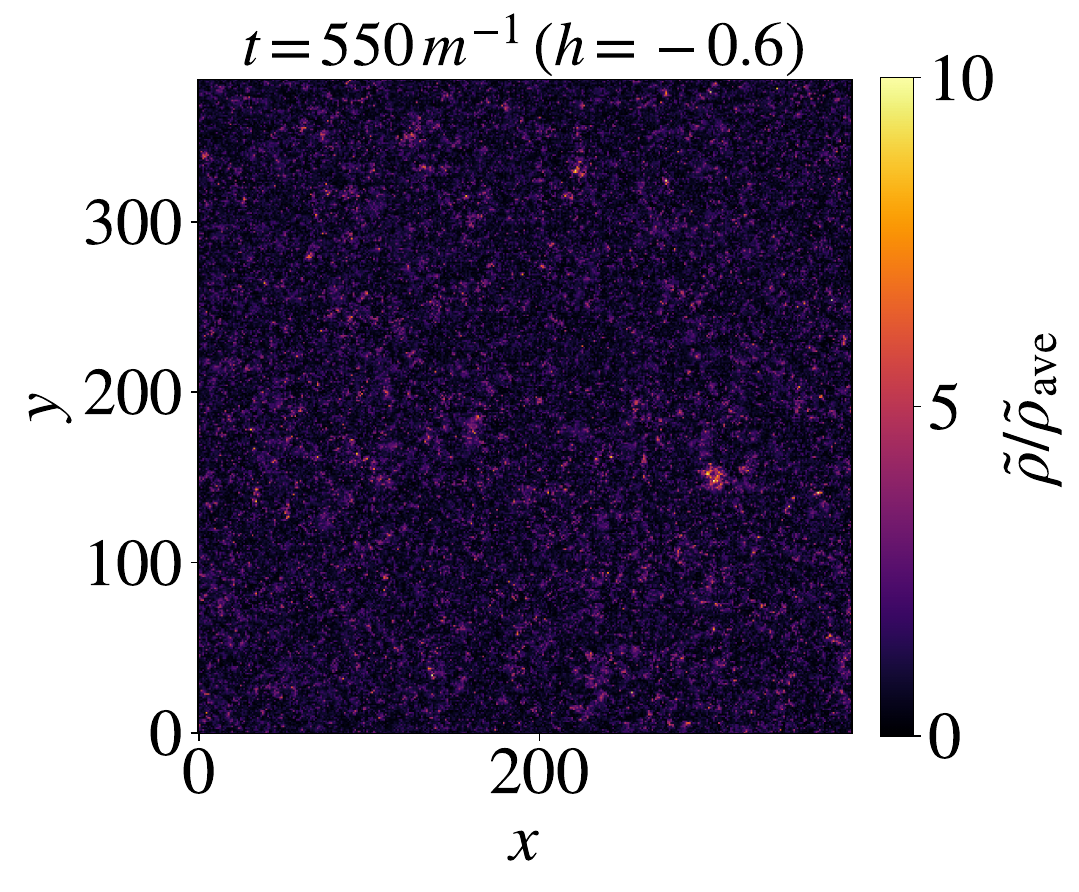}
\end{minipage}
	\caption{Two-dimensional snapshots of overdense objects for $h=0$ and $h=\pm 0.6$. The snapshots correspond to slices at $z=0$ of the simulation grid. The leftmost column corresponds to times close to the end of the resonance stage.}
	\label{snap_h06}
\end{figure*}

It is well known that the gradient energy plays a crucial role in oscillon formation.
During resonance, a significant fraction of the inflaton field energy is transferred into the fluctuations, leading to highly anisotropic spatial distribution of energy. After the resonance ends, provided that the potential has an appropriate shape—typically shallower than a quadratic form around its minimum, which induces an effective attractive self-interaction—the anisotropic energies can coalesce into spatially localized, long-lived, non-topological solitons with oscillatory behavior, known as oscillons. 
In numerical simulations, the formation of oscillons is typically identified by the emergence of spatially localized overdense regions with energy densities exceeding several times the average background value. The threshold energy density $\tilde{\rho}_{\text{th}}$ is commonly set in the range $\tilde{\rho}_{\text{th}}/\tilde{\rho}_{\text{ave}}$ between $3$ and $10$~\cite{Amin:2011hj,Zhou:2013tsa,Antusch:2017flz}, where $\tilde{\rho}_{\text{ave}}$ is the volume averaged total energy density in the box. We emphasize that different threshold criteria select different oscillon populations; consequently, the inferred oscillon properties can vary substantially depending on the adopted threshold. The threshold scan in appendix~\ref{app_threshold} shows that our main qualitative conclusions are stable for $\tilde{\rho}_{\text{th}}/\tilde{\rho}_{\text{ave}}\geq10$, whereas lower thresholds are sensitive to diffuse structures. We therefore adopt $\tilde{\rho}_{\text{th}}=10\tilde{\rho}_{\text{ave}}$ throughout the main text.  Fig.~\ref{snap_h06} presents the two-dimensional snapshots of the oscillon configurations for $h=0$ and $h=\pm 0.6$, respectively. These snapshots reveal distinguishable differences among the oscillons produced under different parameter choices. 

To capture the properties of oscillons, let us study the details of these overdense objects. We define the energy fraction of the overdense objects,
\begin{equation}
	f_{\text{osc}}(\tilde{t}) = \frac{\int_{\tilde{\rho}\geq\tilde{\rho}_{\text{th}}} \mathrm{d}^3 \tilde{x} \, \tilde{\rho}(\tilde{t},\tilde{\mathbf{x}})}
	{\tilde{\rho}_{\text{ave}}\times \text{volume}}.
\end{equation}
{ This definition includes all voxels above the density threshold. To determine the oscillon number $N_{\rm osc}$ and the geometric observables, we group these voxels into connected components and apply the selection criteria specified in appendix~\ref{app_threshold}. Let $\Omega_i$ denote the $i$th region satisfying the oscillon-identification criteria, with comoving volume $V_i^{\rm com}=\int_{\Omega_i}d^3\tilde{x}$. The total comoving volume and the population-averaged equivalent physical radius are
\begin{equation}
	\begin{aligned}
		{V}_{\rm osc}
		&=\sum_{i=1}^{N_{\rm osc}}V_i^{\rm com},\\
		\left\langle R_{\rm eq}^{\rm phys}\right\rangle
		&=\frac{a(t)}{N_{\rm osc}}
		\sum_{i=1}^{N_{\rm osc}}
		\left(\frac{3V_i^{\rm com}}{4\pi}\right)^{1/3}.\label{phys_radii}
	\end{aligned}
\end{equation}}
In the following, we will discuss the properties of the oscillons and the relation with the potential feature.

\subsubsection{Morphology, size, and abundance}
\begin{figure*}[t]
	\centering
	\includegraphics[width=17.5cm]{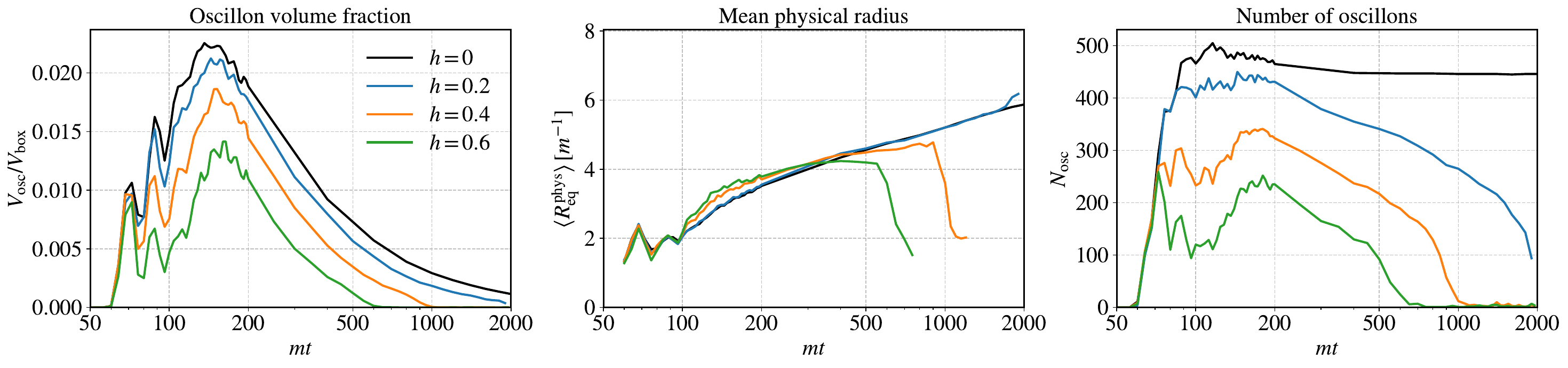}
	\includegraphics[width=17.5cm]{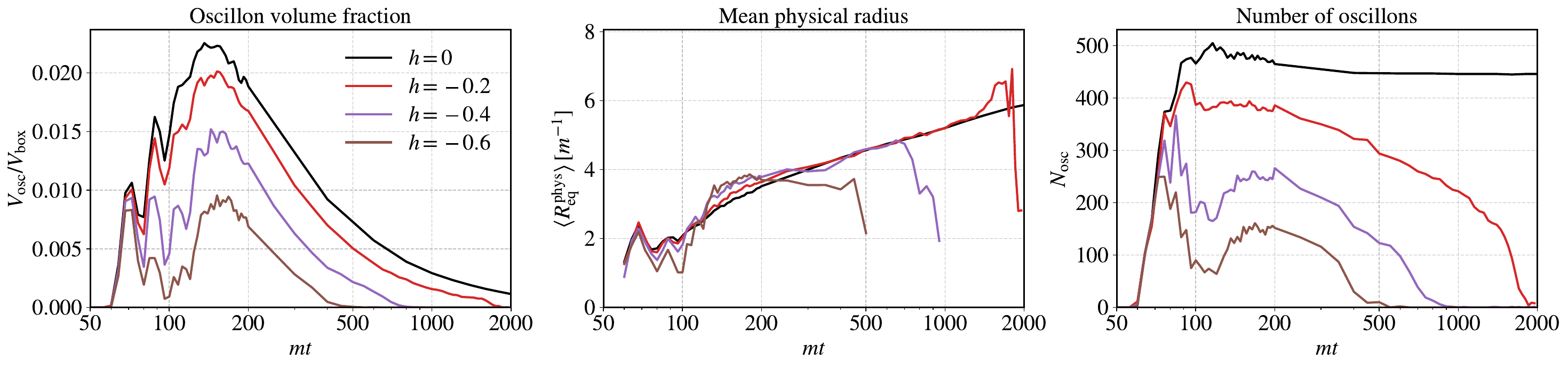}
	\caption{Evolution of the comoving volume fraction (left), the population-averaged equivalent physical radius (middle) {  defined by Eq.~(\ref{phys_radii})}, and the number of identified oscillons (right) for the parameter sets	$P_1$ (upper row) and $P_2$ (lower row). { Candidate regions satisfy $\tilde{\rho}\geq 10\tilde{\rho}_{\rm ave}$ and are identified using 26-neighbor connectivity with periodic boundaries. We retain components containing at least 15 voxels, extending less than half the box size in each direction, and satisfying the sphericity cut $\psi\geq0.6$. The identification procedure and its sensitivity to these cuts are discussed in appendix~\ref{app_threshold}.}
	}
	\label{shape_volume}
\end{figure*}

Figure~\ref{shape_volume} shows the volume fraction, population-averaged physical radius, and number of the identified overdense objects. At the end of resonance ($t_{\rm re}\sim70\,m^{-1}$), the comoving volume fraction selected by $\tilde{\rho}_{\rm th}=10\tilde{\rho}_{\rm ave}$ is approximately $1\%$, with little dependence on $h$. After resonance, the overdense regions coalesce and relax into oscillons around $t_{\rm of}\sim100\,m^{-1}$. The volume fraction reaches its maximum shortly thereafter and then decreases, with both its peak and late-time value increasingly suppressed for larger $|h|$.
	
At $t_{\rm re}$, the identified overdense regions remain irregular and their equivalent radii fluctuate substantially (see Fig.~\ref{snap_h06}). During the following $\sim30\,m^{-1}$, they relax into nearly spherical oscillons with typical physical radii of $1$--$3\,m^{-1}$. After formation, the mean physical radius generally increases and can reach several $m^{-1}$ in the longer-lived cases. This expansion is consistent with the microscopic analysis in Fig.~\ref{oscillon_profile}, where the Gaussian-fitted width increases as the oscillon amplitude decreases. Since $V_{\rm osc}^{\rm com}/V_{\rm box}^{\rm com}\propto
N_{\rm osc}\langle(R_{\rm eq}^{\rm phys})^3\rangle/a^3$, the comoving volume fraction can decrease despite the growth of individual physical radii, because cosmic expansion and the declining oscillon population dominate.

The oscillon abundance also exhibits a clear dependence on $h$. The potential deformation suppresses oscillon formation, with fewer oscillons being produced as its magnitude $|h|$ increases. It subsequently declines as oscillons decay and disappear, with the decrease occurring earlier for larger $|h|$. During decay, small low-density fragments may be counted as additional objects at lower thresholds, producing an apparent increase that does not represent the formation of new compact oscillons, as discussed in appendix~\ref{app_threshold}.

\begin{figure*}[t]
	\centering
			\includegraphics[width=17.0cm]{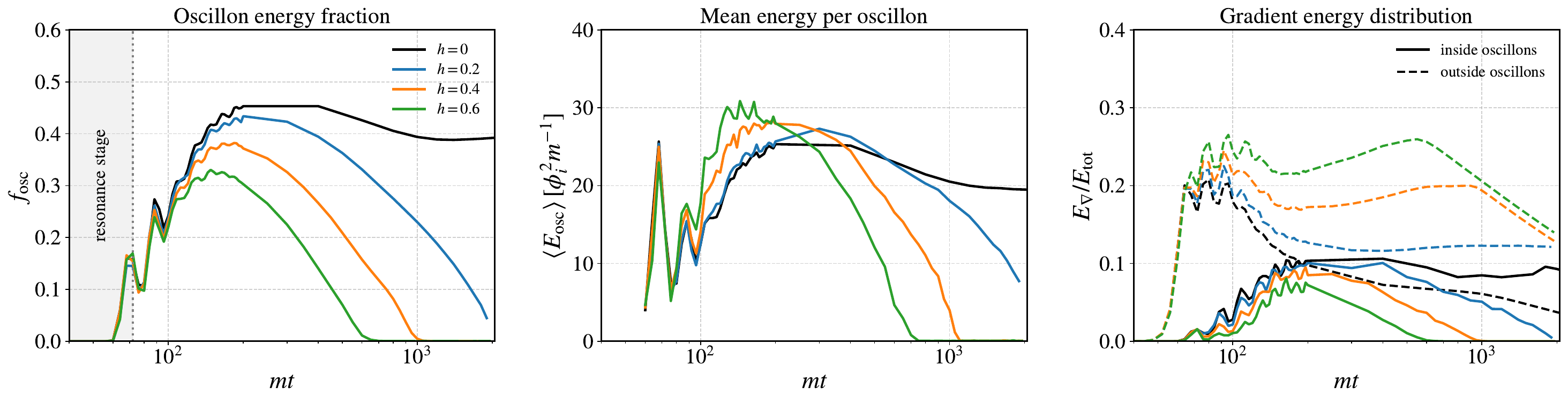}
			\includegraphics[width=17.0cm]{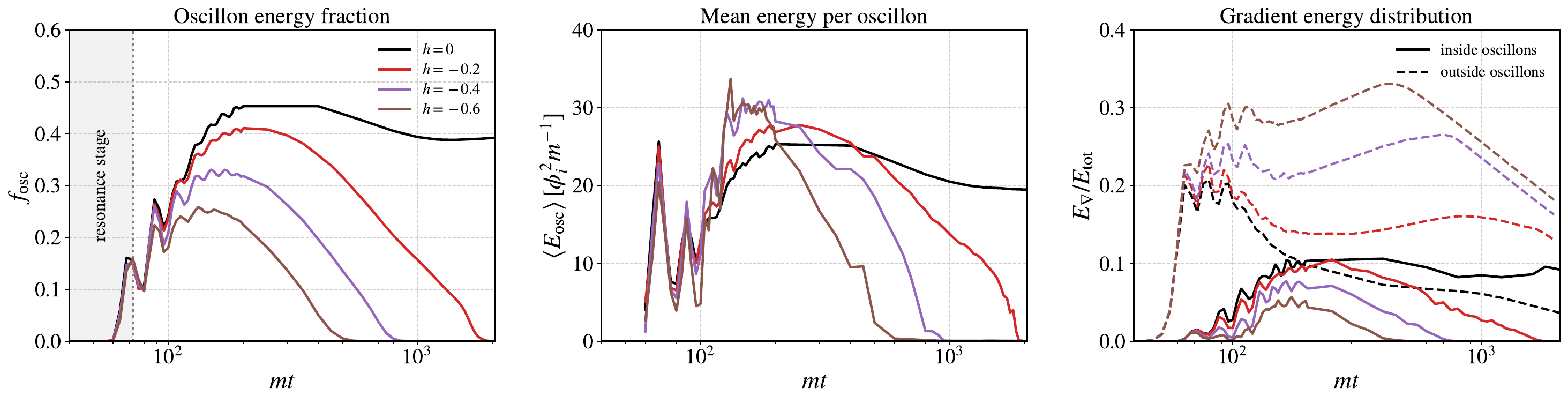}
	\caption{Energy diagnostics for the oscillon populations identified with $\tilde{\rho}_{\rm th}=10\,\tilde{\rho}_{\rm ave}$. The upper and lower rows correspond to the $P_1$ and $P_2$ parameter sets, respectively. From left to right, the panels show the total oscillon energy fraction $f_{\rm osc}=E_{\rm osc}/E_{\rm tot}$, the mean energy per oscillon in units of $\phi_i^2/m$, and the fractions of the total energy in gradient energy inside (solid) and outside (dashed) the oscillons. Colors distinguish the values of $h$ indicated in the legend. The shaded region and vertical dotted line in the left panels mark the resonance stage and its end, respectively.
	}
	\label{energies}
\end{figure*}

\begin{figure*}[t]
	\includegraphics[width=17.0cm]{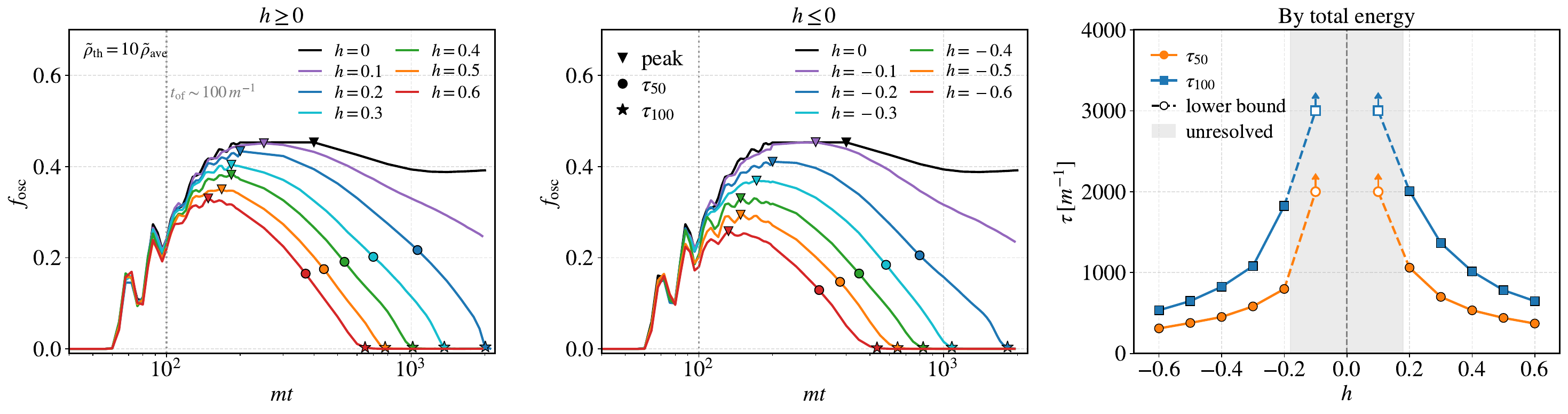}
	\includegraphics[width=17.0cm]{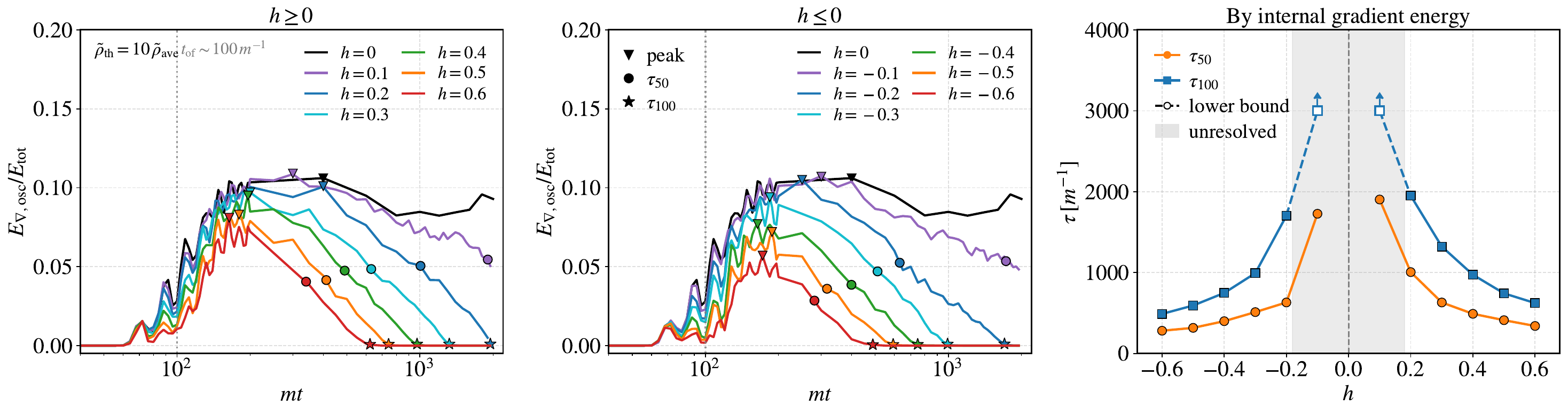}
\caption{
		Evolution of the total oscillon energy fraction (upper row) and the internal gradient-energy fraction (lower row), using the identification threshold $\tilde{\rho}_{\rm th}=10\tilde{\rho}_{\rm ave}$. The left and middle columns show the results for $h\geq0$ and $h\leq0$, respectively. Downward triangles mark the peak values, while circles and stars mark $\tau_{50}$ and $\tau_{100}$, where $\tau_{50}$ and $\tau_{100}$ are defined as the times at which the corresponding energy has decreased by $50\%$ and completely vanished, respectively. The right column shows these decay times as functions of $h$, with orange circles and blue squares denoting $\tau_{50}$ and $\tau_{100}$, respectively. Filled markers denote resolved values, whereas open markers with upward arrows indicate lower bounds. { The $\tau_{100}$ lower bounds at $h=\pm0.1$ are obtained from the simulations with $N^3=512^3$.} Dashed segments adjoining such points are shown only as guides to the eye. The gray bands highlight the small-$|h|$ region in which at least one decay time remains unresolved. No lifetime constraint is assigned at $h=0$ because its lifetime extends far beyond the present numerical reach.
}
\label{energy_lifetime}
\end{figure*}

\subsubsection{Energy and lifetime}
Perhaps the most important properties of oscillons are their energy and lifetime.
The total energy fraction of the overdense objects and the averaged energy contained per object are shown in Fig.~\ref{energies}. Only part of the oscillon energy is inherited from the resonance stage. At the end of resonance, dense objects contain $\sim15\%$ of the total energy for $\tilde{\rho}_{\rm th}=10\,\tilde{\rho}_{\rm ave}$, with this fraction decreasing mildly as $|h|$ increases. Consistent with our earlier discussion of the gradient energy, the potential feature mildly modifies the energy transfer during the resonance stage, while its overall effect remains limited.

Since oscillons are relatively stable and approximately spherically symmetric objects, we examine the energy fraction in dense regions and the average energy per object. Combined with the earlier analysis of their morphology, we identify that oscillons form at $t\sim 100\,m^{-1}$, after which the oscillon energy begins to grow steadily and their morphology becomes essentially stable. In the interval between the end of resonance and oscillon formation, the energy of dense regions continues to increase due to the self-interaction of the potential. The magnitude of this energy growth depends on $h$, with smaller $h$ leading to a larger increase. Consequently, the $h=0$ case exhibits the largest growth, reaching a final energy fraction of $\sim 45\%$, whereas for $h=0.6$ the fraction peaks at only $\sim 30\%$. Larger values of $h$ correspond to a shorter growth phase and an earlier onset of decay. Shortly after formation, the mean energy per oscillon is slightly lower in the $h=0$ model than in the models with $h\neq0$. This initial ordering does not imply greater stability: oscillons in the $h\neq0$ models lose energy and decay both earlier and more rapidly, whereas the $h=0$ population persists for much longer.

Now let us study the lifetime of oscillons. We now focus on a statistical analysis of their collective properties. It was suggested that the oscillon lifetime could be determined from the time at which the gradient energy in the simulation box decays as
$a^{-3}$~\cite{Shafi:2024jig}. However, we argue that this method is not sufficiently accurate. The gradient energy is distributed throughout the simulation volume, with oscillons accounting for only a fraction of it. Because the evolution of gradient energy inside and outside of the oscillons differs, tracking the total gradient energy cannot reliably determine the oscillon lifetime.

To see this more clearly, let us track the evolution of the gradient energy inside and outside of the oscillons. We extract the gradient energy inside and outside of the oscillons from the \texttt{HDF5} data and the results are shown in Fig.~\ref{energies}. Clearly, for $h=0$ the gradient energy contained inside the oscillons exceeds that of outside, whereas for $h\neq0$ the gradient energy is predominantly distributed outside the oscillons. More importantly, we find that the evolution of the gradient energy inside and outside of the oscillons is qualitatively different. For $h=0$,
the fraction of gradient energy contained inside of the oscillons remains approximately constant, indicating that this component redshifts as $a^{-3}$, characteristic of matter-like behavior. In contrast, the gradient energy fraction outside oscillons decreases with time, implying that the total gradient energy does not scale globally as $a^{-3}$.
For $h\neq0$, the gradient energy fraction inside oscillons decays approximately linearly with time, with the decay rate increasing for larger values of $h$; the gradient energy outside oscillons exhibits a more intricate behavior: for larger values of $h$, the external component undergoes a period of growth, partially sourced by the energy released during oscillon decay. Once the energy inside oscillons has decayed to zero, i.e., after the complete decay of oscillons, the total gradient energy fraction in the box exhibits a power-law decay in time, appearing as a straight line on a logarithmic vertical scale in figures. Consequently, in all cases we find that the total gradient energy within the simulation box does not redshift as $a^{-3}$.
These observations indicate that the total gradient energy may not provide a fully reliable measure of the oscillon lifetime and merits further examination. By contrast, the gradient energy evolution within the overdense objects provides a more robust diagnostic. Since there is no unique definition for the lifetime of oscillons, particularly in statistical calculations, we compute the characteristic decay times for the gradient energy inside oscillons, measuring both the time at which the energy decreases to $50\%$ of its peak value and the time for its full dissipation. These two reference criteria are adopted because the $50\%$ decay marks the stage at which the oscillon energy ceases to be dominant, while $100\%$ decay indicates that even the longest-lived oscillons in the box have disappeared. From a statistical perspective, the averaged oscillon lifetime is expected to lie between these two limits. The results are shown in Fig.~\ref{energy_lifetime}.

We also estimate the oscillon lifetime from the total energy contained in the identified oscillon regions, which complements the internal gradient energy as a diagnostic of spatial localization. Following the same procedure, we determine the $50\%$-decay time $\tau_{50}$ and the complete-decay time $\tau_{100}$ for both quantities. The results are summarized in Fig.~\ref{energy_lifetime} and lead to the following conclusions:
\begin{itemize}
	\item The two diagnostics give consistent qualitative dependence on $h$ and nearly the same lifetime hierarchy. The total-energy decay times are generally slightly longer, since the internal gradient energy can decrease as an oscillon loses spatial localization while part of its kinetic and potential energy remains within the identified region. The difference becomes more pronounced for small $|h|$, where the oscillons survive close to or beyond the reliable simulation window and their comoving cores become increasingly difficult to resolve. We therefore regard the results in this region as only partially resolved rather than as a significant disagreement between the two diagnostics.
	
	\item A nonzero $h$ substantially shortens the oscillon lifetime, particularly for $|h|>0.1$, and the dependence on $h$ is strongly nonlinear. For a fixed $|h|$ within the resolved range, negative $h$ generally produces a shorter lifetime than positive $h$. Thus, dip-type features suppress the persistence of oscillons somewhat more efficiently than bump-type features of the same magnitude. This ordering is reproduced by both the total-energy and internal-gradient-energy diagnostics.
	
	\item The threshold scan presented in appendix~\ref{app_threshold} further tests the robustness of these lifetime estimates. The results obtained with $\tilde{\rho}_{\rm th}=3\tilde{\rho}_{\rm ave}$ differ appreciably from those at higher thresholds, while the deviations at $\tilde{\rho}_{\rm th}=5\tilde{\rho}_{\rm ave}$ are smaller but remain visible. These low thresholds also select diffuse envelopes, transient fragments, and surrounding overdense fluctuations. By contrast, for $10\leq\tilde{\rho}_{\rm th}/\tilde{\rho}_{\rm ave}\leq50$, $\tau_{100}$ and the qualitative dependence on $h$ become nearly threshold independent, while $\tau_{50}$ retains a moderate threshold dependence.
\end{itemize}

A shorter oscillon lifetime allows the energy stored in localized configurations to be released earlier. In the presence of efficient couplings to other fields, this may accelerate the approach to radiation domination and shorten the reheating stage, thereby modifying the post-inflationary expansion history and the mapping between inflationary predictions and observables. A quantitative assessment of this possibility requires simulations including the relevant couplings and thermalization processes.

\begin{figure*}[t]
	\includegraphics[width=4.8cm]{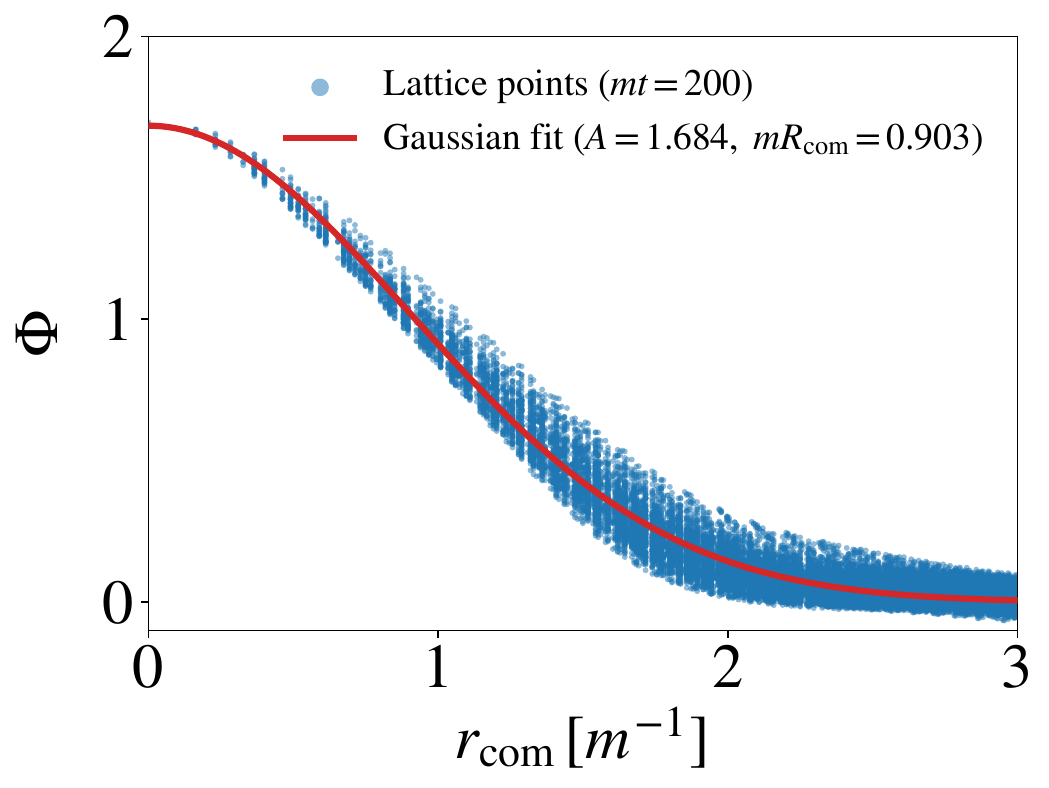}
	\includegraphics[width=4.8cm]{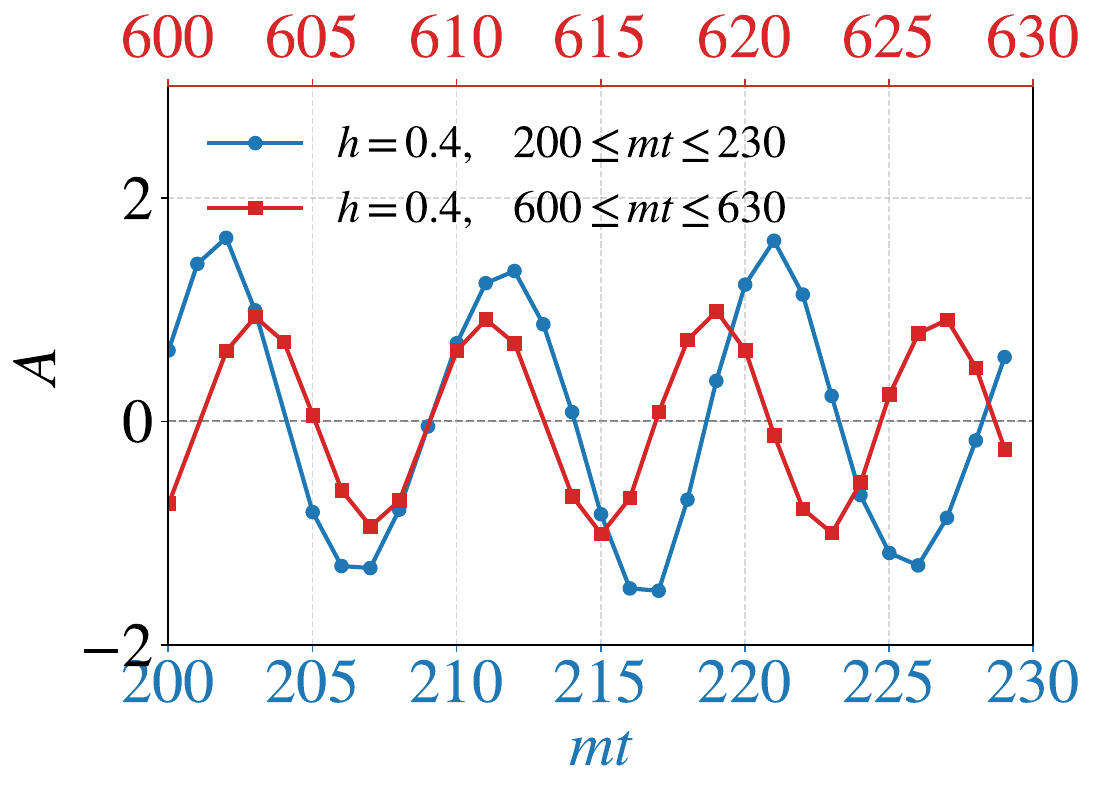}
	\includegraphics[width=4.8cm]{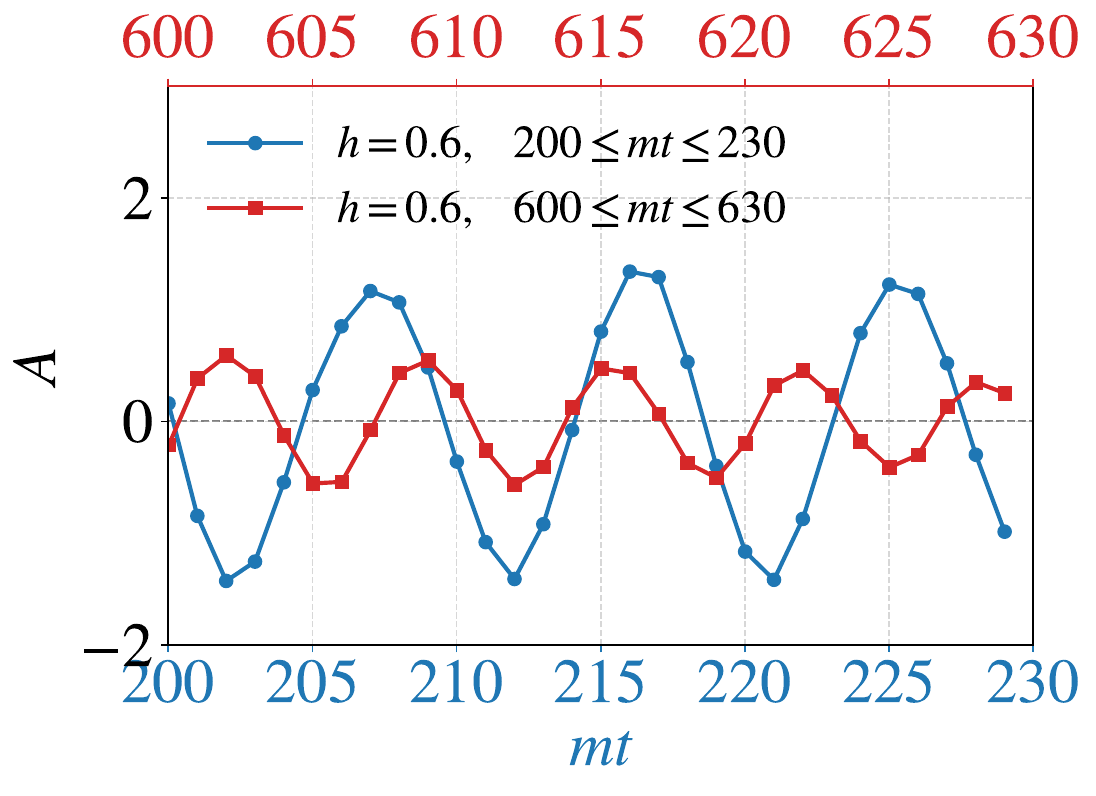}
	\includegraphics[width=4.8cm]{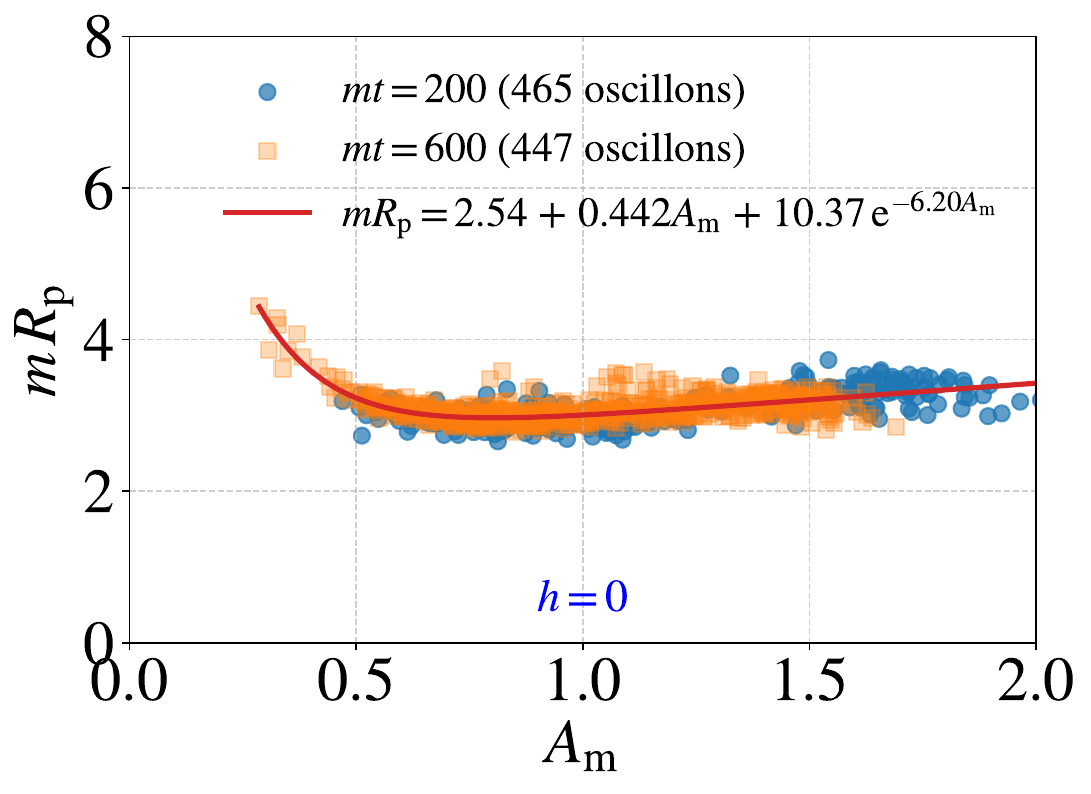}
	\includegraphics[width=4.8cm]{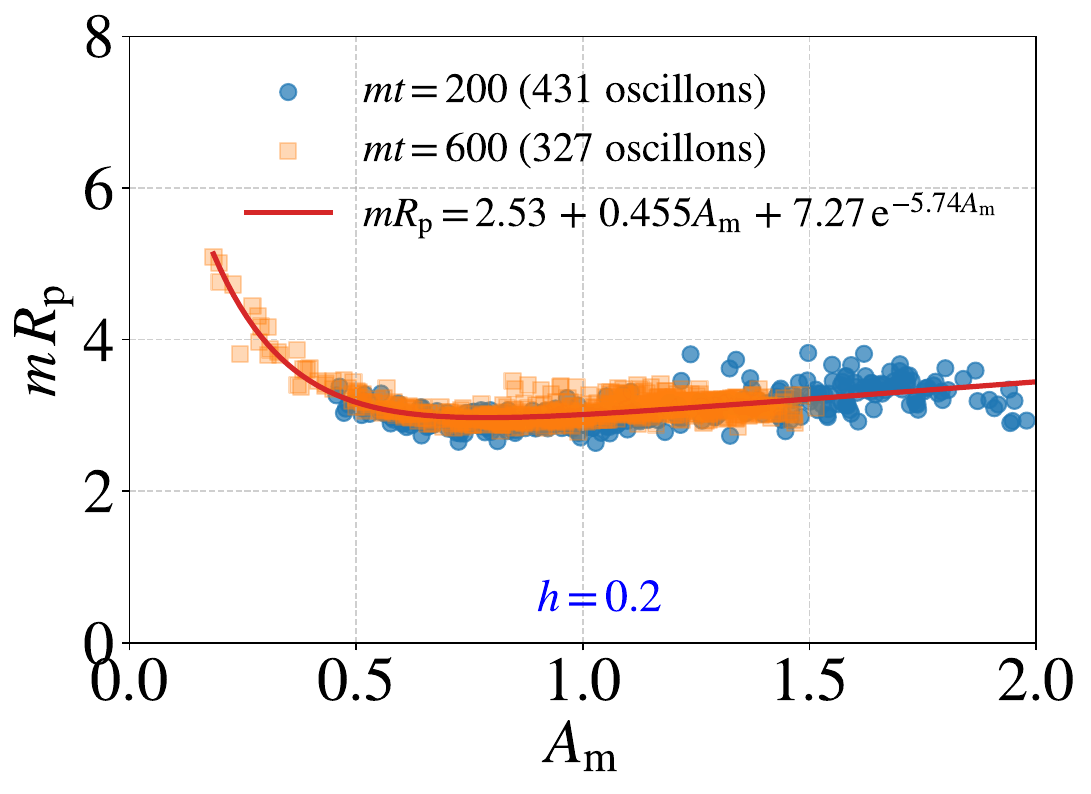}
	\includegraphics[width=4.8cm]{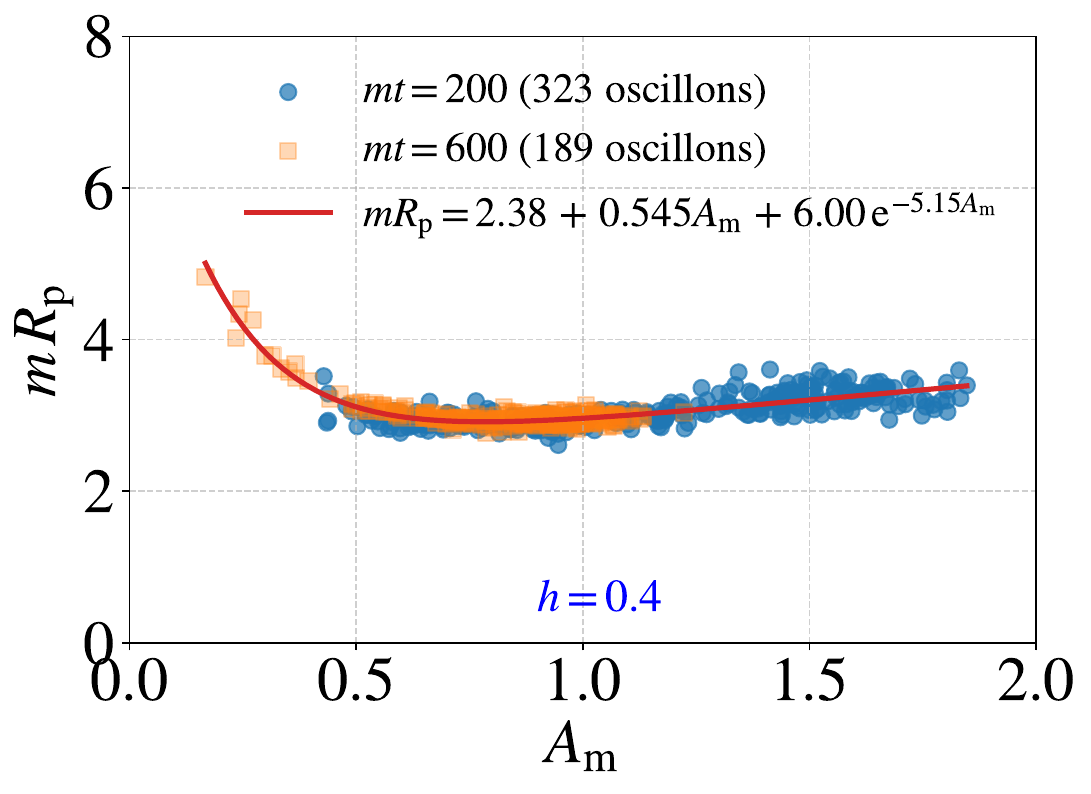}
	\includegraphics[width=4.8cm]{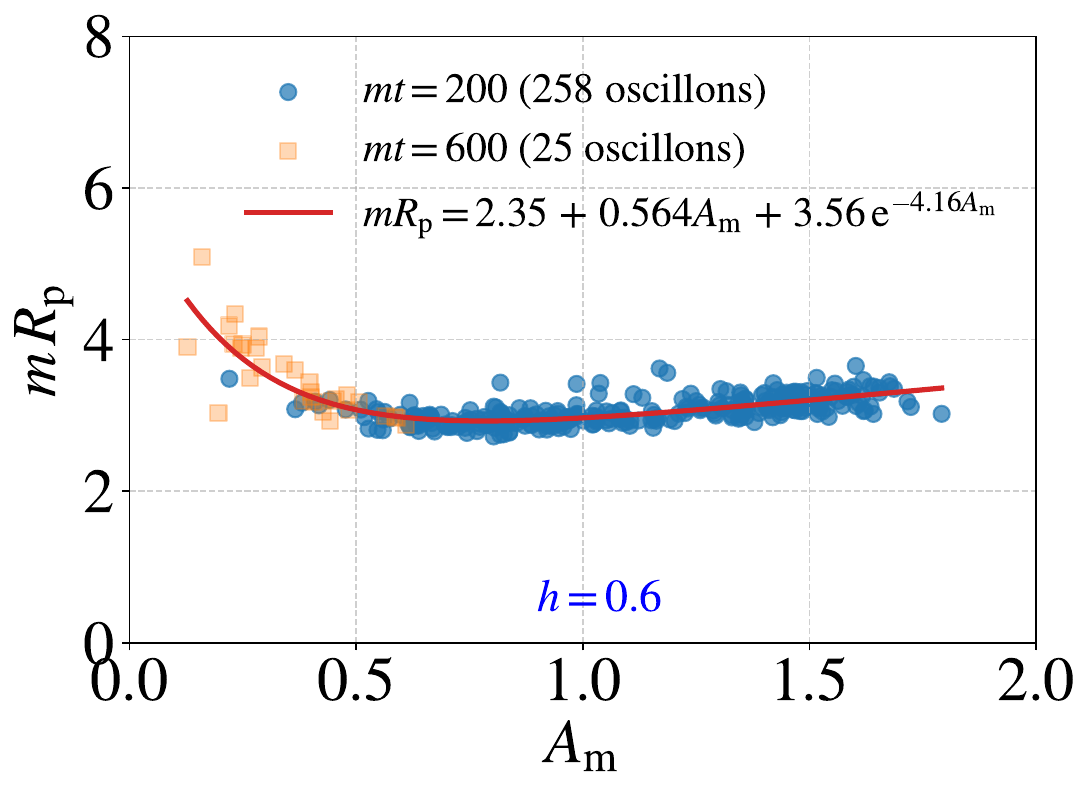}
	\includegraphics[width=4.8cm]{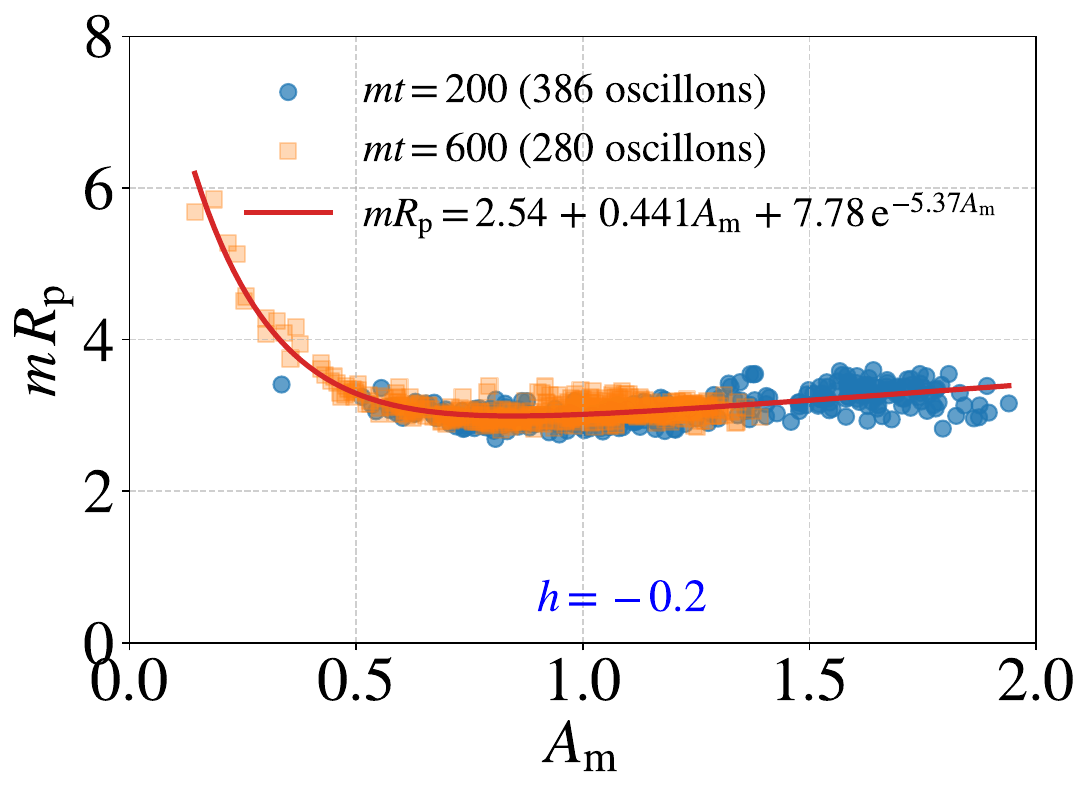}
	\includegraphics[width=4.8cm]{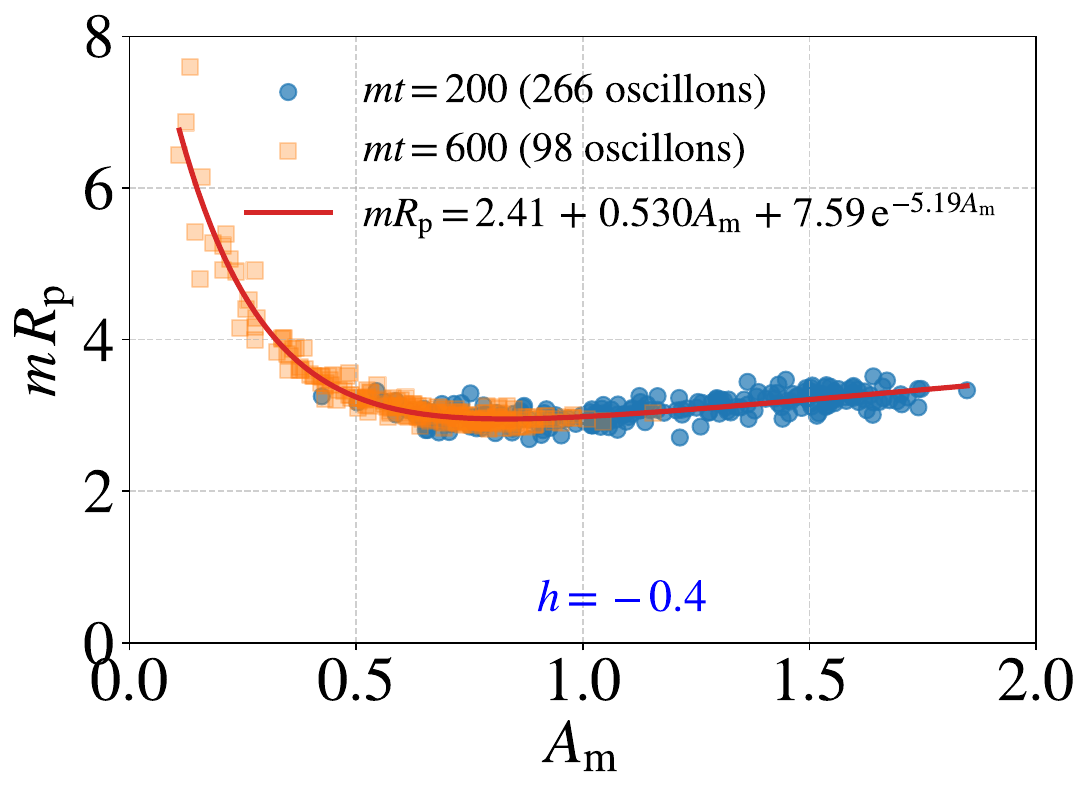}
	\caption{Microscopic properties and dynamics of oscillons. The upper row shows a representative radial field profile with its Gaussian fit (left) and the fitted $A(t)$ obtained from densely sampled outputs over two representative intervals of duration $\Delta(mt)=30$ (middle and right). The second and third rows show the physical width $R_{\rm p}$ as a function of the amplitude $A_{\rm m}$ at $t=200\,m^{-1}$ and $600\,m^{-1}$ for the indicated values of $h$. The solid fitted curves are empirical fits to the combined data at $t=200\,m^{-1}$ and $600\,m^{-1}$, using the phenomenological parametrization of Ref.~\cite{Piani:2025dpy}.}
	\label{oscillon_profile}
\end{figure*}

\subsubsection{Microscopic properties and dynamics of oscillons}
In the preceding discussion, we focused primarily on the global energetics of the system. In this subsection, we turn to a microscopic analysis of the properties and dynamics of individual oscillons. Because oscillons are approximately spherically symmetric, long-lived, quasibreather-like localized configurations, we model their field profiles using the Gaussian ansatz,
\begin{equation}
	\Phi(t,r)=\frac{\phi}{\phi_i}=A(t)\,e^{-{r^2}/{R_{\rm c}^2}}.
\end{equation}
Here, $A(t)$ encodes both the amplitude and the quasiperiodic oscillatory behavior of the oscillon, and we define its peak value, $A_{\rm m}$, as the oscillon amplitude. The parameter $R_{\rm c}$ denotes the comoving width, while the corresponding physical width is $R_{\rm p}=a\,R_{\rm c}$. In the upper-left panel of Fig.~\ref{oscillon_profile}, we show a representative oscillon and its radial field profile, demonstrating that the Gaussian ansatz provides an excellent approximation to its internal configuration. 

However, a fit performed at a single time slice is insufficient to reliably determine the amplitude and radius of an oscillon. Because oscillons undergo quasiperiodic oscillations, a given snapshot may sample an arbitrary oscillation phase, and the instantaneous fitted value, $A(t)$, therefore does not necessarily coincide with the peak amplitude $A_{\mathrm{m}}$. An accurate determination of these quantities consequently requires tracking each oscillon over several complete oscillation cycles. For this purpose, we continuously identify and track the oscillons within two time intervals $200\,m^{-1}\leq t\leq230\,m^{-1}$ and $600\,m^{-1}\leq t\leq630\,m^{-1}$ (see the middle and right panels in the upper row of Fig.~\ref{oscillon_profile} for representative oscillons), performing a Gaussian fit to the profile of every oscillon at each output time ($\Delta t=1\,m^{-1}$). Note that we assume that the oscillon amplitude, $A_{\mathrm{m}}$, and its radius remain approximately constant over the two short time intervals. Using this procedure, we obtain the amplitude–radius distributions of all oscillons at the two time slices $t=200\,m^{-1}$ and $t=600\,m^{-1}$ \footnote{The former corresponds to the early stage following oscillon formation in all models, whereas the latter probes the final stage of the oscillon lifetime in the $h=\pm0.6$ model, which exhibits the shortest-lived oscillons.}, shown in the second row of Fig.~\ref{oscillon_profile}.

These statistical results reveal three main features. First, in each model, the oscillons populate a well-defined amplitude--width relation, consistent with previous results~\cite{Gleiser:2008ty,Gleiser:2009ys,Amin:2010jq,Piani:2025dpy}. Their physical widths remain narrowly clustered around $R_{\rm p}\sim 3\,m^{-1}$ over the intermediate-amplitude range, but increase rapidly once $A_{\rm m}$ becomes sufficiently small, indicating appreciable radial expansion only during the terminal stage of decay.

Second, a rough comparison suggests that the amplitude--width relations are broadly similar across the models considered for $A_{\rm m}\gtrsim0.5$, whereas more noticeable model-dependent differences emerge in the low-amplitude regime. This suggests that the potential feature has only a modest effect on the quasibreather configurations at moderate and large amplitudes, while its influence becomes more pronounced as the oscillons approach the low-amplitude terminal regime. Its primary effect is instead dynamical: at later times, the oscillon populations in the featured models shift toward smaller amplitudes and become increasingly depleted as $h$ increases. This shows that the feature accelerates their evolution toward the low-amplitude terminal branch. Together with the enhanced gradient energy outside the oscillon cores discussed in the preceding subsection, these behaviors are consistent with enhanced continuous radiative energy loss.

Third, the characteristic oscillation period decreases from $T_{\rm osc}\simeq9\text{--}10\,m^{-1}$ shortly after formation toward $T_{\rm osc}\simeq2\pi m^{-1}$ as the amplitude decreases. The $h=0.6$ example at $t\simeq600\,m^{-1}$ is already close to this limit, whereas the $h=0.2$ and $0.4$ examples still exhibit somewhat longer periods. This behavior admits a natural interpretation within the quasibreather picture. As an oscillon radiates energy, its amplitude decreases and its fundamental frequency increases toward the mass threshold $\omega=m$, corresponding to $T_{\rm osc}=2\pi m^{-1}$. Previous studies have shown that this quasistatic evolution may terminate at a critical frequency $\omega_c\lesssim m$, at which the energy of the quasibreather family is minimized. Further radiative energy loss can then no longer be accommodated by adiabatic evolution along the quasibreather branch, and the oscillon rapidly decays~\cite{Saffin:2006yk,Salmi:2012ta}. 
{ These observations support enhanced radiative energy loss as a possible explanation for the shorter lifetimes, although confirming this decay channel would require a direct measurement of the outward energy flux or radiation spectrum of individual oscillons.}

\subsubsection{Role of the $Z_2$ asymmetry in the nonlinear dynamics}
The Floquet analysis presented above shows that the symmetric feature generally produces a stronger linear instability than the one-sided case. It is therefore important to determine whether this enhancement
promotes the formation of oscillons or instead leads to more diffuse nonlinear fragmentation. To address this question, we perform dedicated lattice simulations for the symmetric potential and compare them with
the one-sided results using the same numerical setup, initial conditions, and oscillon-identification criteria.

\begin{figure*}[t]
	\centering
	
	\subfloat[\label{fig:z2-grad}]{
		\includegraphics[width=0.235\textwidth]
		{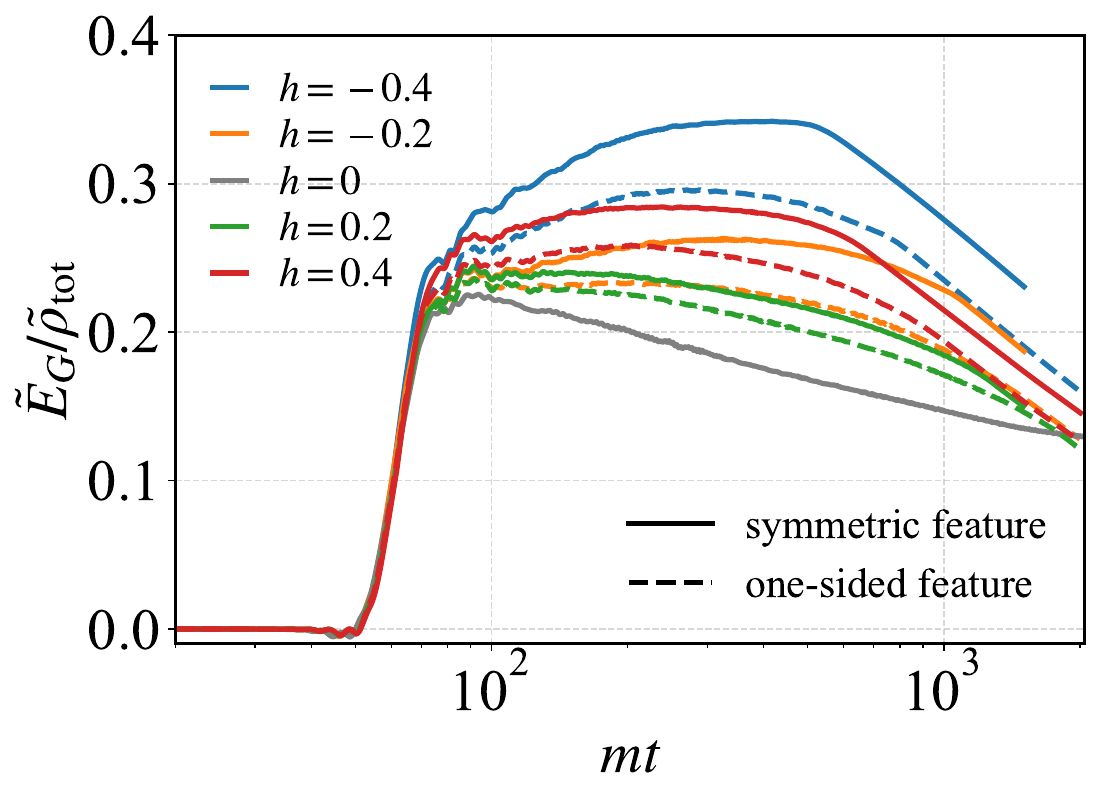}
	}
	\hfill
	\subfloat[\label{fig:z2-gradinout}]{
		\includegraphics[width=0.235\textwidth]
		{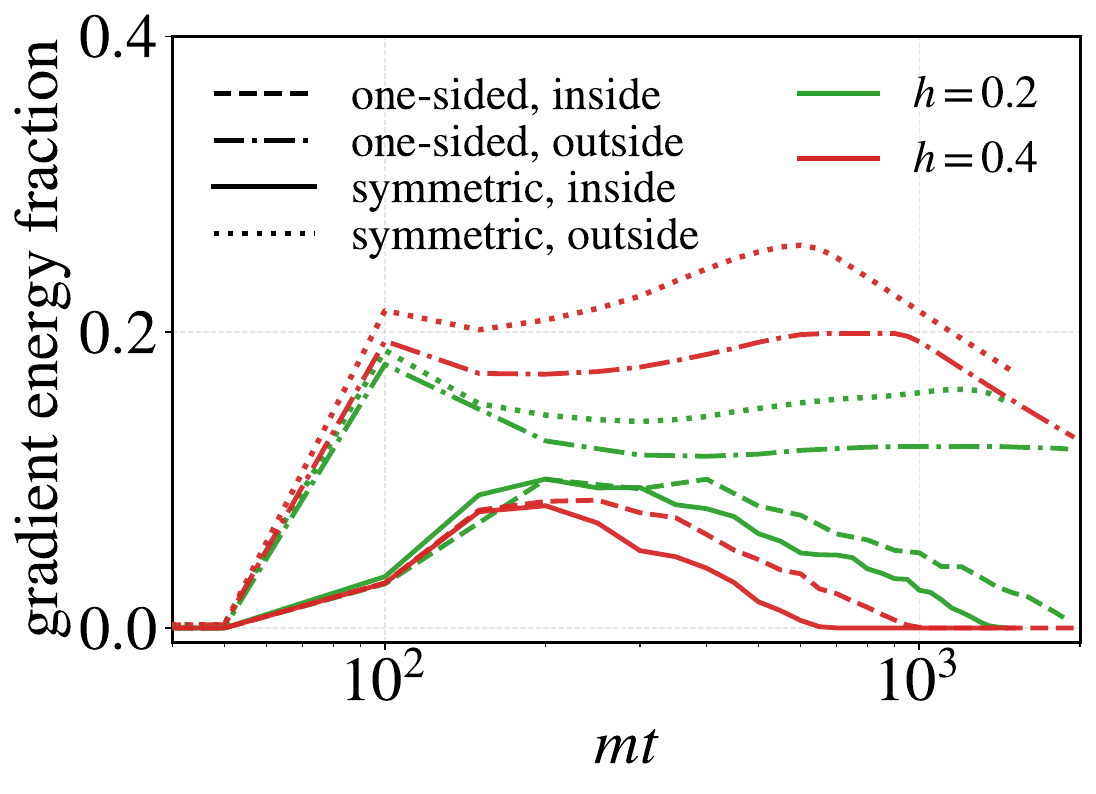}
	}
	\hfill
	\subfloat[\label{fig:z2-eosc}]{
		\includegraphics[width=0.235\textwidth]
		{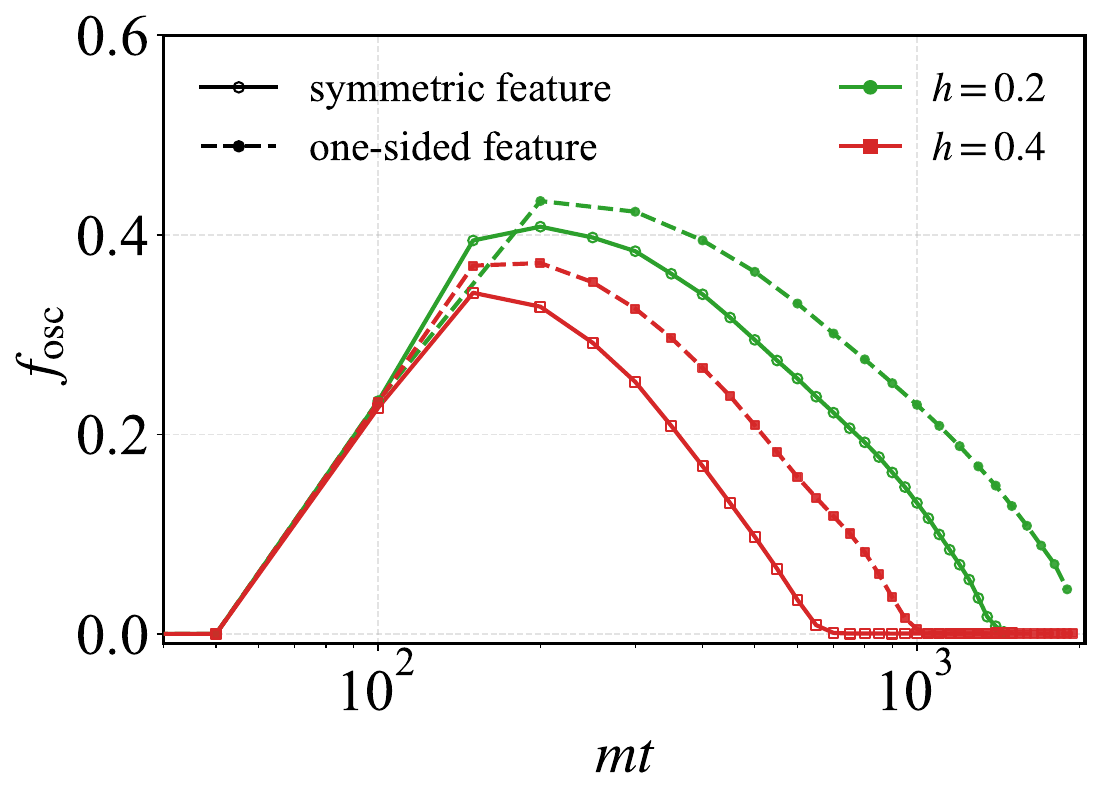}
	}
	\hfill
	\subfloat[\label{fig:z2-ar-h04}]{
		\includegraphics[width=0.235\textwidth]
		{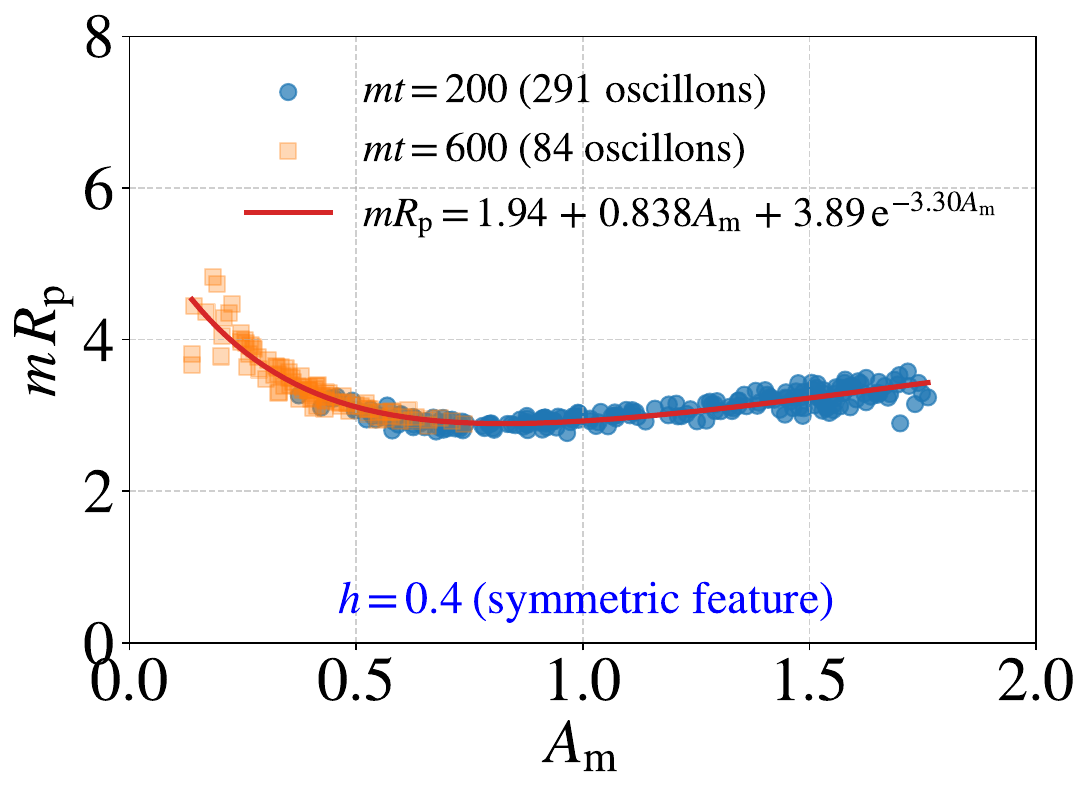}
	}
	
	\caption{
		Comparison of the nonlinear dynamics for the one-sided and $Z_2$-symmetric features. Panel (a) shows the evolution of the total gradient energy, while panel (b) separates the gradient energy into	contributions from inside and outside the identified oscillons. Panel (c) compares the total energy stored in the oscillon population. Panel (d) shows the physical width $mR_{\rm phys}$ as a function of the amplitude $A_{\rm m}$ for the symmetric potential with $h=0.4$, evaluated around $mt=200$ and $600$. The same numerical setup and oscillon-identification criteria are used in all
		comparisons.}
	\label{fig:z2-nonlinear}
\end{figure*}

As shown in Fig.~\ref{fig:z2-nonlinear}(a), the symmetric feature produces a larger fraction of total gradient energy than the one-sided feature. The spatial decomposition presented in Fig.~\ref{fig:z2-nonlinear}(b), however, shows that most of this enhancement is associated with regions outside the identified oscillon cores. In other words, the stronger resonance does not primarily concentrate the gradient energy inside compact, high-density configurations. Instead, it generates a more spatially dispersed distribution of inhomogeneous field fluctuations.

This distinction is further reflected in the evolution of the total energy stored in oscillons, shown in Fig.~\ref{fig:z2-nonlinear}(c). The oscillon energy is appreciably smaller in the symmetric case and decreases more rapidly at late times. According to the same decay criterion employed for the one-sided potential, this behavior corresponds to a shorter lifetime of the oscillon population. Thus, although restoring the $Z_2$ symmetry enhances the initial instability and the subsequent fragmentation, a smaller fraction of the amplified energy is captured and retained by long-lived localized configurations.

Figure~\ref{fig:z2-nonlinear}(d) provides a complementary microscopic view by showing the amplitude--width relation for the symmetric potential with $h=0.4$. The oscillon population approximately follows the same characteristic amplitude--width trajectory at the two extraction times, but shifts toward the small-amplitude regime as the system evolves. Compared with the corresponding one-sided result presented above, this shift occurs more rapidly in the symmetric potential. Moreover, the number of identified oscillons also decreases more rapidly in the symmetric case. These observations are consistent with the faster decrease of the total oscillon energy and indicate an accelerated weakening of the localized configurations.

Taken together, these results demonstrate that restoring the $Z_2$ symmetry enhances the resonance and produces stronger nonlinear fragmentation, but makes the amplified energy less likely to remain localized in oscillon cores. The excess gradient energy is instead predominantly distributed outside the oscillons, while the energy stored in the oscillon population is reduced and dissipates more rapidly. The $Z_2$ asymmetry is therefore not essential for the appearance of the resonance bands or for oscillon formation itself, but it has an important quantitative effect on the efficiency and persistence of the resulting localized structures.

\subsection{Equation of state}
\label{subsec_EoS}
It is well known that during preheating, if the inflaton oscillates about the minimum of a potential of the form $V(\phi)\sim \phi^{n}$, the EoS is $\omega\sim (n-2)/(n+2)$ under the homogeneous field approximation \cite{Turner:1983he,Lozanov:2017hjm,Garcia:2020wiy}. However, once the anisotropic type of gradient energy contributions are taken into account, the EoS would deviate from this estimate. In our models, the copious formation of oscillons contains a significant fraction of gradient energy component, which can appreciably alter the cosmic energy budget and the EoS parameter, potentially modifying the expansion history.

\begin{figure*}[t]
	\centering
	\includegraphics[width=7.0cm]{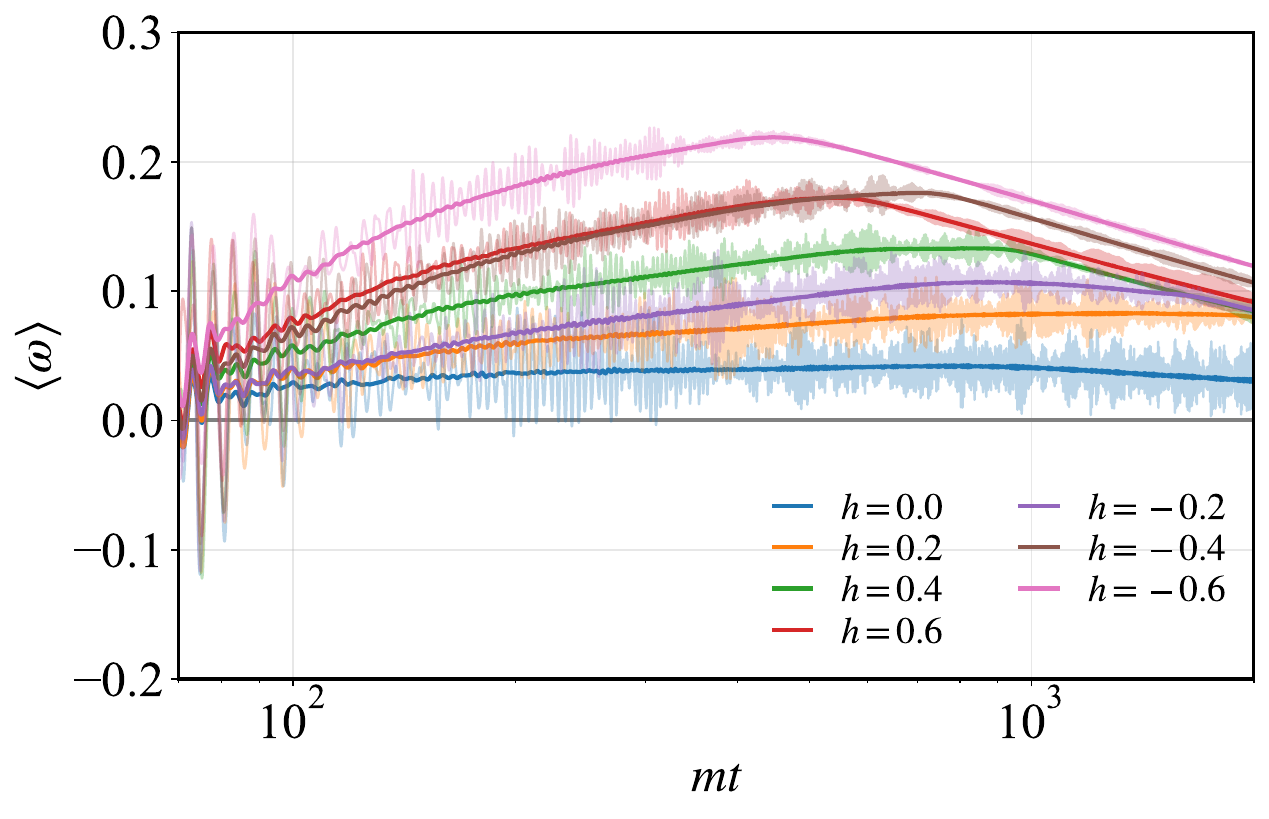}
	\includegraphics[width=7.0cm]{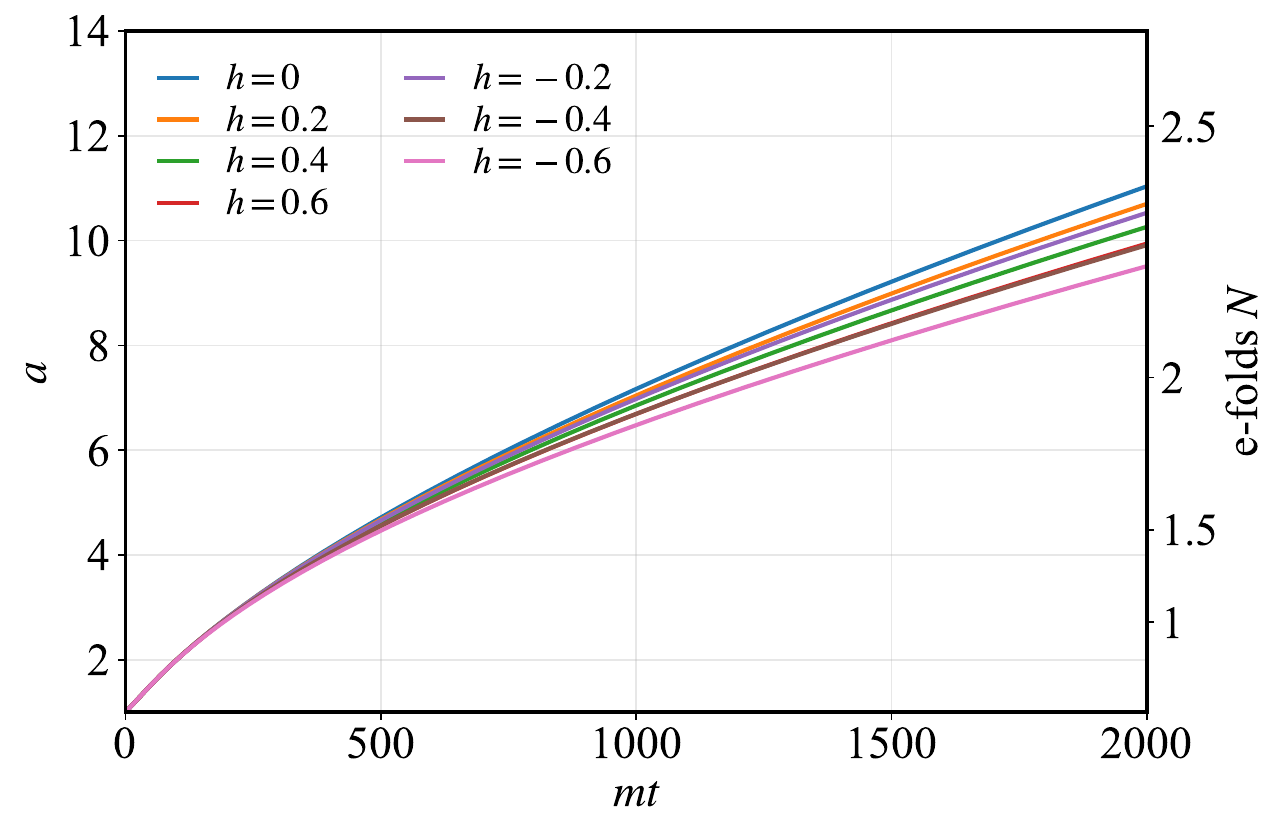}
	\caption{The evolution of the EoS (left) and the scale factors (right).}
	\label{EoS}
\end{figure*}

Now let us study the evolution of the EoS during this process. Using equations~(\ref{energy_density}) and (\ref{pressure}), the EoS parameter can be obtained,
\begin{equation}
	\langle\omega\rangle = 
	{ \frac{\tilde{p}}{\tilde{\rho}}} 
	= \frac{\tilde{E}_{\mathrm{K}}-\frac{1}{3}\tilde{E}_{\mathrm{G}}-\tilde{V}}
	{\tilde{E}_{\mathrm{K}}+\tilde{E}_{\mathrm{G}}+\tilde{V}}.
\end{equation}
We show the simulation results of the EoS and the corresponding scale factors in Fig.~\ref{EoS}. In all models, the EoS parameter remains below $1/3$, yet exhibits different evolutionary behaviors. At early times, the EoS is positive for all models and increases with the magnitude of $h$. These behaviors can be understood by analyzing the relative contributions of the various energy components for different values of $h$. Since the kinetic energy constitutes the dominant fraction (exceeding $45\%$), followed by the potential and gradient contributions, the pressure is primarily controlled by the kinetic component. Although a larger $h$ enhances the gradient energy, it simultaneously increases the kinetic energy while reducing the potential energy. The net effect is an increase in the pressure $\tilde{p}$ with $h$, accounting for the observed EoS behavior.
Within the simulated time interval, the EoS parameter does not yet approach zero. Nevertheless, in the absence of external couplings, one expects that after the subsequent dissipation of gradient energy, the EoS will asymptotically approach zero, in agreement with the prediction of the homogeneous-field approximation.

In summary, larger $|h|$ results in a larger averaged EoS during preheating, thereby reducing the cosmic expansion rate. Short-lived oscillons imply that energy remains trapped in localized clump-like structures for shorter duration, thereby accelerating the transition of the Universe to a radiation-dominated era. These effects substantially influence the observable of the inflationary process~
\cite{Liddle:2003as,Lozanov:2017hjm,Antusch:2020iyq,Antusch:2025ewc}. Quantifying them, however, requires a complete reheating process that accounts for inflaton couplings to external field(s), its full decay, and thermalization leading to radiation domination---aspects beyond the scope of this work.  A quantitative assessment of these effects requires more detailed calculations; the more detailed investigation is left for future work. 

\subsection{Gravitational waves}
It is well known that the anisotropic energy emerging during preheating would source a stochastic background of gravitational waves. To investigate the mechanism and dynamical properties of the gravitational wave production, we begin by introducing the tensor perturbations of the metric that characterize gravitational waves:
\begin{equation}
	\mathrm{d}s^2 = -\mathrm{d}t^2 + a^2(t)(\delta_{ij} + h_{ij})	\mathrm{d}x^i \mathrm{d} x^j.
\end{equation}
The equation of motion for the tensor perturbations $h_{ij}$ is
\begin{equation}
	\ddot{h}_{ij} + 3H\dot{h}_{ij} - \frac{{\nabla}^2}{a^2}{h}_{ij} =
	\frac{2}{M_P^2a^2} \Pi_{ij}^{TT},
\end{equation}
where $\Pi_{ij}^{TT}$ represents the transverse-traceless part of the effective anisotropic stress tensor $\Pi_{ij} = \partial_i\phi\partial_j\phi$. The energy density of the produced stochastic gravitational waves is 
\begin{eqnarray}
	&&\rho_{\text{GW}}(t) = \frac{M_P^2}{4}\langle \dot{h}_{ij}(\textbf{x},t) \dot{h}_{ij}(\textbf{x},t) \rangle
	\\ \nonumber
	&&\approx \frac{M_P^2}{4V} \int _V \frac{\mathrm{d}^3 \textbf{k}}{(2\pi)^3} 
	\dot{h}_{ij}(\textbf{k},t) \dot{h}^{*}_{ij}(\textbf{k},t)
	= \int \frac{\mathrm{d}\rho_{\text{GW}}}{\mathrm{d} \log k} \mathrm{d} \log k,
\end{eqnarray}
where the symbol $\langle...\rangle$ represents a spatial or temporal average. Since our simulation is performed within a box with finite volume, we adopt the spatial average throughout. The normalized gravitational wave energy density power spectrum (per logarithmic momentum interval) is then given by
\begin{equation}
	\Omega(k,t) = \frac{1}{\rho_c} \frac{\mathrm{d}\rho_{\text{GW}}}{\mathrm{d} \log k},
\end{equation}
where $\rho_c$ is the critical energy density. 

\begin{figure*}[t]
	\centering
	\includegraphics[width=7cm]{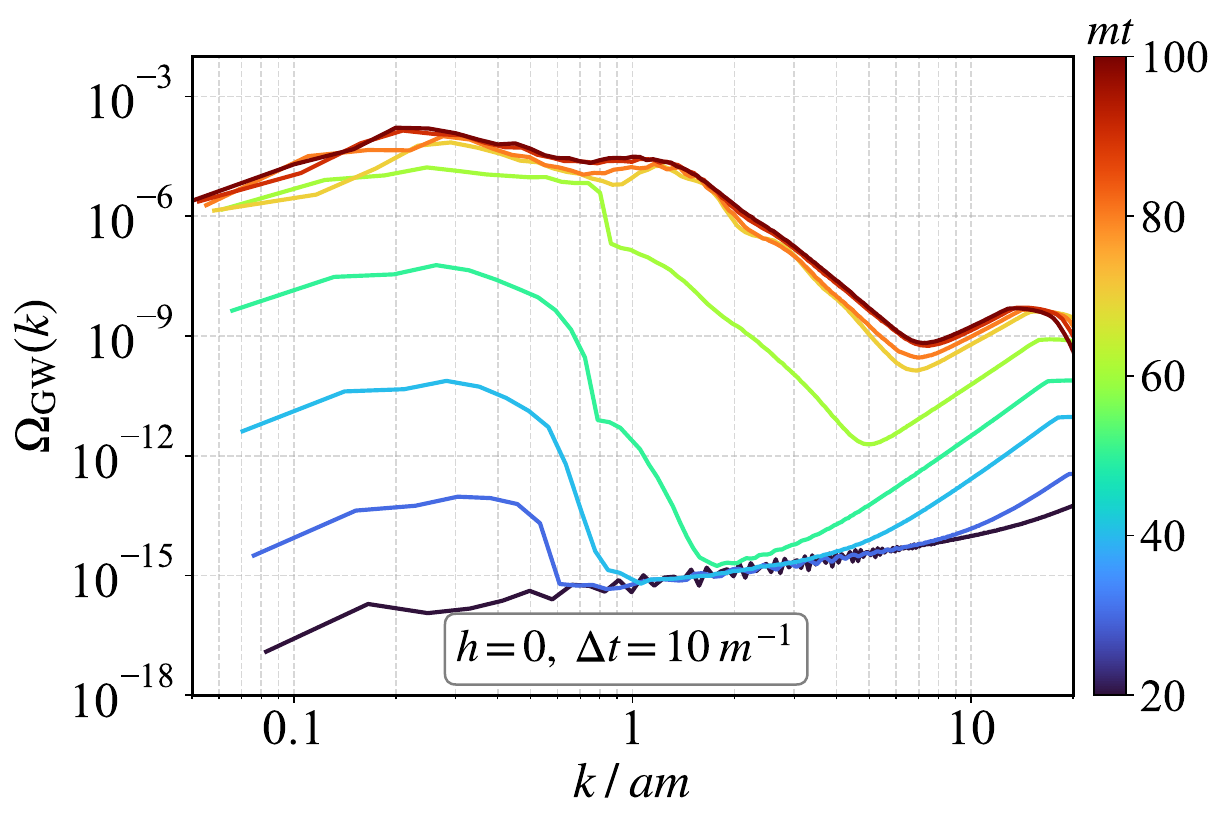}	
	\includegraphics[width=7cm]{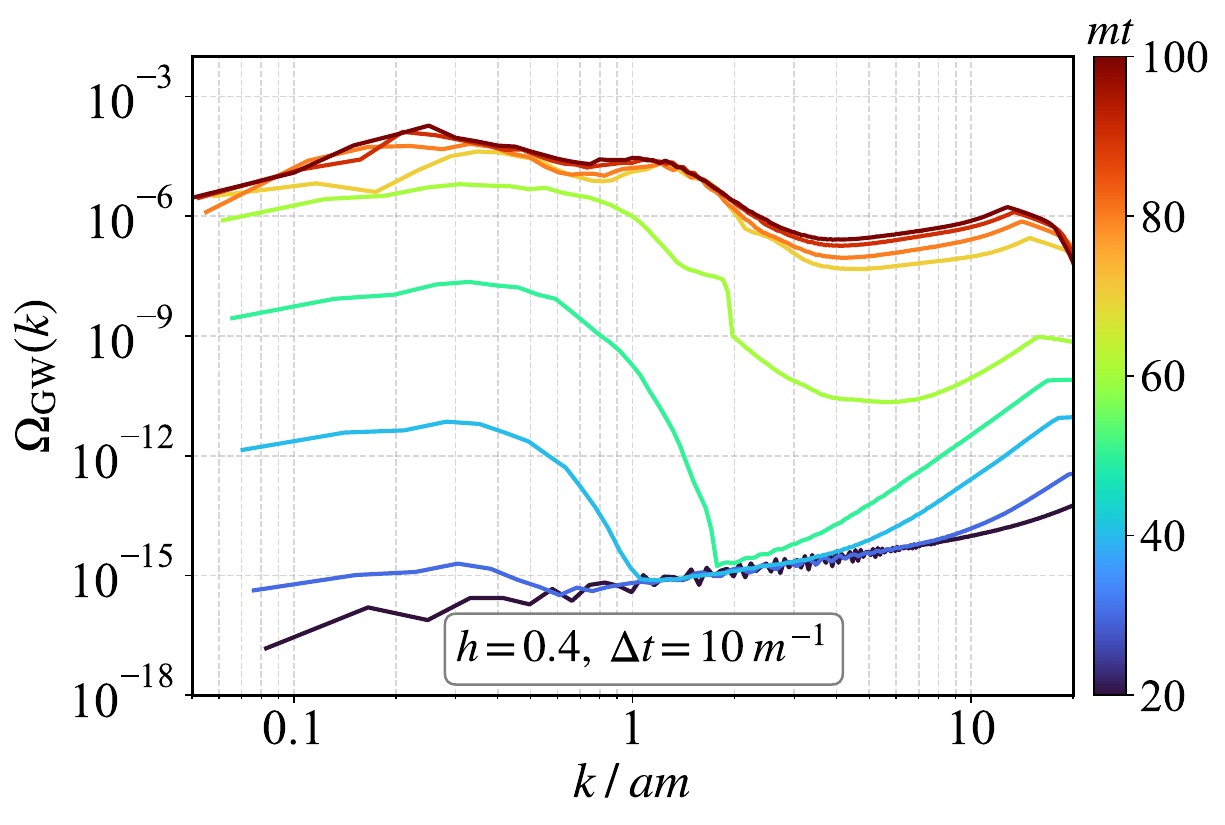}
	\includegraphics[width=7cm]{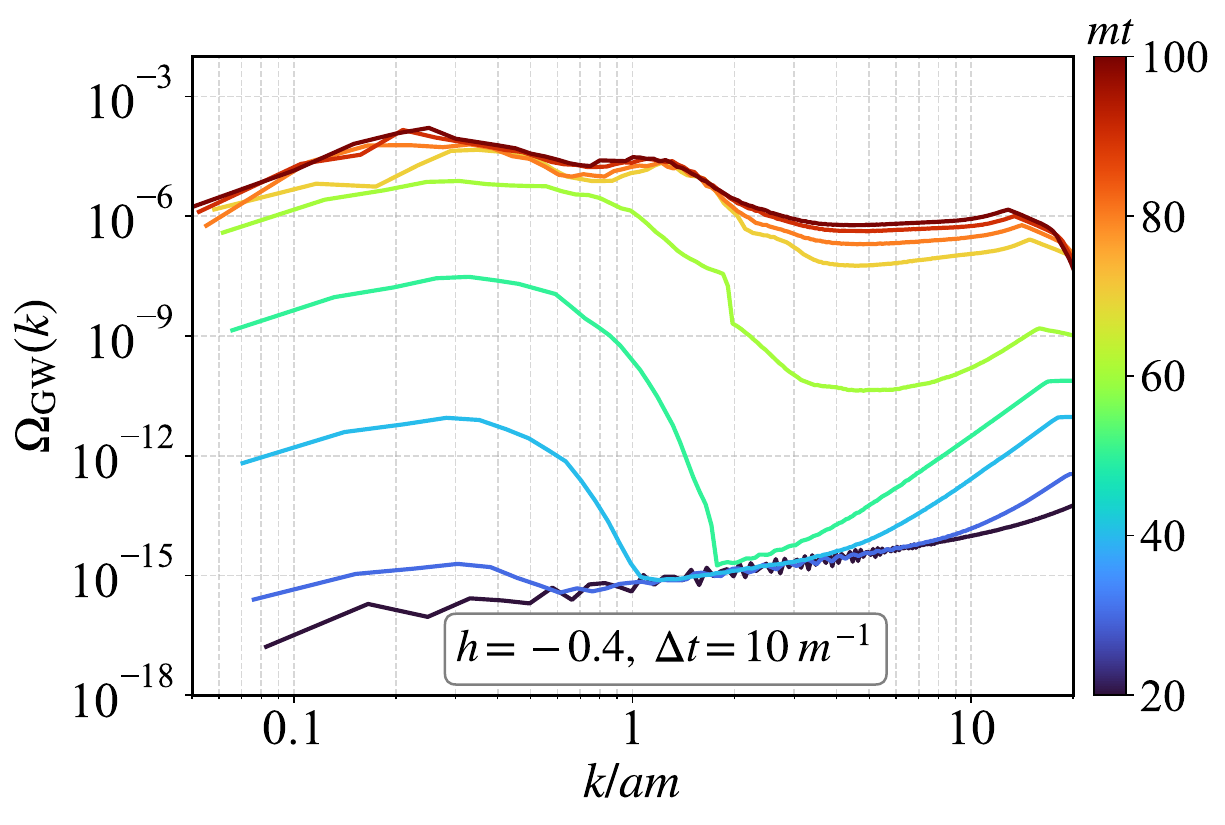}
	\includegraphics[width=7cm]{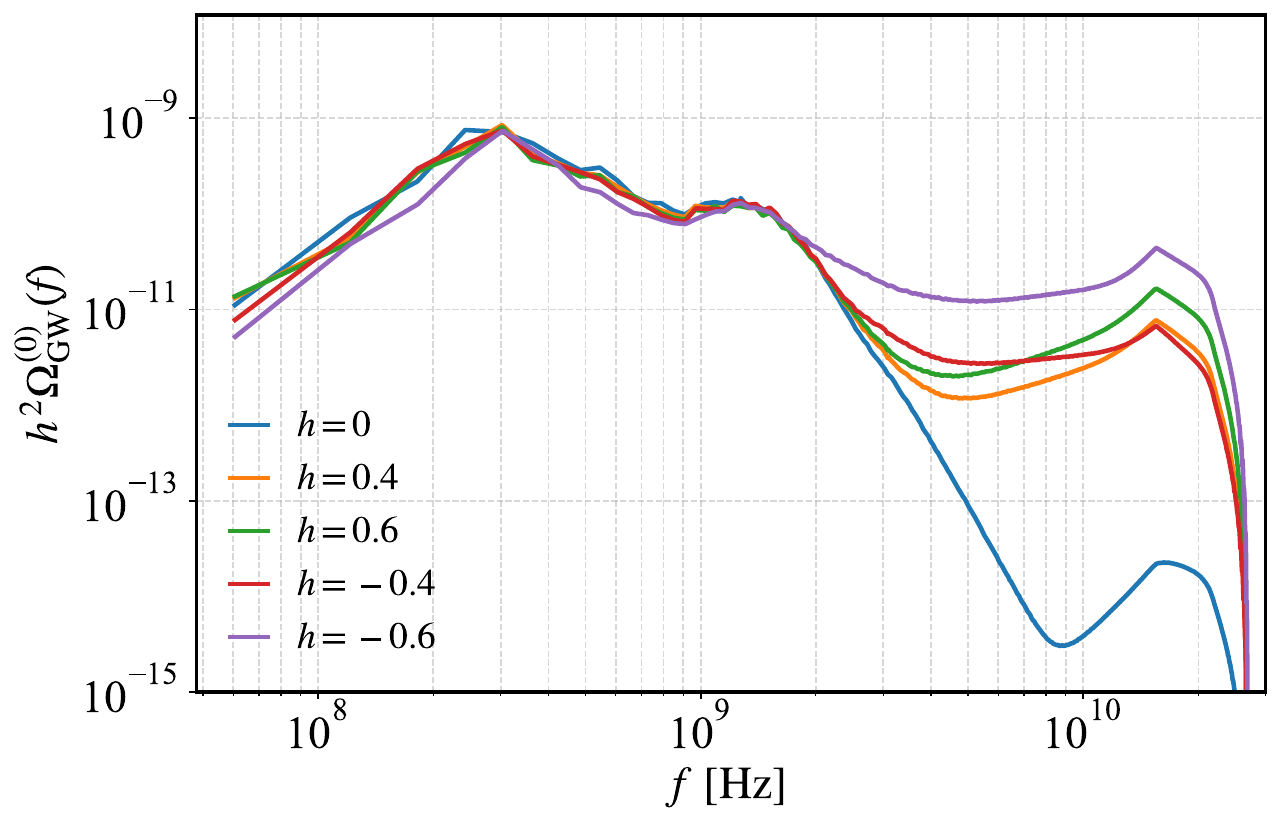}
	\caption{The gravitational wave spectrum at production and at the present day for selected models. To obtain the present-day gravitational wave spectrum, we identify the emission time $t_e$ as the moment when the spectrum saturates and ceases to grow; this time is model dependent. For simplicity, we further assume that the effective EoS parameter remains $\omega_{\rm eff}=1/3$, which is equivalent to assuming that the universe enters a radiation dominated era immediately after gravitational wave production.}
	\label{GW_spectra}
\end{figure*}

We present in Fig.~\ref{GW_spectra} representative results of the spectrum at production and at present day, including the standard T-model and the model with bump/dip. For the standard T-model, we find that
the gravitational wave emission concentrates in the low-momentum modes and are mainly produced during the linear stage of parametric resonance, while the high-momentum modes are largely suppressed. For the parameters considered in this paper, gravitational waves are most strongly amplified in modes with momenta $k\lesssim 2\,m$. This behavior can be correlated with the power spectrum of the scalar field.
For models with $h\neq 0$, gravitational waves in high-momentum modes are also significantly excited, but they are weaker than those in the low-momentum modes. This behavior may be attributed to the decay of short-lived oscillons, but this requires further confirmation. This constitutes the main difference from the gravitational wave spectra of the standard T-model, in which the high-momentum modes are rather weak. 

Because of cosmic expansion, the energy density of the gravitational waves—which behave as radiation—is diluted, and their frequencies are redshifted. The energy density power spectrum and the redshifted frequency at present day are \cite{Dufaux:2007pt}
\begin{eqnarray}
	h^2 \Omega_{\text{GW}}^{(0)} (f) &=& h^2\Omega_{\text{rad},0} \left(\frac{a_{\mathrm{e}}}{a_{*}}\right)^{1-3\omega_{\text{eff}}}
	\left(\frac{g_{*}}{g_0}\right)^{-1/3}\Omega_{\text{GW}}^{\text{(e)}}(k),
	\nonumber  \\ 
	f &=& 4\times 10^{10} \left(\frac{a_{\mathrm{e}}}{a_{*}}\right)^{\frac{1-3\omega_{\text{eff}}}{4}}
	\frac{k}{a_{\mathrm{e}} {\rho_{\mathrm{e}}}^{1/4}} \text{Hz},
\end{eqnarray}	
where $h^2\Omega_{\text{rad},0} = 4.3\times 10^{-5}$ is the abundance of radiation today and the subscripts ``e'', ``*'', and ``0'' denote the time of gravitational wave emission, the end of reheating, and present day, respectively. The emission time is chosen to be the time when the gravitational-wave spectrum becomes stationary, leading to a model-dependent choice of $t_e$.
We set the ratio of the relativistic degrees of freedom at $t_*$ and $t_0$ to $g_{*}/{g_0}=100$. The results are shown in lower right of Fig.~\ref{GW_spectra}. Note that, the reheating process has not yet been completed at gravitational wave emission time $t_{\rm e}$. Subsequent inflaton decay and thermalization are still required for the universe to reach thermal equilibrium and enter the radiation dominated era. For simplicity, we adopt a rough approximation in which the universe is assumed to become radiation dominated immediately after $t_{\rm e}$, corresponding to an effective EoS $\omega_{\rm eff}=1/3$ and hence $(a_{\rm e}/a_{\rm *}) ^{1-3\omega_{\text{eff}}}=1$. Taking into account the full reheating dynamics would introduce an additional suppression factor, reducing the resulting gravitational wave spectrum by a factor of a few.
It is evident that these gravitational waves lie far beyond the sensitivity of current detectors~\cite{Li:2009zzy,Akutsu:2008qv,Holometer:2016qoh,Patra:2024eke}. However, one may constrain the gravitational waves indirectly, such as using the cosmological constraints on the effective number of relativistic degrees of freedom, see Refs.~\cite{Pagano:2015hma,Caprini:2018mtu,Clarke:2020bil,Saha:2024lil,Lozanov:2026dgu}.

\section{Conclusion and discussion}\label{sec4}
Reheating serves as the transitional stage connecting inflation to the radiation dominated era, during which the energy stored in the inflaton field must be transferred to produce Standard Model particles through specific mechanisms. Due to the lack of direct observational probes, the detailed microphysics of reheating remains unclear, despite the existence of several well-motivated scenarios, such as parametric resonance.

In this work, we investigated self-resonance preheating in a small-$\alpha$ attractor T-model containing a localized Gaussian feature near the potential minimum. By combining Floquet analysis with fully nonlinear lattice simulations, we studied how the sign and magnitude of the feature parameter $h$ affect parametric resonance, oscillon formation and decay, the effective equation of state, and gravitational wave production.

The Floquet analysis shows that the feature deforms the principal instability band and generates additional resonance bands extending toward higher momenta. By separately examining the effects of the modified background trajectory and the feature-dependent effective mass, we find that the deformation of the resonance structure is driven predominantly by the localized variation of $V''(\phi)$. Changes in the background motion have a comparatively minor effect, mainly influencing low-momentum modes at small oscillation amplitudes. We also considered a $Z_2$-symmetric potential with identical features at $\pm\phi_s$. Restoring the symmetry strengthens the resonance and makes the additional high-momentum bands more prominent, demonstrating that the new resonance structure does not originate from the explicit $Z_2$ breaking of the one-sided potential.

The nonlinear simulations show that the feature has only a modest influence on energy transfer during the resonance stage, whereas the subsequent evolution becomes strongly dependent on $h$. Larger $|h|$ generally produces stronger transient fragmentation and excites more high-momentum fluctuations. However, a smaller fraction of the amplified energy is retained in compact oscillon cores. Using the identification threshold $\tilde{\rho}_{\rm th}=10\tilde{\rho}_{\rm ave}$, we find that increasing $|h|$ suppresses oscillon formation, leading to fewer identified oscillons and smaller total energy and comoving volume fractions. The mean physical radius remains of order a few $m^{-1}$ after formation and generally increases as the oscillon amplitude decreases. 

To characterize the oscillon lifetime, we tracked both the total energy and the internal gradient energy contained in the identified regions. We defined $\tau_{50}$ as the time at which the corresponding energy decreases to one half of its peak value and $\tau_{100}$ as the disappearance time of the compact oscillon population. The two diagnostics give consistent qualitative dependence on $h$, although the total energy decay times are generally slightly longer. Within the directly resolved range, particularly for $|h|\gtrsim0.2$, the oscillon lifetime decreases nonlinearly with increasing $|h|$. At fixed $|h|$, negative $h$ generally produces a shorter lifetime than positive $h$, indicating that dip-type features suppress the persistence of oscillons more efficiently than bump-type features of the same magnitude. The $h=\pm0.1$ cases remain only partially resolved, while no finite lifetime is assigned to the exceptionally long-lived $h=0$ case within the present numerical reach. The systematic threshold scan further shows that the properties inferred using $\tilde{\rho}_{\rm th}=3\tilde{\rho}_{\rm ave}$, and to a lesser extent those obtained using $5\tilde{\rho}_{\rm ave}$, are affected by diffuse oscillon envelopes, surrounding overdensities, and low-density decay fragments. For $10\leq\tilde{\rho}_{\rm th}/\tilde{\rho}_{\rm ave}\leq50$, $\tau_{100}$ and the qualitative dependence on $h$ become nearly threshold independent, while $\tau_{50}$ retains a moderate threshold dependence. We therefore adopted $\tilde{\rho}_{\rm th}=10\,\tilde{\rho}_{\rm ave}$ as the lowest threshold that reliably isolates the compact oscillon cores.

The microscopic analysis shows that the identified oscillons retain regular Gaussian-like radial profiles and follow similar amplitude--width relations at moderate and large amplitudes. The potential feature therefore does not primarily shorten their lifetimes through an immediate and strong deformation of their cores. Instead, the enhanced gradient energy outside the compact regions and the faster migration of the oscillon population toward smaller amplitudes { are consistent with enhanced radiative energy loss in the featured models}. As their amplitudes decrease, the fundamental frequencies approach the mass threshold and the oscillons evolve more rapidly toward the terminal branch of the quasibreather family, where rapid decay occurs. The $Z_2$-symmetric feature provides further support for this interpretation: although it enhances the initial resonance and fragmentation, most of the additional gradient energy remains outside the compact cores, resulting in less energy stored in oscillons and a faster decay of the population.

We also find that gravitational waves are generated predominantly during the resonance stage, while their production becomes strongly suppressed after oscillon formation. The potential feature leaves the low-frequency part of the spectrum approximately unchanged but enhances its high-frequency tail through the additional small-scale fluctuations. The feature-induced differences in energy distribution and oscillon persistence also modify the transient averaged equation of state and the post-inflationary expansion history.

Our simulations describe self-resonance in the absence of external couplings and therefore do not constitute a complete treatment of reheating. In the presence of efficient couplings to other fields, the earlier release of energy from short-lived oscillons may accelerate the approach to radiation domination and shorten the reheating stage. Quantifying this possibility requires simulations including inflaton couplings, subsequent particle production, and thermalization. Higher-resolution simulations extending to later times will also be necessary to determine the lifetimes of the small-$|h|$ models more accurately and to obtain quantitative predictions for the cosmological consequences of the potential feature.

\section{Acknowledgments}
We are grateful to the High-Performance Computing Platform of the Basic Physics Experiment Center at Nanchang University for supporting part of the lattice simulations. We thank Shao-Wen Wei, Ke Yang, Wen-Di Guo, and Peng Cheng for valuable discussions.  B.-M. Gu thanks Lanzhou Center for Theoretical Physics for its hospitality during his visit, where this work was originally initiated. This work is partially supported by the National Natural Science Foundation of China (Grants No. 12165013, 12605109, 12375049, 12575055, 12475056, and 12247101). Y.-P. Zhang is supported by ``Talent Scientific Fund of Lanzhou University''. F.-W. Shu is supported by the Key Program of the Natural Science Foundation of Jiangxi Province under Grant No. 20232ACB201008, and the Ganpo High-Level Innovative Talent Program.

\newpage
\appendix 

\section{Model parameters of T-model}\label{appsec}
\begin{figure}[t]
		\centering
		\includegraphics[width=7.0cm]{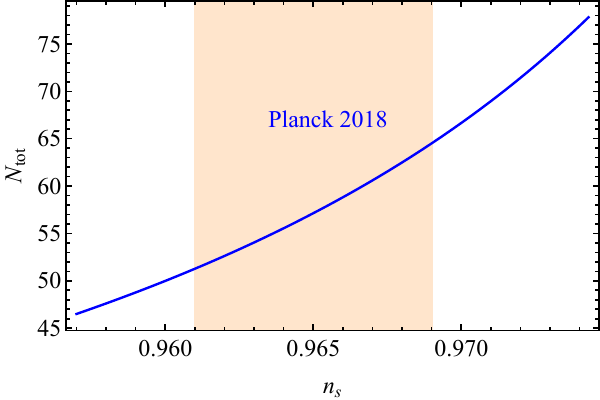}
		\caption{The plot of $N_{\text{tot}}$ versus $n_s$ for $\alpha=10^{-4}M_P^2$ ($\lambda=50\sqrt{\frac{2}{3}}$). The dashed regions are the constraints from Planck 2018.}
		\label{N-ns}
	\end{figure}

	The model parameters can be constrained by the CMB constraints on the spectral index $n_s$ and the amplitude of the scalar power spectrum $A_s$. For $n=1$ of~(\ref{T_model_pot}), it is straightforward to obtain
	\begin{equation}
		n_s=1-6\epsilon_V+2\eta_V=
		1-8\lambda^2 \mathrm{csch} ^2\left(\lambda \frac{\phi}{M_P}\right).
	\end{equation}
	Thus the field value at horizon crossing is 
	\begin{equation}
		\phi_*=\frac{1}{\lambda}\mathrm{arcsinh} \left(2\lambda\sqrt{\frac{2}{1-n_s}}\right).
		\label{field_HC}
	\end{equation}
	The amplitude of the power spectrum at horizon crossing is 
	\begin{equation}
		A_s=\frac{H^2(\phi_*)}{8\pi^2\epsilon_V(\phi_*)}=\frac{4\lambda^2V_0}{3\pi^2(1-n_s)^2}=2.1\times 10^{-9}.
	\end{equation}
	For a given $\lambda$, the observational constraints on $n_s$ and $A_s$ limits $V_0$.
	The field value at the end of inflation is defined by $\epsilon_V(\phi_{\mathrm{end}})=1$, which gives
	\begin{equation}
		\phi_{\mathrm{end}}=\frac{1}{2\lambda}\mathrm{arcsinh}\left(2\sqrt{2}\lambda\right).
		\label{field_end}
	\end{equation}
	Using equations~(\ref{field_HC}) and (\ref{field_end}) we immediately get the total e-folds of inflation
	\begin{equation}
		N_{\mathrm{tot}}=\int _{\phi_{\mathrm{end}}}^{\phi_*}\frac{\mathrm{d}\phi}{\sqrt{2\epsilon_V}}=\frac{2}{1-n_s}-\frac{\sqrt{1+8\lambda^2}-1}{8\lambda^2}.
	\end{equation} 
	We show the plot of \textit{e}-folds in Fig.~\ref{N-ns}. In this work we set $\alpha=10^{-4}M_P^2$  (corresponds to $\lambda=50\sqrt{\frac{2}{3}}$) and $V_0=1.14\times 10^{-14} M_P^4$, yielding $n_s=0.965$ and $N_{\text{tot}}\simeq 57$.

\section{Derivation of the Floquet exponent}\label{Floquet_computation}
We compute the Floquet exponent following the procedure outlined in~\cite{Amin:2014eta}. We first recast equation~(\ref{pert_EoM}) into a first-order formalism,
\begin{equation}
	\dot{x}(t)=U(t)x(t), 
	\quad U(t)=\left(\begin{array}{cc}
		0	&  1\\
		-k^2-V''(\phi(t)) & 0
	\end{array}\right),
\end{equation}
where $x(t)=[\delta\phi_k,\delta\pi_k]^T$ with $\delta\pi_k=\delta\dot{\phi}_k$. Next, we need to calculate the time period $T$ of $U(t)$. The period is defined by
\begin{equation}
	T=2\int_{\phi_{\mathrm{min}}}^{\phi_{\mathrm{max}}}\frac{\mathrm{d}\phi}{\sqrt{2V(\phi_{\mathrm{max}}) - 2V(\phi)}},
\end{equation}
where  $\phi_{\mathrm{max}}$ is the chosen positive turning point of the background oscillation, and $\phi_{\mathrm{min}}$ is obtained by solving $V(\phi_{\mathrm{min}})=V(\phi_{\mathrm{max}})$. Then, we solve the static version ($H=0$ and $a=1$) of the scalar field equation~(\ref{phi_EoM}) and the perturbed equation~(\ref{pert_EoM}) with two sets of initial conditions $\{\delta\phi_k^{(1)}(0)=1, \delta\pi_k^{(1)}(0)=0\}$ and $\{\delta\phi_k^{(2)}(0)=0, \delta\pi_k^{(2)}(0)=1\}$, from 0 to $T$. At last, the Floquet exponent can be obtained by
\begin{equation}
	\mathfrak{Re}[\mu_k^{\pm}]=\frac{1}{T} |\ln o_k^{\pm}|,
\end{equation}
where 
\begin{equation}
	o_k^{\pm}=\frac{\delta\phi^{(1)}_k + \delta\pi^{(2)}_k}{2} 
	\pm \frac{	\sqrt{\left(\delta\phi_k^{(1)} - \delta\pi_k^{(2)}\right)^2 
			+ 4\delta\phi_k^{(2)}\delta\pi_k^{(1)} }}{2} 
	,  
\end{equation}
with all quantities evaluated at $t=T$. Since (\ref{Floquet_solution}) contains both of the growing and decaying branches of solution, it exhibits exponential growth whenever $\mathfrak{Re}[\mu_k] \neq 0$. For convenience, we therefore restrict ourselves to the case $\mathfrak{Re}[\mu_k]>0$. 

\begin{figure*}[t]
	\centering
	\includegraphics[width=15cm]{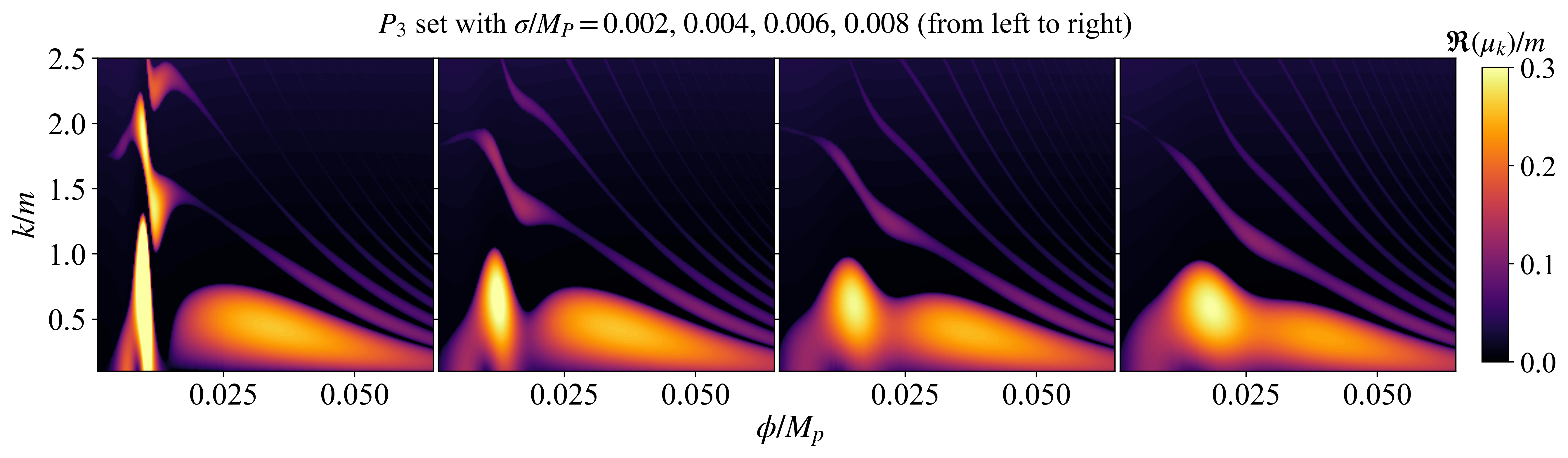}
	\includegraphics[width=15cm, trim={0 0 0.3cm 0}, clip]{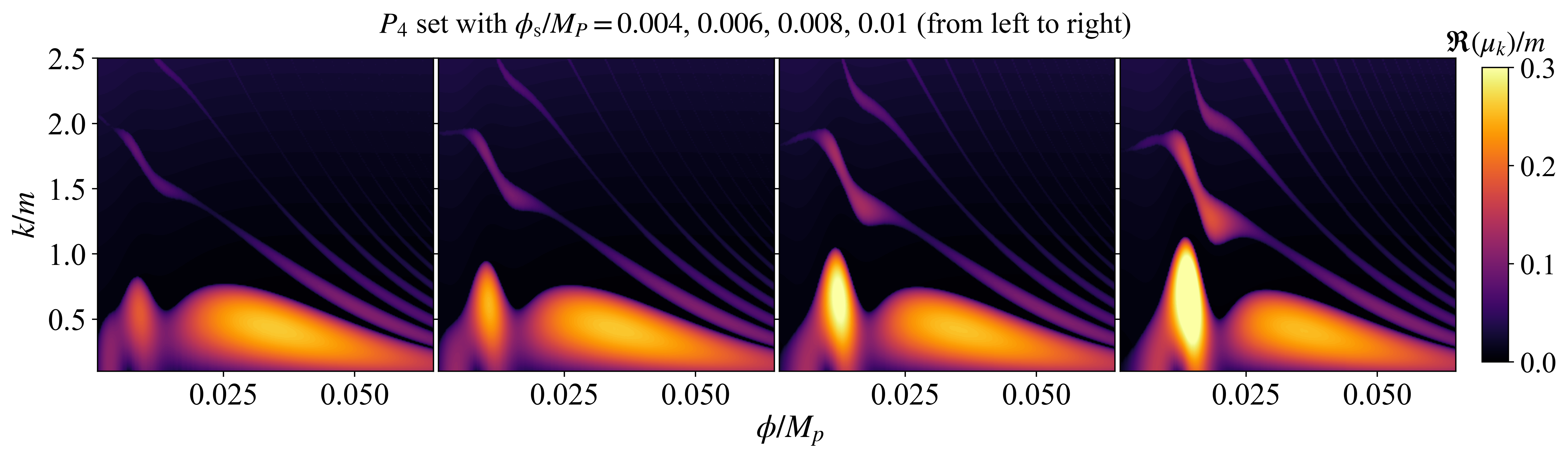}
	\caption{The Floquet charts showing the instability bands of the deformed T-model for $P_3$ (upper) and $P_4$ (lower) set parameters, with cosmic expansion neglected. Note that the Floquet exponent is scaled by the inflaton mass $m$ and $H_{\rm i}\backsimeq 0.01\,m$ for $\alpha=10^{-4}\,M_P^2$.}
	\label{Deformed_chart}
\end{figure*}

\begin{figure*}[t]
	\centering
	\includegraphics[width=7cm]{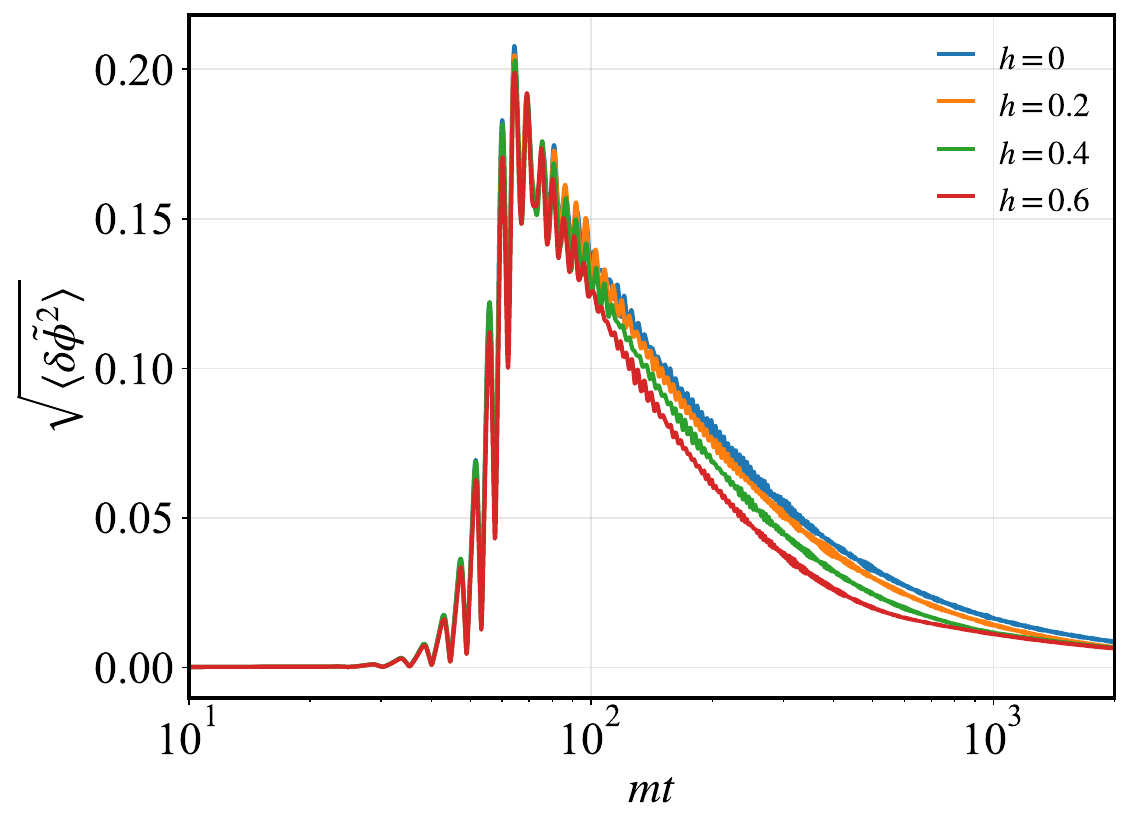}
	\includegraphics[width=7cm]{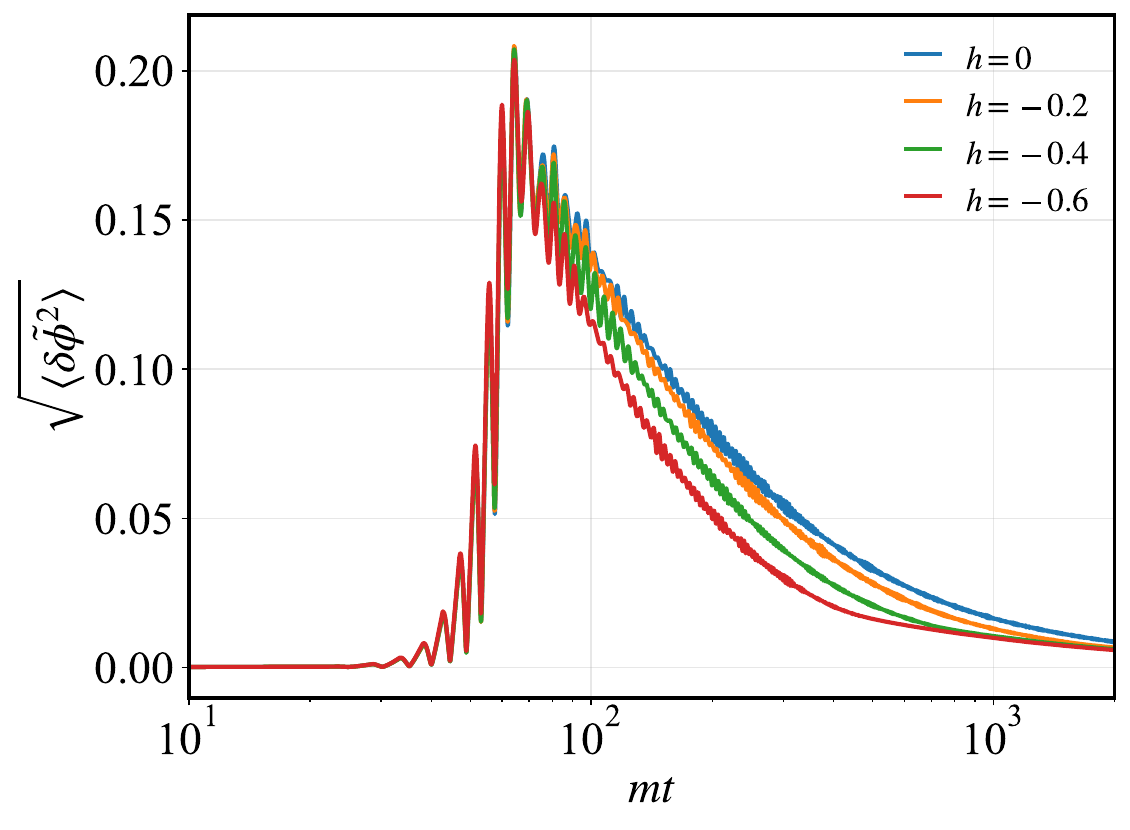}
	\caption{Evolution of the fluctuations, corresponding to the parameter sets $P_1$ (left) and $P_2$ (right).}
	\label{6variance}
\end{figure*}

\section{Convergence test}\label{CT}
It is important to emphasize that the simulation is performed in comoving coordinates, so that the simulation volume remains fixed throughout the evolution. As the universe expands, objects with fixed physical size therefore appear progressively smaller in the box. Consequently, although oscillons are initially well resolved with characteristic scales much larger than the lattice spacing, their comoving size eventually becomes comparable to a single lattice cell. At this point, the lattice simulation can no longer capture oscillon dynamics, marking the breakdown of the lattice simulation. Assuming the physical size of an oscillon is $L_{\rm osc}$ (typically of order several $m^{-1}$), the breakdown of the simulation occurs when the lattice spacing becomes comparable to the comoving size of the oscillon
\begin{equation}
	\,{\rm d}x \equiv \frac{2\pi}{N\,k_{\rm IR}} \backsimeq \frac{L_{\rm osc}}{a(t_{\rm br})},
\end{equation}
where ${\rm d}x $ is the lattice spacing in comoving coordinate.
Consequently, the simulations remain reliable only for 
\begin{equation}
a\lesssim \frac{N}{2\pi}k_{\rm IR}L_{\rm osc}.
\end{equation}
A careful choice of the lattice grid number $N$ and the infrared cutoff $k_{\rm IR}$ is therefore essential for sufficient resolution. Larger $N$ and higher cutoff wavenumbers generally extend the reliable simulation time. However, as can be seen from the Floquet charts in Fig.~\ref{Floquet_h}, the resonant modes are mainly confined to $k<m$, $k_{\rm IR}$ cannot be too large, as excluding these modes would degrade the description of the preheating dynamics. Hence, in practice, obtaining more reliable simulations over longer time scales generally requires increasing the lattice grid number $N$.

\begin{figure}[t]
	\centering
	\includegraphics[width=6cm]{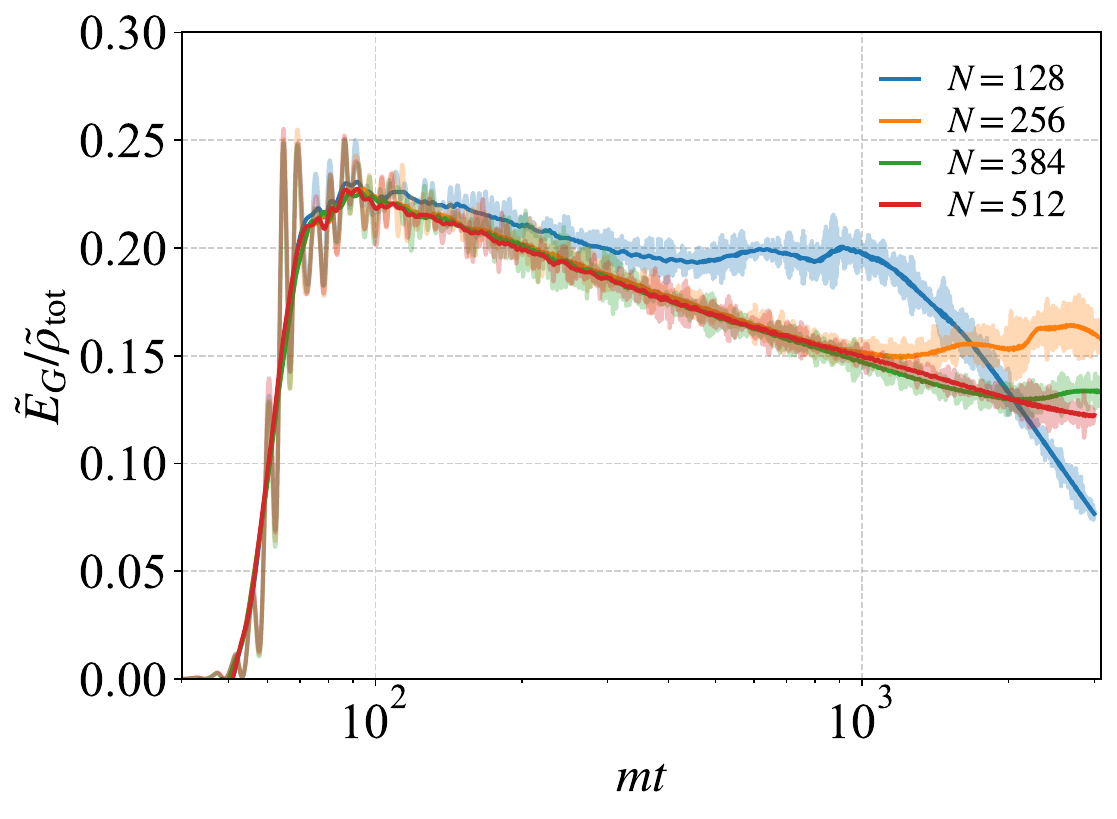}
	\includegraphics[width=6cm]{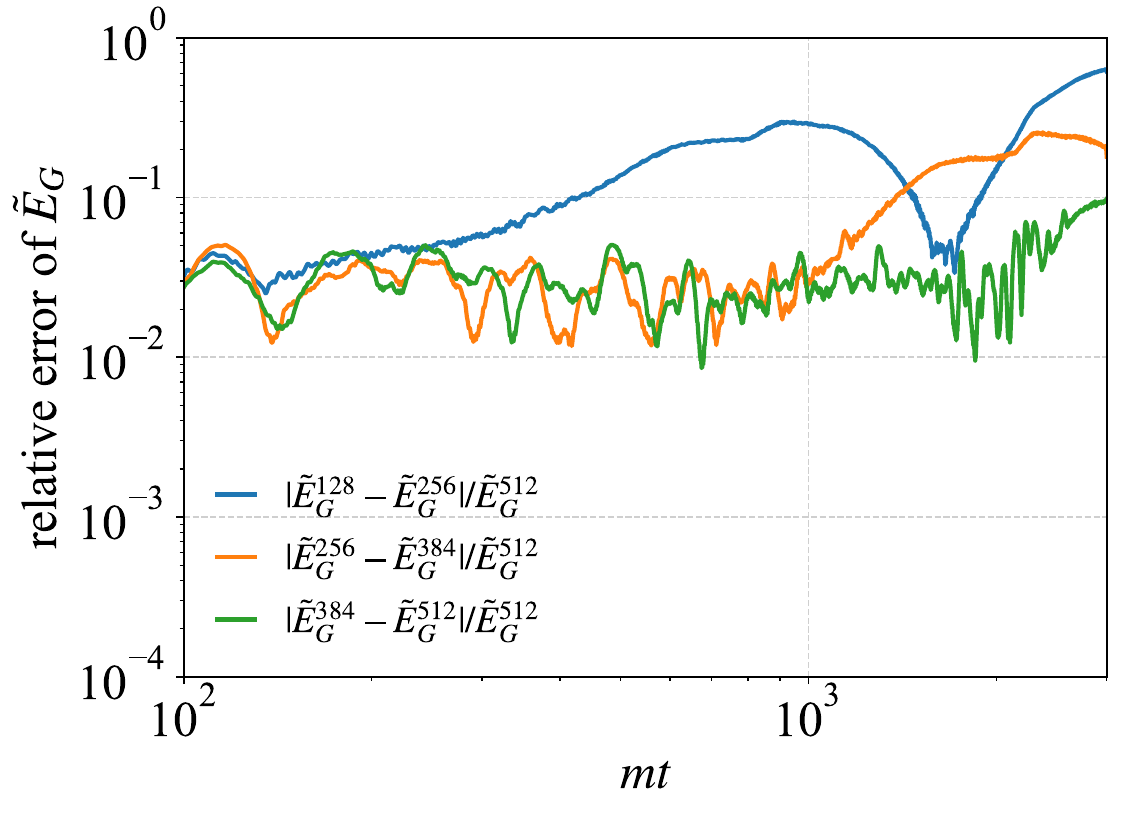}
	\caption{The gradient energy fraction for the T-model ($h=0$) with different resolutions (left) and the relative errors (right). We find that, for $k_{\rm IR}=0.1\,m$, simulations with $N=256$ remain reliable up to $t\lesssim 1000\,m^{-1}$, while increasing the resolution to $N=384$ extends the trustworthy regime to approximately $t\sim 2000\,m^{-1}$, in both cases, the relative errors remain below $\sim3\%$ within these intervals.}
	\label{convergence_test}
\end{figure}

To assess the robustness of our numerical results, we performed convergence tests with $k_{\rm IR}=0.1\,m$ and using grid points of 
$N^3=128^3$, $256^3$, $384^3$, and $512^3$. As an illustration, we compare the gradient energy computed at different lattice resolutions, with the results shown in Fig.~\ref{convergence_test}. We expect that increasing the grid number $N$ leads to more accurate results. As shown, the lattice with $N^3=128^3$ grids are insufficient to faithfully capture the formation and subsequent evolution of oscillons; at this resolution, the evolution of the gradient energy is only qualitatively captured up to $t\sim 300\,m^{-1}$, beyond which the numerical errors become very large.
For $N^3=256^3$, the simulation remains reliable up to $t_{\rm br} \backsimeq 1000\,m^{-1}$ ($a_{\rm br}\backsimeq 7.1$). Based on this result, we obtain a conservative estimate for the characteristic oscillon scale, $L_{\rm osc}\sim 1.75\,m^{-1}$. Using this characteristic oscillon scale, we predict that the simulation breaks down at $t_{\rm br} \backsimeq 1900\,m^{-1}$ ($a_{\rm br}\backsimeq 10.7$) for $N^3=384^3$, while for $N^3=512^3$ the corresponding breakdown occurs at $t_{\rm br} \backsimeq 2900\,m^{-1}$ ($a_{\rm br}\backsimeq 14.3$). These estimates are consistent with the convergence tests presented in Fig.~\ref{convergence_test}. As shown in the figure, the results for $N^3=384^3$ are reliable for $t\lesssim 2000\,m^{-1}$, whereas for $N=512$ the reliable regime extends to $\sim 3000\,m^{-1}$. Since, in most cases of our model, the oscillon lifetime is shorter than $2000\,m^{-1}$, we therefore adopt $N^3=384^3$ as a compromise that ensures sufficient accuracy while significantly reducing the computational cost. { Figure 19 presents the convergence test for $h=0$. We also reran the $h=\pm 0.1$ models on $512^3$ grids. The $\tau_{100}$ lower bounds near $mt\simeq3000$ in Fig.~11 are based on these higher-resolution runs, rather than on extrapolations of the main $384^3$ simulations, whose reliable window ends near $mt\simeq2000$.}

\begin{figure}[t]
	\centering
	\includegraphics[width=6cm]{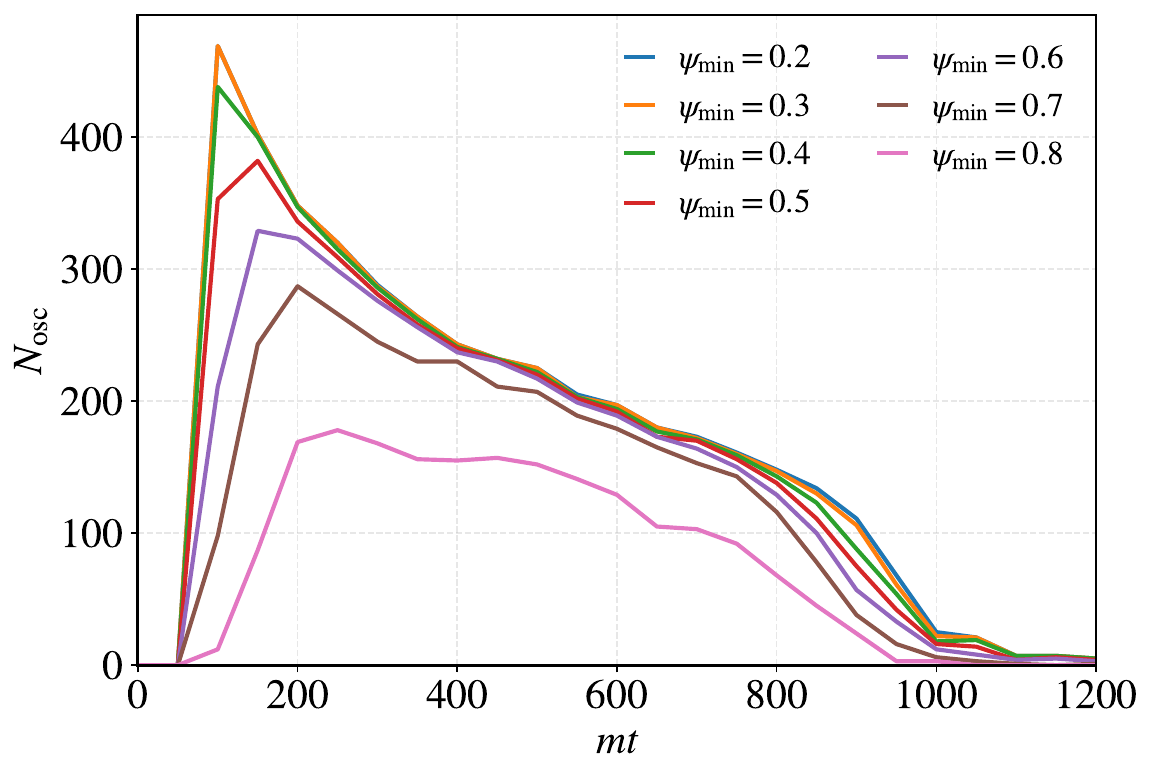}
	\includegraphics[width=6cm]{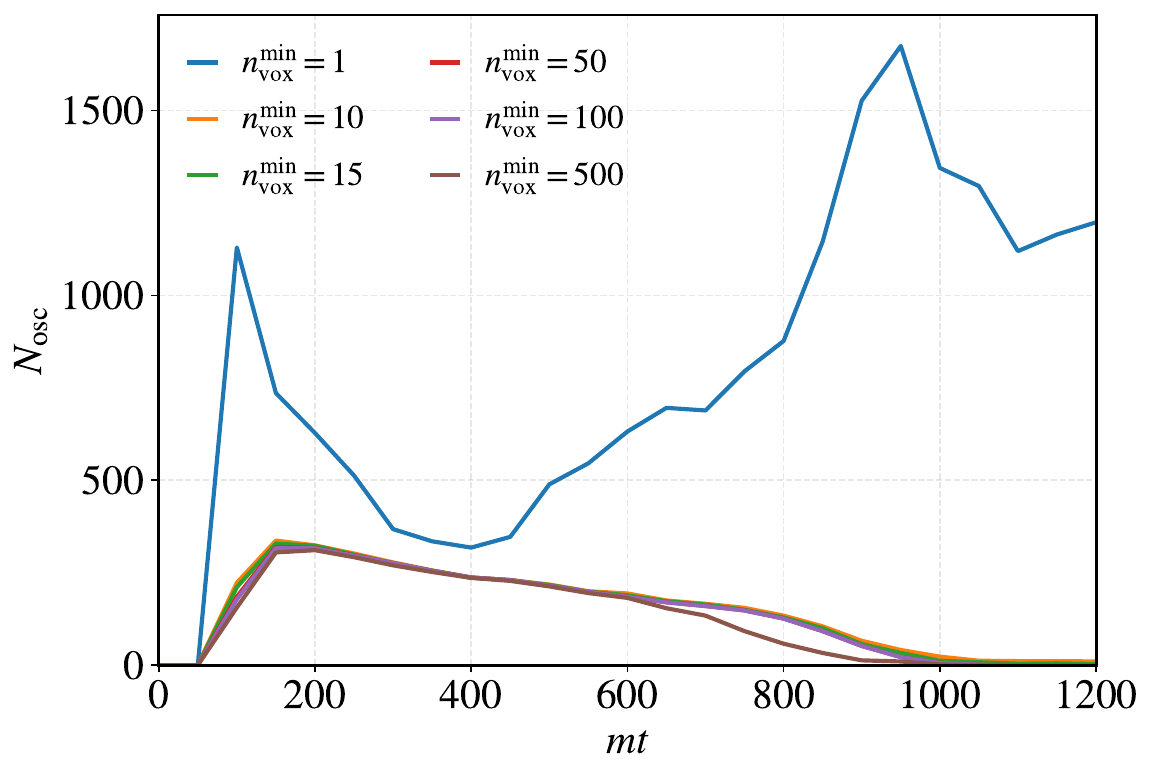}
	\caption{{ Sensitivity of the identified oscillon number to the  sphericity cut (upper) and minimum-size cut (lower) for $h=0.4$ (using HDF5 snapshots saved at intervals of $\Delta(mt)=50$). The density threshold is $\tilde{\rho}_{\rm th}=10\tilde{\rho}_{\rm ave}$ in both panels. We fix $\psi_{\min}=0.6$ in lower panel and $n_{\rm vox}^{\min}=15$ in upper panel.}}
	\label{N_osc_stats}
\end{figure}

\section{The role of parameters $\sigma$ (feature width) and $\phi_{\mathrm{s}}$ (location)}
\label{sigma_phis}

We consider the parameter sets listed in table~\ref{tab_app}, which correspond to different widths and locations of potential feature. The Floquet charts are shown in Fig.~\ref{Deformed_chart}. The parameter $\sigma$ determines the width of the new resonance band. Generally, a sharp feature in the potential narrows the field range of the emerging resonance band while markedly enhancing its Floquet exponent. At the same time, it extends the coverage in momentum space, allowing a broader set of fluctuation modes to resonate. For example, when $\sigma=0.002\,M_P$, nearly all modes in $0<k<2.5m$ undergo resonance around the potential feature. The results for different feature locations show that, for a fixed $h$, moving the feature farther from the minimum produces a larger deviation from the original T-model.

\begin{table}[]
	\centering
	\setlength{\tabcolsep}{5pt}
	\begin{tabular}{c | c | c |  c | c | c}
		\hline
		\hline
		Sets & $V_0~[M_p^4]$ & $\alpha~[M_p^2]$ & $h$ & $\sigma~[M_p]$  & $\phi_s~[M_p]$ \\
		\hline
		~ &~ & ~ & ~ & $0.002$ & $0.008$ \\
		$P_3$ &$1.14\times 10^{-14}$ & $10^{-4}$ & 0.6 & $0.004$ & $0.008$ \\
		~ &~ & ~ & ~ & $0.006$ & $0.008$ \\
		~ &~ & ~ & ~ & $0.008$ & $0.008$ \\
		\hline 
		~ &~ & ~ &~ & $0.004$ & $0.004$ \\
		$P_4$ &$1.14\times 10^{-14}$ & $10^{-4}$ & 0.6 & $0.004$ & $0.006$ \\
		~ &~ & ~ & ~ & $0.004$ & $0.008$ \\
		~ &~ & ~ & ~ & $0.004$ & $0.010$ \\
		\hline
		\hline
	\end{tabular}
	\caption{The parameter sets $P_3$ and $P_4$.}\label{tab_app}
\end{table}

\section{{ Sensitivity to oscillon-identification criteria}}
\label{app_threshold}
\begin{figure*}[t]
	\centering
	\includegraphics[width=17cm]{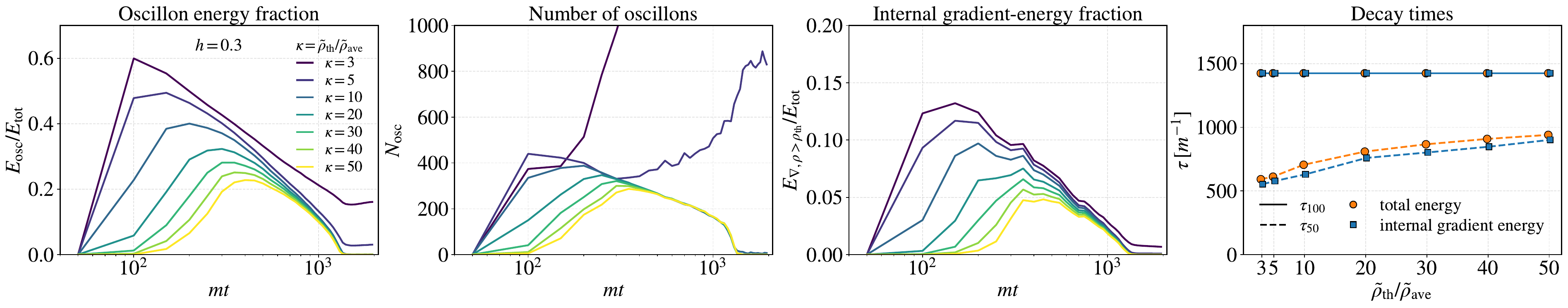}
	\includegraphics[width=17cm]{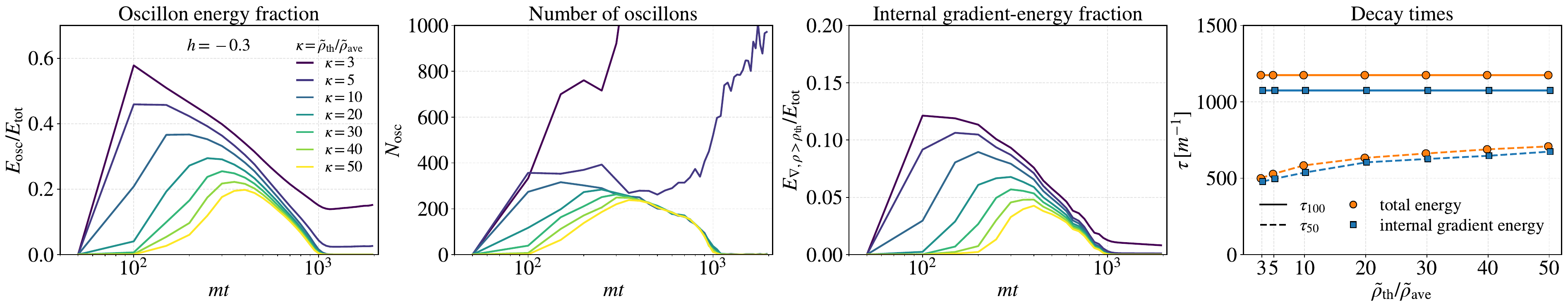}
	\caption{Dependence of the identified overdense structures on the density threshold for $h=0.3$ ($P_1$, upper row) and $h=-0.3$ ($P_2$, lower row). Here, $\kappa\equiv\tilde{\rho}_{\rm th}/\tilde{\rho}_{\rm ave}$, with $\kappa=3,5,10,20,30,40,$ and $50$. From left to right, the panels show the energy fraction contained in dense objects satisfying $\tilde{\rho}\geq\tilde{\rho}_{\rm th}$, the number of identified objects, the internal gradient energy fraction in these regions, and the corresponding decay times. In the last column, solid and dashed curves denote $\tau_{100}$ and $\tau_{50}$, respectively, while orange circles and blue squares represent results obtained from the total energy and internal gradient energy. For clarity, only the sparsely sampled outputs with $\Delta(mt)=50$ are shown; the densely sampled outputs at $mt<200$ are not included.}
	\label{threshold_dependence}
\end{figure*}

{ 
We first examine how the minimum-size and sphericity cuts affect the identified oscillon number. At each snapshot, voxels satisfying $\tilde{\rho}(\mathbf{x},t)\geq 10\tilde{\rho}_{\rm ave}(t)$ are grouped using 26-neighbor connectivity, including links across periodic faces, edges, and corners. A component is retained if it contains at least $n_{\rm vox}^{\min}$ voxels, its extent in each direction after periodic unwrapping is less than half the box size, and its sphericity satisfies
	\begin{equation}
		\psi_i=\frac{\pi^{1/3}(6V_i)^{2/3}}{A_i}\geq\psi_{\min},
	\end{equation}
	where $V_i$ is its voxel volume and $A_i$ is the surface area of its triangulated boundary. We impose no additional axis-ratio or temporal-persistence cut.
	
As an  example, we show the dependence of oscillon number on $\psi_{\min}$ and $n_{\rm vox}^{\min}$ in Fig.~\ref{N_osc_stats} for $h=0.4$. The counts for $10\leq n_{\rm vox}^{\min}\leq100$ nearly coincide. Setting $n_{\rm vox}^{\min}=1$ admits numerous single-voxel regions, particularly at late times, whereas $n_{\rm vox}^{\min}=500$ excludes additional smaller components. We therefore set $n_{\rm vox}^{\min}=15$. For $0.2\leq\psi_{\min}\leq 0.6$, the identified counts nearly coincide after the oscillon population has formed and stabilized ($mt\gtrsim200$). A larger $\psi_{\min}$ imposes a stricter morphology cut, excluding newly formed candidates that have not yet become nearly spherical and thereby reducing the count during formation. We therefore adopt the relatively conservative choice $\psi_{\min}=0.6$. We have checked the other values of $h$ and find the same post-formation agreement; the trend of fewer oscillons forming at larger $|h|$ also persists under these cuts.}

{ We next examine the sensitivity of oscillon identification to the density threshold.} The main results in this work are based on the density threshold $\tilde{\rho}_{\rm th}=10\tilde{\rho}_{\rm ave}$. Since an oscillon has a smooth spatial profile rather than a sharply defined boundary, its identified energy, size, and number are necessarily threshold-dependent operational quantities. To quantify this dependence, we define $\kappa=\tilde{\rho}_{\rm th}/\tilde{\rho}_{\rm ave}$ and perform a systematic scan over $\kappa=3,5,10,20,30,40,$ and $50$. All other geometrical and numerical identification criteria are kept unchanged.

Figure~\ref{threshold_dependence} presents representative results for $h=0.3$ and $h=-0.3$. As expected, the energy and internal gradient energy fractions decrease as $\kappa$ increases, since progressively smaller and denser regions are selected. The results at $\kappa=3$ and, to a lesser extent, $\kappa=5$ differ appreciably from those obtained at higher thresholds. These low thresholds include diffuse oscillon envelopes, surrounding overdensities, and low-density fragments. The number of identified objects is particularly sensitive to these contributions: lowering the threshold may either connect neighboring regions into a single component or count transient fragments as separate objects. For $\kappa\geq10$, the selected regions predominantly trace the compact oscillon cores, and their qualitative evolution becomes considerably more stable.

For each threshold, we determine the decay times using both the total energy and the internal gradient energy contained in the identified objects. Here, $\tau_{50}$ denotes the time after the energy maximum at which the corresponding energy has decreased to one half of its peak value, while $\tau_{100}$ denotes its complete disappearance within the numerical resolution. For $\kappa=3$ and $5$, the selected energy does not vanish completely because the density mask also captures numerous diffuse, low-density clumps. In these cases, $\tau_{100}$ is operationally assigned when the energy fraction settles onto its late-time plateau, indicating the complete decay of the compact oscillon component.
We find that the value of $\tau_{50}$ retains a moderate threshold dependence because the diffuse envelopes and compact cores decay at different rates. In contrast, $\tau_{100}$ is nearly insensitive to $\kappa$ for the representative models shown. The total energy and internal gradient energy diagnostics also give the same qualitative behavior. In particular, the oscillon lifetime for $h=-0.3$ remains shorter than that for $h=0.3$ throughout the threshold scan. Thus, although the absolute energy fraction and object number depend on the operational definition, the lifetime hierarchy and the main dependence on $h$ are robust. We therefore adopt $\kappa=10$ in the main analysis, as the lowest threshold that reliably isolates the compact cores while retaining sufficient spatial information.

\newpage
\twocolumngrid

\end{document}